\documentclass[twocolumn,amssymb, amsmath,nobibnotes, aps, prb, floatfix,superscriptaddress]{revtex4-1}
\usepackage{bm}%
\usepackage{graphicx}
\usepackage[toc,page]{appendix}
\usepackage{perpage}
\usepackage{dsfont}
\usepackage{amsmath,amssymb,mathrsfs}
\usepackage{latexsym}
\usepackage{graphicx} 
\usepackage{epstopdf}
\usepackage{tikz}
\usetikzlibrary{matrix}
\usepackage{graphicx,epstopdf,color}
\usepackage{amsfonts}
\usepackage{amsmath,amssymb,mathrsfs}
\usepackage{hyperref}
\usepackage{amsmath} 
\usepackage{bm}

\allowdisplaybreaks[1]

\begin{document}

\onecolumngrid
\author{Lo\"{i}c Henriet}
\affiliation{Centre de Physique Th\'{e}orique, \'{E}cole Polytechnique, CNRS, Universit\'{e} Paris-Saclay, F-91128 Palaiseau, France}
\author{Karyn Le Hur}
\affiliation{Centre de Physique Th\'{e}orique, \'{E}cole Polytechnique, CNRS, Universit\'{e} Paris-Saclay, F-91128 Palaiseau, France}

\title{Quantum sweeps, synchronization, and Kibble-Zurek physics in dissipative quantum spin systems}
\date\today

\begin{abstract}
We address dissipation effects on the non-equilibrium quantum dynamics of an ensemble of spins-1/2 coupled via an Ising interaction. Dissipation is modeled by a (ohmic) bath of harmonic oscillators at zero temperature and correspond either to the sound modes of a one-dimensional Bose-Einstein (quasi-)condensate or to the zero-point fluctuations of a long transmission line. We consider the dimer comprising two spins and the quantum Ising chain with long-range interactions, and develop a (mathematically and numerically) exact stochastic approach to address non-equilibrium protocols in the presence of an environment. For the two spin case, we first investigate the dissipative quantum phase transition induced by the environment through quantum quenches, and study the effect of the environment on the synchronization properties. Then, we address Landau-Zener-Stueckelberg-Majorana protocols for two spins, and for the spin array. In this latter case, we adopt a stochastic mean-field point of view and present a Kibble-Zurek type argument to account for interaction effects in the lattice. Such dissipative quantum spin arrays can be realized in ultra-cold atoms, trapped ions, mesoscopic systems, and are related to Kondo lattice models.
\end{abstract}

\maketitle

\section{Introduction}
Spin-boson models play a major role in various branches of physics, from condensed-matter physics, quantum optics, quantum dissipation, to quantum computation \cite{haroche,leggett,weiss,KLH}. A large collection of harmonic oscillators (bosons) can simulate dissipation, resulting in the celebrated Caldeira-Leggett model \citep{Caldeira_Leggett}, giving rise to dissipation-induced quantum phase transitions observed in various contexts \cite{Pierre,Finkelstein}. For example, a ohmic bosonic bath can be engineered through a long transmission line or a one-dimensional Luttinger liquid \cite{InesSaleur,KLH2}. An environment can also affect the critical exponents associated with a phase transition such as the disordered-ordered transition in the quantum Ising chain \citep{De_Gennes,Pfeuty,sachdev,Pankov,sachdev_werner_troyer,werner_volker_troyer_chakravarty}. \\

An impurity spin embedded in an environment also emerges as an effective model for strongly correlated quantum matter within dynamical mean-field theory\cite{DMFT}. The spin-boson model can be seen as a variant of the Caldeira-Leggett model where the quantum particle is a spin-1/2. The spin-boson model with an Ohmic bath exhibits a variety of rich phenomena such as a dissipative quantum phase transition separating an unpolarized (delocalized) and a polarized (localized) phase for the spin, as well as a coherent-incoherent crossover in the dynamical Rabi-type properties\cite{leggett,weiss}. This model is also intimately related to Ising models with long-range forces and to Kondo physics\cite{Anderson_Yuval_Hamann,Blume_Emery_Luther}. \\

Several theoretical methods have been devised to study the dissipative spin dynamics for one spin in an ohmic bath such as the non-interacting blip approximation\cite{leggett,weiss}, Quantum Monte Carlo (QMC) methods on the Keldysh contour\cite{Gull_Millis_Lichtenstein_Rubtsov,Schmidt_Werner_Muhlbacher_Komnik,Schiro,Millis}, and the time-dependent (TD) Numerical Renormalization Group (NRG) approach\cite{TDNRG,TDNRG_Anders_Schiller,TDNRG_1,Peter_two_spins}, with recent progress done concerning the treatment of driving and quenches \cite{TDNRG_quench}. Stochastic approaches have been developed both in the context of stochastic wavefunction approaches\cite{Dalibard_Castin_Molmer} or Stochastic Schr\"{o}dinger Equation (SSE) methods on the density matrix\cite{2010stoch,stochastic,Rabi_article}. Stochastic Liouville equations were obtained for the density matrix in Refs. \onlinecite{Stockburger_Mac,Stockburger,Stockburger_2,Schrodinger_langevin,Koch_morse}. \\

In this paper, we first consider a cluster of two spins in such a ohmic bosonic bath. The two spins are coupled through an Ising interaction. This model, which can be realized in ultra-cold atoms\cite{recati_fedichev,orth_stanic_lehur,Carlos_scientific_reports}, reveals a dissipative quantum phase transition similar to the one-spin situation, but occurring at a smaller dissipation strength\cite{Garst_Vojta,Peter_two_spins,sougato,Winter_Rieger}, which facilitates the application of numerical methods such as the SSE method in a large window of the phase diagram. Using the Rabi-type dynamics of the spin system, we reproduce the phase diagram obtained using the NRG approach\cite{Peter_two_spins} and QMC\cite{Winter_Rieger}, showing the trustability of the SSE method. We also compute spin-spin correlations induced by the bath at long time, and compare our results with those obtained with a variational approach\cite{sougato}. We quantitatively address the occurrence of synchronization between the two spins, in relation with the spin-spin correlation function. Then, we investigate non-equilibrium quenched dynamics far in the polarized phase, which has not been discussed previously in the literature, and also Landau-Zener-Stueckelberg-Majorana\cite{Landau,Zener,Stueckelberg,Majorana} type interferometry for the dimer model. Next, we consider a quantum Ising spin chain with long-range forces allowing a mean-field treatment for the spin dynamics. The main aspect we explore concerns the extension of Kibble-Zurek type physics\cite{Kibble,Zurek,dzarmaga,damski,review_KBZ} induced by magnetic field gradients in time (Landau-Zener sweeps) in the case of an interacting spin ensemble subject to dissipation. Applying the stochastic procedure as well as a physical argument, we describe the interplay between interactions between spins and dissipative effects from the bath on the well-known Landau-Zener formula\cite{Landau,Zener,Stueckelberg,Majorana}.We note that recent theoretical works have addressed similar questions regarding the effect of macroscopic dissipation on the dynamical properties of quantum spin arrays\cite{Garry_Camille,Clerk}. We also note recent experiments in ultra-cold atoms addressing Kibble-Zurek type physics \cite{Hadzibabic}.

\subsection{Model} 
Hereafter, we focus on a system of $M$ interacting spins (for the dimer $M=2$ and for a spin array $M\rightarrow +\infty$), which are coherently coupled to one common bath of harmonic oscillators:

\begin{align}
H =&\frac{\Delta}{2}\sum_{p=1}^M  \sigma_p^x+\sum_{p=1}^M \sum_{k} \lambda_{k} e^{ik x_p} \left(b^{\dagger}_{-k}+b_k \right) \frac{\sigma_p^z}{2} \notag \\
&-\frac{K}{M}\sum_{p \neq r}\sigma^z_p \sigma^z_{r}+\sum_{k} \omega_{k} b^{\dagger}_{k} b_{k}.
\label{ising_1}
\end{align}

Here,  $\sigma^\nu_p$ with $\nu=\{x,y,z\}$ are Pauli matrices related to the spatial site $p$ and the Planck constant $\hbar$ is set to unity. At each site, the states $|\pm_{z,p}\rangle$, corresponding to the two eigenstates of  $\sigma^z$ with eigenvalues $\pm1$, define the two possible orientations of the spin. The long-range ferromagnetic Ising interaction can be engineered in systems of trapped ions \citep{schaetz,Cirac_spinboson,monroe} and ultra-cold atoms \citep{recati_fedichev,orth_stanic_lehur,Scelle,Sortais}. It can also be the result of the Van der Waals interaction in Rydberg media \citep{rydberg_1,rydberg_2,rydberg_3}. This model can also be seen as an example of Kondo lattices in one dimension through bosonization \cite{Giamarchi} (for a review on Kondo lattices, see for example Ref. \onlinecite{Si}).

\subsection{Bath effects} 

The interaction with the bath plays an important role and affects both the equilibrium and the dynamical properties of the system. The spin-bath interaction is fully characterized by the spectral function $J(\omega)=\pi \sum_k \lambda_k^2 \delta(\omega-\omega_k)$, where we assume $\omega_k=v_s |k|$. Here, $v_s$ represents the velocity of the sound modes of a one-dimensional Bose-Einstein condensate or a long transmission line.
Hereafter, we shall focus on the case of ohmic dissipation at zero temperature, where the spectral function reads $J(\omega)=2 \pi \alpha \omega \exp \left(-\frac{\omega}{\omega_c}\right)$. Here, $\omega_c$ is a high energy cutoff and the dimensionless parameter $\alpha$ quantifies the strength of the interaction with the bath. These parameters can be derived microscopically for an ultra-cold atom setting \cite{recati_fedichev,orth_stanic_lehur,Carlos_scientific_reports}.

The bath induces both a renormalization of the tunneling element $\Delta$, and a strong Ising-type {\it ferromagnetic} interaction $K'_{|j-p|}$ between the spins $j$ and $p$, which is mediated by an exchange of bosonic excitations at low wave vectors\citep{orth_stanic_lehur}. This interaction is reminiscent of the Ruderman-Kittel-Kasuya-Yosida interaction for Kondo lattices \cite{RKKY}. The bosonic induced-coupling has been observed in light-matter systems\cite{Senellart,Majer,Kontos}, for example. This interaction can be exemplified by applying an exact unitary transformation $\tilde{H}=V^{-1}HV$ on the Hamiltonian (\ref{ising_1}), with $V=\exp \left\{\frac{1}{2}\sum_{k}\sum_{j=1}^M\sigma_j^z e^{ikx_j}\frac{\lambda_{k}}{\omega_{k}}   (b_k-b_{-k}^{\dagger}) \right\}$. The transformed Hamiltonian indeed reads:
\begin{align}
\tilde{H}=&\ \sum_{j=1}^M \frac{\Delta}{2} \left( \sigma^{+}_j e^{i \Omega_j}+\sigma^{-}_j e^{-i \Omega_j} \right)- \sum_{j \neq r} K^r_{|j-p|} \sigma^z_j \sigma^z_{r} \notag \\
+&\sum_{k} \omega_{k} b^{\dagger}_{k} b_{k},
\label{N_spins}
\end{align}
where $\Omega_j=i\sum_{k} \frac{\lambda_{k}}{\omega_{k}}  e^{ik x_j} (b_{k}-b_{-k}^{\dagger})$.  Note that $K^r_{|j-p|}=\frac{K}{M}+K'_{|j-p|}$ explicitly denotes the renormalized Ising coupling between the spins $j$ and $p$,  with
\begin{align}
K'_{|j-p|}=\frac{\alpha \omega_c}{2} \frac{1}{1+\frac{\omega_c^2 (x_j-x_p)^2}{v_s^2}}.
\label{renormalized_coupling}
\end{align}
The excitation of the spin $j$ comes with a simultaneous polarization of the neighboring bath into a coherent state $|\Omega_j\rangle=e^{i\Omega_j}|0\rangle$, resulting in a renormalization of the tunneling element. This argument can be made rigorous by an adiabatic renormalization procedure, developed in Refs. \onlinecite{leggett,weiss}. In the regime $\Delta/\omega_c \ll 1$, one can indeed assume that the high frequency modes of the bath (above a given frequency $\omega_l(\Delta)$ corresponding to several units of $\Delta$) adjust instantaneously to the value of the spin. The tunneling element is then dressed by the bath, and is renormalized to $\tilde{\Delta}<\Delta$. This procedure can be iterated and converges in the ohmic case and for $\alpha<1$, to a renormalized value of the bare tunneling element $\Delta$ to $\Delta_r=\Delta (\Delta/\omega_c)^{\alpha/(1-\alpha)}$. This result sheds light on the mechanism at the origin of the dissipative quantum phase transition induced by the bath\cite{Garst_Vojta,Peter_two_spins,Winter_Rieger}. At strong coupling, the bath entirely polarizes the spins, by analogy to a ferromagnetic phase. For one spin, the quantum phase transition belongs to the Kosterlitz-Thouless class, where the order parameter at equilibrium $\langle \sigma^z_j \rangle$ exhibits a jump\cite{KLH}. In this case, the critical value $\alpha_c$ of the coupling is $\alpha_c=1$. The universality class of the transition is unchanged when the number $M$ of spins is increased (and remains finite) \cite{Winter_Rieger}, and the associated critical value $\alpha_c$ decreases with (finite) $M$  \cite{sougato,Winter_Rieger}, due to the strong ferromagnetic interaction between the spins induced by the bath. The case $M=2$ was systematically studied in Ref. \onlinecite{Peter_two_spins}.

The rest of the paper is organized as follows. In Sec. II, we summarize the general methodology used to compute the spin dynamics, in the case of one single spin coupled to bosonic degrees of freedom. These groundings will allow us to expose the extension of the methology to the case of two spins ($M=2$) in Sec. III. We will then investigate the quantum phase transition displayed by the two-spin system and present several results concerning the spin dynamics both in the unpolarized and in the polarized phase. We find that the spin dynamics in the polarized phase exhibits an universal behaviour, in the sense that it becomes independent of the coupling strength $\alpha$. Then we study the effect of the bath on the synchronization properties of the two spins in relation with spin-spin correlation functions. We also present Landau-Zener-Stueckelberg-Majorana interferometry \cite{Landau,Zener,Stueckelberg,Majorana} protocols using the interaction mediated by the bath. In Sec. IV, we extend the methodology to the case of an infinite array ($M \to \infty$) at a mean-field level. We present results concerning the dynamics as well as Landau-Zener sweeps. In this case, we apply a Kibble-Zurek type argument to account for the mean-field dynamics. Finally, Appendices will be devoted to some mathematical derivations. 

\section{Methodology for spin dynamics} 

In this Section, we re-derive the real-time spin dynamics in the case of $M=1$ spin and introduce the notations that will be used in the next sections. All the developments are based on different steps related to Refs. \onlinecite{FV,leggett,weiss,2010stoch,stochastic,Rabi_article}, which will be exposed in detail below.


\subsection{Feynman-Vernon influence functional}
The original reference for this technique introduced by Feynman and Vernon is Ref. \onlinecite{FV}. 

To compute the dynamics of the spin in contact with the bosonic environment, we focus on the different elements of the spin reduced density matrix. Let $\{ |\sigma\rangle\}=\left\{|+_{z}\rangle,|-_{z}\rangle \right\}$ be a basis of the Hilbert spin state $\epsilon_S$ and $\{|u_{n}\rangle\}$ be a basis of the bath Hilbert space $\epsilon_B$. The total density matrix of the system is denoted by $\rho$, and $\rho_S$ is the spin reduced density matrix. More precisely, $\rho_S$ is the partial trace of the total density matrix over the bosonic degrees of freedom. The evolution of the total density matrix can be expressed with the unitary time-evolution operator of the whole system $U$. At a given time $t$, the elements of the spin reduced density matrix read

\begin{align}
\langle \sigma_f | \rho_S (t) | \sigma_f'\rangle&=\sum_{n}   \langle  u_{n}, \sigma_f | U(t) \rho(t_0) U^{\dagger}(t) |  u_{n},\sigma_f '\rangle.
\label{eq:densitymatrix}
\end{align}
We have $|\sigma_f\rangle, |\sigma_f'\rangle  \in \{ |+_{z}\rangle,|-_{z}\rangle\}$. Next, we express the propagators thanks to a path-integral description, but we need another hypothesis in order to go further in the calculations: we assume that spin and bath are uncoupled at the initial time $t_{0}$ when they are brought into contact, so that the total density matrix can be factorized, $\rho (t_0)=\rho_B(t_0)\otimes \rho_S(t_0)$. For the remaining of the article, we will assume such factorising initial conditions, but the Feynman-Vernon influence functionnal approach can be generalized for a general initial condition, as shown in Refs. \onlinecite{Grabert_Schramm_Ingold,weiss}. The initial state of the bath will always be a thermal state at inverse temperature $\beta$. We start with the spin initially in the state $|+_z\rangle$ so that 
\begin{equation}
\rho_{S} (t_{0})=| +_{z} \rangle\langle +_{z} |=\left(\begin{array}{cc}
1 & 0   \\
0 & 0  \end{array} \right).
\label{eq:initial_condition}
\end{equation}

 The time-evolution of the spin reduced density matrix can be then re-expressed as, 
\begin{equation}
\langle \sigma_f | \rho_S (t) | \sigma_f'\rangle=  \int D\sigma D\sigma' \mathcal{A} [\sigma]  \mathcal{A}^* [\sigma'] \mathcal{F}_{[\sigma, \sigma']}.
\label{eq:densitymatrixelement}
\end{equation}
The integration runs over all spin paths $\sigma$ and $\sigma'$ such that $|\sigma(t_0)\rangle=|\sigma'(t_0)\rangle=|+_z\rangle$, $|\sigma(t)\rangle=|\sigma_f\rangle$ and $|\sigma'(t)\rangle=|\sigma_f'\rangle$. The term $\mathcal{A} [\sigma]$ denotes the amplitude to follow one given spin path $\sigma$ in the sole presence of the transverse field term in Eq. (1). The effect of the environment is fully contained in the so-called Feynman-Vernon influence functional $\mathcal{F}_{[\sigma, \sigma']}$ which reads\cite{FV,weiss}:

\begin{widetext}
\begin{equation}
\mathcal{F}[\sigma,\sigma']=\exp \left\{-\frac{1}{\pi} \int_{t_0}^t ds \int_{t_0}^s ds'\left[-i L_1(s-s')\frac{ \sigma (s)-\sigma '(s) }{2} \frac{ \sigma (s')+\sigma '(s') }{2} +L_2(s-s')\frac{\sigma (s)-\sigma'(s) }{2} \frac{ \sigma (s')-\sigma'(s')}{2}\right] \right\},
\label{eq:influence}
\end{equation}
\end{widetext}
where a spin path jumps back and forth between the two values $\sigma(s)=\pm 1$. The functions $L_1$ and $L_2$ read
\begin{align} &L_1(t)=\int_0^{\infty} d \omega J(\omega) \sin \omega t ,  \notag \\
&L_2(t)=\int_0^{\infty} d \omega J(\omega) \cos \omega t \coth \frac{\beta \omega}{2}.
\label{Ls_2}
\end{align}
For an ohmic bath in the zero-temperature limit $\beta\rightarrow +\infty$, the functions $L_1$ and $L_2$ explicitly read,
\begin{align} &L_1(t)= 4\pi\alpha \omega_c^2 \frac{\omega_c t}{(1+\omega_c^2t^2)^2} \notag \\
&L_2(t)=2\pi\alpha \omega_c^2 \frac{1-\omega_c^2 t^2}{(1+\omega_c^2t^2)^2}.
\label{Ls_3}
\end{align}
A derivation of Eq.~(\ref{eq:influence}) is done in the Appendix A. 

From Eq. (\ref{eq:influence}), we see that the bosonic environment couples the symmetric and anti-symmetric spin paths $\eta(t)=1/2[\sigma(t)+\sigma'(t)] $ and $ \xi(t)=1/2[\sigma(t)-\sigma'(t)]$ at different times. These spin variables take values in $\{-1,0,+1\}$ and are the equivalent of the classical and quantum variables in the Schwinger-Keldysh representation. We have then integrated out the bosonic degrees of freedom, which no longer appear in the expression of the spin dynamics, but the prize to pay is the introduction of a spin-spin interaction term which is not local in time. This long range interaction in time is reminiscent of the quantum Ising model with long range forces\citep{Anderson_Yuval_Hamann,Blume_Emery_Luther} in $1/r^2$. Dealing with such terms is difficult at a general level. The spin dynamics at a given time $t$ depends on its state at previous times $s<t$: the dynamics is said to be non-Markovian.

\subsection{ ``Blips'' and ``Sojourns''}

The next step is the rewriting of the spin path in the language of ``Blips'' and ``Sojourns'', following the work of Ref. \onlinecite{leggett}. 

\begin{figure}[t!]
\center
\includegraphics[scale=0.30]{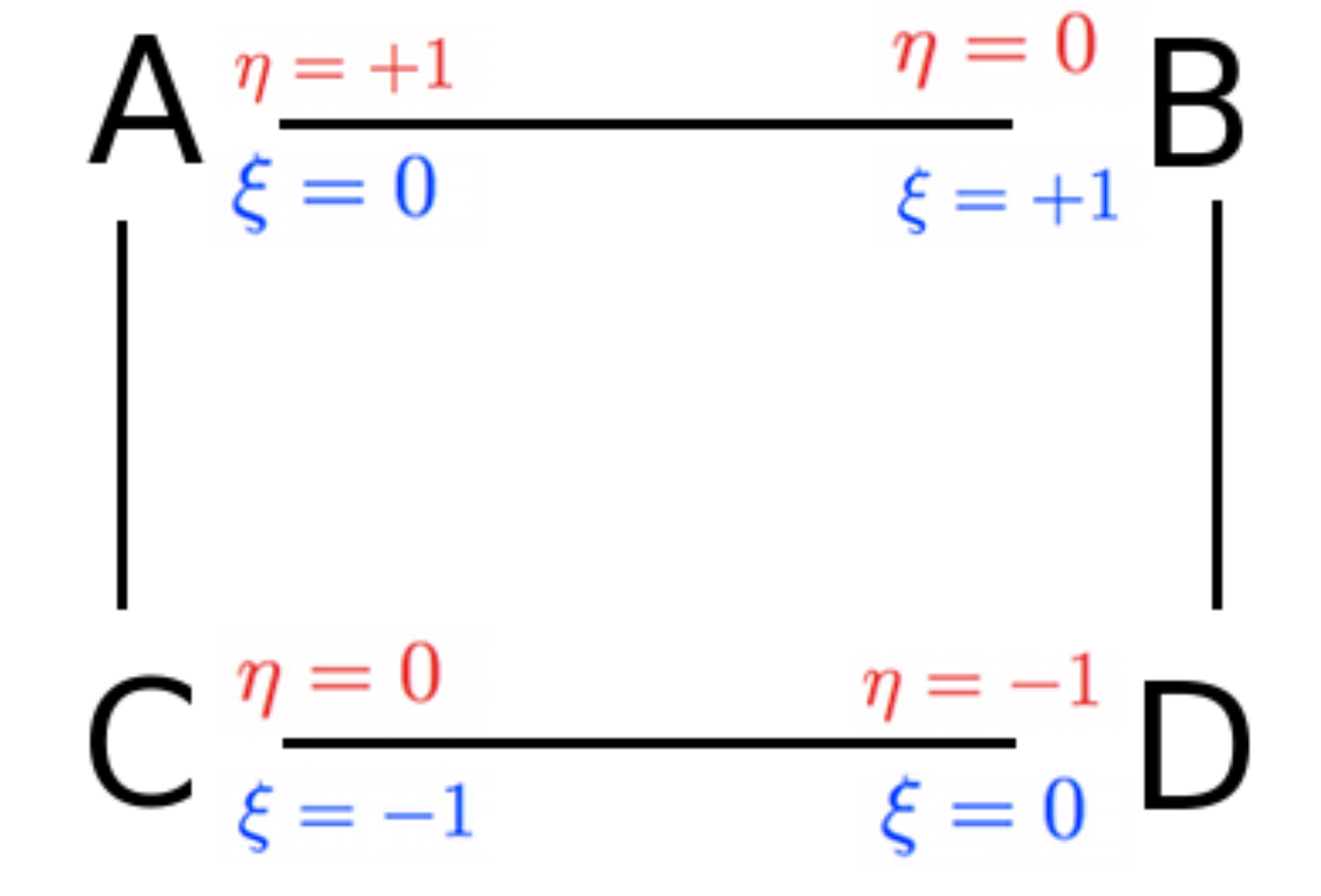}  
\caption{(Color online) Spin states.}
\label{etats}
\end{figure}

The double path integral  in Eq. (\ref{eq:densitymatrixelement}) can be viewed as one single path that visits the four states A (for which $\eta=1$ and $\xi=0$), B (for which $\eta=0$ and $\xi=1$), C (for which $\eta=0$ and $\xi=-1$) and D (for which $\eta=-1$ and $\xi=0$). States A and D correspond to the diagonal elements of the density matrix (also named `sojourn' states) whereas B and C correspond to the off-diagonal ones (also called `blip' states) \cite{leggett,weiss}. The four states are depicted in Fig.~\ref{etats}. 

\begin{figure}[t!]
\center
\includegraphics[scale=1.05]{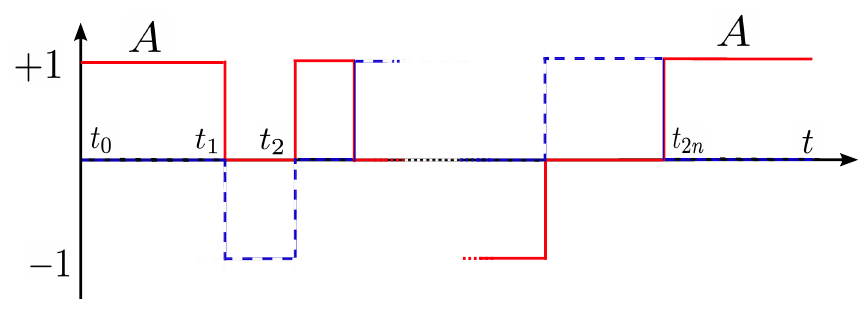}  
\caption{(Color online) Spin path- $\eta(t)=\sum_{j=0}^{2n} \Upsilon_j\theta(t-t_j)$ in red; $\xi(t)=\sum_{j=1}^{2n} \Xi_j\theta(t-t_j)$ in dashed blue.}
\label{spin_path_1}
\end{figure}

As stated previously, the spin is initially in the state $|+_z\rangle$, so that the double spin path is initially constrained in the diagonal state A, which can be seen as the element top left element of the spin density matrix. We will first focus on the computation of the upper left diagonal element of the density matrix, describing the probability
\begin{align}
p_0(t)=\langle +_z| \rho_S(t) |+_z\rangle =(1+\langle \sigma^z (t) \rangle) /2,
\end{align}
 to find back the system in the state $|+_z\rangle $ at time $t$. We consider then spin paths that end in the sojourn state A. Such a path makes $2n$ transitions along the way at times $t_i$, $i \in \{1,2,..,2n\}$ with $t_0<t_1<t_2<...<t_{2n}$. We can write this spin path as $\xi(t)=\sum_{j=1}^{2n} \Xi_j\theta(t-t_j)$ and $\eta(t)=\sum_{j=0}^{2n} \Upsilon_j\theta(t-t_j)$ where the variables $\Xi_i$ and $\Upsilon_i$ take values in $\{-1,1\}$. Such a path is visualised in Fig.~\ref{spin_path_1}. The variables $\Xi$ (in blue) describe the blip parts, and the variables $\Upsilon$ (in red) on the other hand characterize the sojourn parts. 

After the introduction of these variables, $p_0$ can be expressed as a series in $\Delta^2$, as shown in Refs. \onlinecite{leggett,weiss} :
\begin{equation}
p_0(t)=\sum_{n=0}^{\infty} \left(\frac{i\Delta}{2} \right)^{2n} \int_{t_0}^{t} dt_{2n} ... \int_{t_0}^{t_2} dt_{1} \sum_{\{\Xi_j\},\{\Upsilon_j\}' } \mathcal{F}_{n}.
\label{eq:p(t)_1}
\end{equation}
 The prime in $\{\Upsilon_j\}'$ in Eq. (\ref{eq:p(t)_1}) indicates that the initial and final sojourn states are fixed according to the initial and final conditions. More precisely we have $\Upsilon_0=\Upsilon_{2n}=1$. The influence functional reads: 
 
 \begin{align}
 &\mathcal{F}_{n}= \mathcal{Q}_1 \mathcal{Q}_2, \label{10}
  \end{align}
 with
  \begin{align}
  &\mathcal{Q}_1 =\exp \left[ \frac{i}{\pi} \sum_{k=0}^{2n-1}\sum_{j=k+1}^{2n} \Xi_j \Upsilon_k  Q_1(t_j-t_k) \right] \label{Q_1} \\
  &\mathcal{Q}_2 =\exp \left[ \frac{1}{\pi} \sum_{k=1}^{2n-1}\sum_{j=k+1}^{2n} \Xi_j \Xi_k  Q_2(t_j-t_k) \right].\label{Q_2} 
 \end{align}
 
The functions $Q_1$ and $Q_2$, which describe the feedbacks of the dissipative environment, are directly obtained from the spectral function $J(\omega)$ (they are second integrals of the $L_1$ and $L_2$ functions). At zero temperature, we have:
\begin{align}
 Q_1(t)&=\int_0^{\infty} d\omega\frac{J(\omega)}{\omega^2}\sin \omega t, \label{q1} \\
  Q_2(t)&=\int_0^{\infty} d\omega\frac{J(\omega)}{\omega^2}\left(1-\cos \omega t\right).\label{q2}
\end{align} 
For a ohmic spectral density at zero temperature, we have
\begin{align}
 Q_1(t)&=2 \pi  \alpha \tan^{-1} (\omega_c t),  \label{q1_explicit} \\
  Q_2(t)&=\pi \alpha \log (1+\omega_c^2 t^2). \label{q2_explicit}
\end{align} 

From Eq.~(\ref{Q_1}) and Eq.~(\ref{Q_2}), we see that the term $\mathcal{Q}_1$ couples the blips to all the previous sojourns, while $\mathcal{Q}_2$ couples the blips to all the previous blips (including self-interaction). A derivation of these expressions is provided in the Appendix B. 

\subsection{Stochastic decoupling}

At this point, the main difficulty is to treat the long range correlation in time induced by the bath in the quantum limit. The Non Interacting Blip Approximation (NIBA) greatly simplifies the problem and permits to compute some dynamical quantities, but does not allow to investigate the strong coupling regime or the treatment of driving terms\cite{leggett,weiss}. To decouple the spin-spin interaction(s) in time, we use Hubbard-Stratonovitch
variables following previous works from collaborators and us \cite{stochastic,Rabi_article}. Some efforts in this direction were also done in Refs. \onlinecite{Stockburger_Mac,Stockburger,Stockburger_2}. This stochastic unravelling of the influence functional will allow us to write the dynamics of the spin-reduced density matrix as a solution of a stochastic differential equation. Let $h$ and $k$ be two complex gaussian random fields which verify\cite{Rabi_article}
\begin{align}
 \overline{ h(t) h(s)} = & \frac{1}{\pi} Q_2(t-s) + l_1, \label{height_1} \\
 \overline{ k(t) k(s)} = &\  l_2,    \label{height_2}   \\
 \overline{ h(t) k(s) } = & \frac{i}{\pi}  Q_1(t-s) \theta(t-s) + l_3. \label{height_3}
\end{align}
The overline denotes statistical average, $\theta(.)$ is the Heaviside step function and $l_1$, $l_2$ and $l_3$ are arbitrary complex constants. Making use of the identity $\overline{\exp(X)}=\exp(\overline{X^2}/2 )$ , Eqs. (\ref{10}), (\ref{Q_1}) and (\ref{Q_2}) can then be reexpressed as:
 \begin{align}
\mathcal{F}_{n}= \overline{  \prod_{j=1}^{2n} \exp\left[h(t_j) \Xi_j+k(t_{j-1}) \Upsilon_{j-1}    \right]}.
\label{functionnal_1}
\end{align}
The complex constants $l_p$ do not contribute to the average because $\sum_{k=0}^{2n-1} \Upsilon_k=\sum_{j=1}^{2n} \Xi_j=0$. This step was done in Refs. \onlinecite{2010stoch,stochastic} with the introduction of one stochastic field (which is valid in a certain limit, as we will see later), and with two fields in Ref. \onlinecite{Rabi_article}. The summation over blips and sojourn variables $\{\Xi_j\}$ and $\{\Upsilon_j\}$ can be incorporated by considering a product of matrices of the form 

\begin{equation}
V_{0}= \left( \begin{array}{cccc}
0&e^{-h+k }&-e^{h+k }&0 \\
e^{h-k }&0&0&-e^{h+k }\\
-e^{-h-k }&0&0&e^{-h+k }\\
0&-e^{-h-k }&e^{h-k }&0
\end{array} \right),
\label{eq:spin_hamiltonian}
\end{equation}
in the four dimensional vector space of states $\{\textrm{A},\textrm{B},\textrm{C},\textrm{D}\}$. This rewriting was originally introduced in Ref. \onlinecite{Lesovik}. Then, we get
\begin{equation}
p_0(t)=\overline{\sum_{n=0}^{\infty} \left(\frac{i\Delta}{2} \right)^{2n} \int_{t_0}^{t} dt_{2n} ... \int_{t_0}^{t_2} dt_{1} \prod_{j=1}^{2n} V_{0}(t_j)}.
\label{eq:p(t)_12}
\end{equation}

We remark that Eq. (\ref{eq:p(t)_12}) has the form of a time-ordered exponential, averaged over stochastic variables, so that we finally have:
\begin{equation} 
p_0(t)=\overline{\langle \Phi_f | \Phi (t) \rangle},
\label{scalar_prod_SSE}
\end{equation}
where $\langle \Phi_f |=(e^{-k(t_{2n})},0,0,0)$ and  $| \Phi \rangle$ is the solution of the Stochastic Schr\"{o}dinger Equation (SSE),

\begin{equation} 
i \partial_t | \Phi \rangle = V_0 (t) | \Phi \rangle
\label{SSE}
\end{equation}
with initial condition $|\Phi_i \rangle=(e^{k(t_0)},0,0,0)^T$.\\

The vector $|\Phi(t) \rangle$ represents the double spin state which characterizes the spin density matrix. The vectors $|\Phi_i \rangle$ and $|\Phi_f \rangle$ are related to the initial and final conditions of the paths. As spin paths start and end in the sojourn state A, only the first component of these vectors is non-zero. The choice of the phases is linked to the asymmetry between blips and sojourns (see Eq.~(\ref{Q_1}) and Eq.~(\ref{Q_2})). The contribution from the first sojourn is encoded in $|\Phi_i \rangle$, and we artificially suppress the contribution of the last sojourn via $|\Phi_f \rangle$. This final vector depends on an intermediate time, but we can notice that replacing $(e^{-k(t_{2n})},0,0,0)$ by $(e^{-k(t)},0,0,0)$ does not add any contribution on average. The numerical procedure requires a large number of realizations of the fields $h$ and $k$. For each realization, we solve the stochastic equation and $\langle \sigma^z(t) \rangle$ is obtained by averaging over the results of all the realizations. In general we use Fourier series decomposition for the sampling of the fields $h$ and $k$. Details about the sampling can be found in the Appendix C. 

This framework refers to as the Stochastic Schr\"{o}dinger Equation (SSE) method.

It is also possible to incorporate driving effects in the framework of the SSE method in an exact manner. In the present article, we will consider specifically a driving term acting on the spin, which reads $\epsilon(t) \sigma^z$. From the path integral approach and the blip-sojourn decomposition, we see that this effect can be incorporated by adding a deterministic part to the stochastic field $h$ of the SSE. The resulting field $h^d$ reads\cite{2010stoch,stochastic}
\begin{equation} 
h^d(t)=h(t)+\int_{t_0}^t ds \epsilon(s).
\end{equation}

\subsection{Previous results and discussion}

As can be seen in Eq. (\ref{eq:spin_hamiltonian}), the effective Hamiltonian for the spin density matrix is not Hermitian (in general $h$ and $k$ have both a real and an imaginary part). The complexness of $h$ and $k$ may give rise to numerical convergence problems at a general level, and accessing the regime of strong coupling between spin and bath requires a special attention on this issue. For the ohmic spin-boson model, simplifications occur in the regime $\Delta / \omega_c \ll 1$, as shown in Refs. \onlinecite{2010stoch,stochastic}. In this regime the function $Q_1$ in Eq. (\ref{q1_explicit}) can be considered as a constant ($\tan^{-1} (\omega_c t)\simeq \pi/2$), allowing us to use only one stochastic field  $h$ which is purely imaginary. As presented in the references mentioned above, the SSE method then leads to a correct prediction of the dynamical behavior for $0<\alpha <1/2$.

The method with two stochastic fields was then used to compute the dynamics of the driven dissipative Rabi model\cite{Rabi_article}, for which the use of complex fields were not problematic. We could in particular access large values of the light-matter coupling. In the previous subsections, we focused on the computation of the top-left diagonal element of the density matrix $\langle +_z| \rho_S(t) |+_z\rangle $, with the initial condition $\rho_{S} (t_{0})=| +_{z} \rangle\langle +_{z} |$ given by Eq. (\ref{eq:initial_condition}). It is possible to either compute off-diagonal elements of the density matrix or consider another initial state for the spin in the framework of this method, by considering other initial and final vectors $| \Phi_i \rangle$ and $| \Phi_f \rangle$. Such developments are presented in the subsections B and C of the Sec. II of Ref. \onlinecite{Rabi_article}. It is also possible to incorporate driving effects on the bosonic degrees of freedom, as shown in the subsection D of the Sec. II of Ref. \onlinecite{Rabi_article}.\\

Some authors did not express the spin paths in the language of blips and sojourns, but rather reached an effective stochastic Liouville equation for the density matrix, see Refs. \onlinecite{Stockburger_Mac,Stockburger,Stockburger_2,Schrodinger_langevin}. This technique has notably been used to compute the dynamics for the Morse oscillator\cite{Koch_morse}. Non-Markovian master equations\cite{Tu_Zhang,Zhang_Nori} were derived thanks to the same Feynman-Vernon influence functional starting point. A review of the different path-integral methods developped to tackle the non-Markovian dynamics in spin-bath systems is provided in Ref.~\onlinecite{de_Vega_review}.\\

Next, we go further and present other applications of the method to the case of two spins (Sec. III), and the case of the array (Sec. IV). Several applications we will focus on have not been yet addressed in the literature
using an alternative approach.

\section{Two spins}

In this Section, we focus on the case of $M=2$ spins. In this case, it is possible to reach an exact linear stochastic differential equation describing the dynamics of the spin reduced density matrix, as shown in the subsection A below. In this case, the spin-reduced density matrix has a dimension $16$ and it is possible to develop the same formalism as in the one spin case, in an exact manner. The case of two spins is particularly interesting as the quantum phase transition from the unpolarized phase to the polarized phase occurs for a smaller value of $\alpha$ \cite{Peter_two_spins}. While the quantum phase transition was not accessible with the SSE method in the case of one spin ($\alpha_c=1$), it will be possible to investigate this regime for two spins ($\alpha_c \simeq 0.2$), as shown in the subsection B. Synchronization is studied in the subsection C. We finally investigate Landau-Zener-Stueckelberg-Majorana protocols in the subsection D.

\subsection{Exact method for two spins}

For two spins, we will neglect the spatial separation between sites $x_1=x_2=0$. We proceed as in the one-spin case and follow the steps exposed in Sec. II. As before, the two spins initially in the state $|+_z\rangle$ so that $\rho_S (t_0)=|+_z,+_z\rangle \langle +_z +_z |$. The time-evolution of a given element $x=\langle \sigma_{1,f},\sigma_{2,f} | \rho_S (t) | \sigma_{1,f}',\sigma_{1,f}'\rangle$ of the spin reduced density matrix can be then re-expressed as, 

\begin{align}
x=\int \prod_{p=1}^2 \left( D\sigma_p D\sigma_p'\right) \prod_{p=1}^2 \left(\mathcal{A}[\sigma_p] \mathcal{A}^* [\sigma_p']\right) \mathcal{F}_{[\sigma_1,\sigma_2,\sigma_1',\sigma_2']}\notag \\
\times \exp\left\{i \int_{t_0}^{t} ds K \left[ \sigma_1(s) \sigma_2(s)- \sigma_1'(s) \sigma_2'(s)\right]\right\}.
\label{eq:densitymatrixelement_two_spins}
\end{align}

The integration runs over all spin paths $\sigma_1$, $\sigma_2$, $\sigma_1'$ and $\sigma_2'$ such that $|\sigma_p(t_0)\rangle=|\sigma_p'(t_0)\rangle=|+_z\rangle$, $|\sigma_p(t)\rangle=|\sigma_{f,p}\rangle$ and $|\sigma_p'(t)\rangle=|\sigma_{f,p}'\rangle$. As in the one-spin case, the terms of the form $\mathcal{A} [\sigma_p]$ denote the amplitude to follow one given spin path $\sigma_p$ in the sole presence of the transverse field term acting on the spin $p$. The last term of the right hand side of Eq. ~(\ref{eq:densitymatrixelement_two_spins}) comes from the Ising interaction between the two spins. The influence functional $\mathcal{F}_{[\sigma_1,\sigma_2,\sigma_1',\sigma_2']}$ reads :

\begin{widetext}
\begin{equation}
\mathcal{F}_{[\sigma_1,\sigma_2,\sigma_1',\sigma_2']}=e^{-\frac{1}{\pi} \int_{t_0}^t ds \int_{t_0}^s ds' \sum_{i,j=1}^2\left\{-i L_1(s-s')\frac{ \sigma_i (s)-\sigma_i '(s) }{2} \frac{ \sigma_j (s')+\sigma_j '(s') }{2} +L_2(s-s')\frac{\sigma_i (s)-\sigma_i '(s) }{2} \frac{ \sigma_j (s')-\sigma_j '(s')}{2}\right\}}\times \mathcal{G}[\sigma_1,\sigma_2,\sigma_1',\sigma_2'].
\label{eq:influence_two_spins}
\end{equation}
\end{widetext}

The additional term $\mathcal{G}$ in Eq. (\ref{eq:influence_two_spins}) reads : 

\begin{align}
\mathcal{G}[\sigma_1,\sigma_2,\sigma_1',\sigma_2'] &=& e^{ i \frac{\mu}{2} \int_{t_0}^t ds \left[\sum_{j=1}^2 \frac{\sigma_j (s)}{2} \right]^2-\left[\sum_{j=1}^2 \frac{\sigma_j ' (s)}{2} \right]^2},
\label{eq:influence_3_two_spins}
\end{align}
with  $\mu=2/\pi \int_0^{\infty} J(\omega)/\omega=4 \alpha \omega_c$. We recover in Eq.~(\ref{eq:influence_3_two_spins}) that the bath renormalizes the direct Ising interaction between the spins. The term above is indeed similar to the one coming from the direct Ising interaction $K$ (last term of the right hand side of Eq. ~(\ref{eq:densitymatrixelement_two_spins})). In the following we gather these two contributions in a functionnal $\tilde{\mathcal{G}}$ which reads
\begin{align}
\tilde{\mathcal{G}}[\sigma_1,\sigma_2,\sigma_1',\sigma_2'] &=& e^{i \int_{t_0}^{t} ds K_r \left[ \sigma_1(s) \sigma_2(s)- \sigma_1'(s) \sigma_2'(s)\right]},
\end{align}
with $K_r=K+\alpha \omega_c$ the renormalized Ising interaction. 

\begin{figure}[t!]
\center
\includegraphics[scale=1.05]{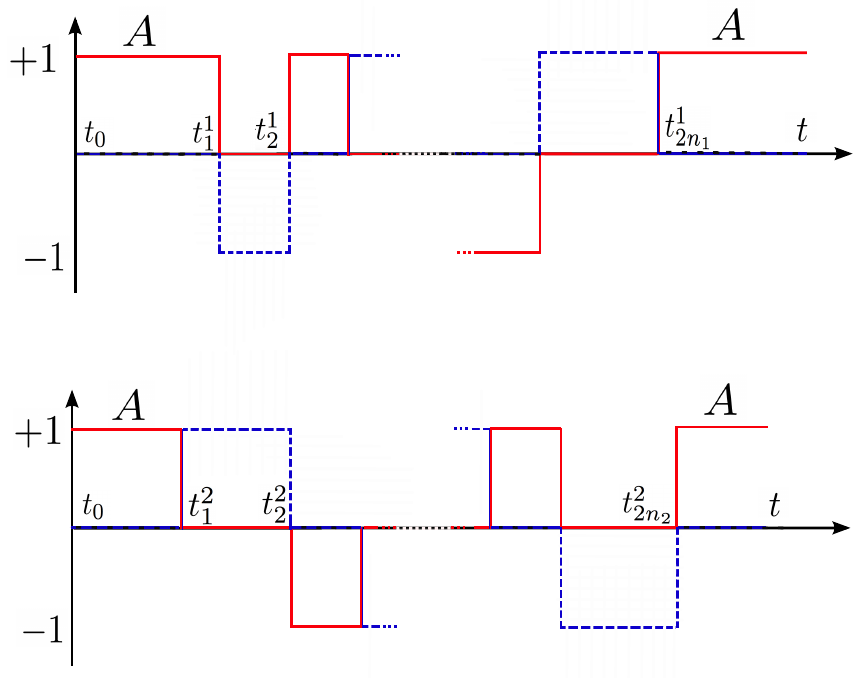}  
\caption{(Color online) Spin path for the dimer problem- The upper part shows the spin path in terms of blips and sojourns for the first spin, while the lower part shows the spin path of the second spin. $\eta^p(t)=\sum_{j=0}^{2n} \Upsilon_j^p\theta(t-t_j)$ in red; $\xi^p(t)=\sum_{j=1}^{2n} \Xi^p_j\theta(t-t_j)$ in dashed blue. The system starts in the state AA, jumps to the state AB at $s_1=t_1^2$, then to the state CB at $s_2=t_1^1$. It finally ends in the state AA at $t$.}
\label{spin_path_2}
\end{figure}

The paths introduced in Eq. (\ref{eq:densitymatrixelement_two_spins}) can be viewed as one single path that visits the sixteen states corresponding to the matrix elements of the spin-reduced density-matrix. We will note $\mathcal{E}=\{$AA, AB, AC, AD, BA, BB, BC, BD, CA, CB, CC, CD, DA, DB, DC, DD$\}$ the set of these states - the states A, B, C and D have been defined for one spin in Sec. II. The four states AA, AD, DA and DD correspond to the diagonal elements of the canonical density matrix, while other states correspond to off-diagonal elements. As before, we consider the case where the spin subsystem starts in the state $| +_z,+_z\rangle$, and we intend to compute the probability 
\begin{align} 
p_1(t)=\langle +_z,+_z | \rho_S (t) | +_z,+_z\rangle,
\end{align}
 to come back in the same state $| +_z,+_z \rangle$ at time $t$. Then, both the first and the second spin path make an even number of transitions along the way at times $t^p_{j}$, $j \in \{1,2,..,2n_p\}$ for $p \in \{1,2\}$ such that $t_{0}<t^p_1<t^p_2<...<t^p_{2n_p}<t$. We can write these spin paths as $\xi^p(t)=\sum_{j=1}^{2n_p} \Xi^p_j\theta(t-t^p_j)$ and $\eta^p(t)=\sum_{j=0}^{2n_p} \Upsilon^p_j\theta(t-t^p_j)$ where the variables $\Xi^p_j$ and $\Upsilon^p_j$ take values in $\{-1,1\}$. Such a path can be visualized in Fig.~\ref{spin_path_2} as a couple of one-spin paths. \newline

The probability $p_1(t)$ is given by a series in $\Delta^2$:

\begin{equation}
p_1(t)=\sum_{\substack{n_1,n_2\\
                  \{\Xi^p_j\},\{\Upsilon^p_j\}'}}\left(\frac{i\Delta}{2} \right)^{2N}  \int_{s_0}^{t} ds_{2N} .. \int_{s_0}^{s_2} ds_{1}  \mathcal{F}_{n_1,n_2},
\label{eq:p(t)_two_spins}
\end{equation}
where $N=n_1+n_2$ and $\{s_0, s_1, ..., s_{2(n_1+n_2)}\}$ is the ordered reunion of the two sequences $\{t^1_j\}$ and $\{t^2_j\}$. The summation over $n_1$ and $n_2$ goes from $0$ to infinity. The prime in $\{\Upsilon^p_j\}'$ in Eq. (\ref{eq:p(t)_two_spins}) indicates that the initial and final states are fixed according to $\Upsilon^1_0=\Upsilon^2_0=\Upsilon^1_{2n_1}=\Upsilon^2_{2n_2}=1$. The influence functional can be written explicitely in terms of $\Xi^p_j$ and $\Upsilon^p_j$ variables:

  \begin{align}
 &\mathcal{F}_{n_1,n_2}= \left( \prod_{p=1}^2  \mathcal{Q}^p_1 \mathcal{Q}^p_2 \mathcal{M}^p_1 \mathcal{M}^p_2 \right) \tilde{\mathcal{G}}[\sigma_1,\sigma_2,\sigma_1',\sigma_2'] \label{10_two_spins},
   \end{align}
   with
   \begin{align}
  &\mathcal{Q}^p_1 =\exp \left[ \frac{i}{\pi} \sum_{k=0}^{2n_p-1}\sum_{j=k+1}^{2n_p} \Xi^p_j \Upsilon^p_k  Q_1(t^p_j-t^p_k) \right], \label{Q_1_2} \\
  &\mathcal{Q}^p_2 =\exp \left[ \frac{1}{\pi} \sum_{k=1}^{2n_p-1}\sum_{j=k+1}^{2n_p} \Xi^p_j \Xi^p_k  Q_2(t^p_j-t^p_k) \right],\label{Q_2_2}   \\
   &\mathcal{M}^p_1 =\exp \left[ \frac{i}{\pi} \sum_{k=0}^{2n_{\overline{p}}-1}\sum_{j: t^p_j>t^{\overline{p}}_k} \Xi^p_j \Upsilon^{\overline{p}}_k Q_1(t^p_j-t^{\overline{p}}_k) \right], \label{Q_1_m_2} \\
  &\mathcal{M}^p_2 =\exp \left[ \frac{1}{\pi} \sum_{k=1}^{2n_{\overline{p}}-1}\sum_{j: t^p_j>t^{\overline{p}}_k} \Xi^p_j \Xi^{\overline{p}}_k  Q_2(t^p_j-t^{\overline{p}}_k) \right].\label{Q_2_m_2}
 \end{align}

In Eqs. (\ref{Q_1_m_2}) and (\ref{Q_2_m_2}), ${\overline{p}}=2$ if $p=1$ and ${\overline{p}}=1$ if $p=2$. The terms $\mathcal{M}^p_1$ and $\mathcal{M}^p_2$ account for retarded interactions between the two spins, mediated by the bath. Their expression in terms of blip and sojourn variables is very similar to the ones of $\mathcal{Q}^p_1$ and $\mathcal{Q}^p_2$ and the principle of their derivation is the same as in the case of one spin (see Appendix B). The situation differs however slightly since the blip variables corresponding to one spin and the sojourn variable corresponding to the other one can be simultaneously both non-zero. A detailled derivation in this particular case is provided in Appendix D. The (renormalized) Ising interaction (in $\tilde{\mathcal{G}}[\sigma_1,\sigma_2,\sigma_1',\sigma_2']$) can be expressed in a convenient way in this description, as we have

\begin{equation}
 \sigma_1(s) \sigma_2(s)- \sigma_1'(s) \sigma_2'(s)=2\left[ \eta^1 (s) \xi^2(s)+ \eta^2 (s) \xi^1(s) \right].
\label{eq:F1}
\end{equation}

 As for the one-spin case, we can proceed to a stochastic unravelling of the influence functional, and we have

  \begin{align}
\mathcal{F}_{n_1, n_2}=& \overline{  \prod_{i=1}^{2n_1} \exp\left[ h(t^1_i) \Xi_j^1+k(t^1_{i-1}) \Upsilon^1_{i-1}    \right]}\notag \\
&\times \overline{  \prod_{j=1}^{2n_2} \exp\left[ h(t^2_j) \Xi_j^2+k(t^2_{j-1}) \Upsilon^2_{j-1}    \right]}\notag \\
&\times  \tilde{\mathcal{G}}[\sigma_1,\sigma_2,\sigma_1',\sigma_2'].
\label{functionnal_2}
\end{align}
The fields $h$ and $k$ verify the correlations of Eqs. (\ref{height_1}), (\ref{height_2}), and (\ref{height_3}). Eq.~(\ref{eq:p(t)_two_spins}) together with Eq.~(\ref{functionnal_2}) has now the form of a time ordered product, averaged over the noise variables.\\

 The summation over the variables $\{\Xi^p_j\}$ and $\{\Upsilon^p_j\}'$ for $p\in \{1,2\}$ can be incorporated by considering an effective Hamiltonian $H_{1}(t)$ for the spin density matrix, acting on the space $\mathcal{E}$. It can be written as a sum of two terms $H_{1}(t)=U_{1}+V_{1}(t)$. The (renormalized) Ising interaction is contained in the first term $U_{1}$, while the second term $V_{1}(t)$ accounts for tunneling events. 
 
 $U_{1}$ is a diagonal matrix, whose elements are $\left(U_{1}\right)_{i,i}=2K_r (\eta^1_i \xi^2_i+ \eta^2_i  \xi^1_i )$, where $\eta^p_i$ and $\xi^p_i$ are the value of $\eta^p$ and $\xi^p$ for the state in the position $i$ in the set $\mathcal{E}=\{$AA, AB, AC, AD, BA, BB, BC, BD, CA, CB, CC, CD, DA, DB, DC, DD$\}$. We sequence $\left(U_{1}\right)_{i,i}$ gives explicitely $(0,k,-k,0,k,0,0,-k,-k,0,0,k,0,-k,k,0)$ with $k=2 K_r$.

The 16 by 16 matrix $V_{1}(t)$ accounts for tunneling elements and has the following form,
 
 \begin{equation}
V_{1}(t)=\frac{\Delta}{2} \left( \begin{array}{cccc}
W&D_{\textrm{B}\to \textrm{A}}&D_{\textrm{C}\to \textrm{A}}&(0 )\\
D_{\textrm{A}\to \textrm{B}}&W&(0)&D_{\textrm{D}\to \textrm{B}}\\
D_{\textrm{A}\to \textrm{C}}&(0)&W&D_{\textrm{D}\to \textrm{C}}\\
(0)&D_{\textrm{B}\to \textrm{D}}&D_{\textrm{C}\to \textrm{D}}&W
\end{array} \right).
\label{eq:system_two_spins}
\end{equation}
 Each term of this matrix corresponds to a transition from one state in $\mathcal{E}$ to another, induced by one spin-flip. It is written in Eq. (\ref{eq:system_two_spins}) in a block structure. Each block is a 4 by 4 matrix that can be given a physical interpretation. The diagonal matrices correspond to flips of the second spin, the first one left unchanged. As a result the matrix $W(t)$ has the same structure as in the one-spin case,

\begin{equation}
W(t)= \left( \begin{array}{cccc}
0&e^{- h+k }&-e^{ h+k }&0 \\
e^{h-k }&0&0&-e^{h+k }\\
-e^{- h-k}&0&0&e^{- h+k }\\
0&-e^{- h-k }&e^{h-k }&0
\end{array} \right).
\end{equation}
All the elements of the 4 by 4 matrices on the diagonal running from the lower left to the upper right are zero, because the corresponding states are not coupled by one single spin-flip. The eight matrices $D_{\textrm{B}\to \textrm{A}}$, $D_{\textrm{C}\to \textrm{A}}$, $D_{\textrm{A}\to \textrm{B}}$, $D_{\textrm{D}\to \textrm{B}}$, $D_{\textrm{A}\to \textrm{C}}$, $D_{\textrm{D}\to \textrm{C}}$, $D_{\textrm{B}\to \textrm{D}}$ and $D_{\textrm{C}\to \textrm{D}}$ describe spin flips of the first spin (the precise transition corresponds to the subscript), the second one left unchanged. They read respectively $e^{- h+k }\times I_4$, $-e^{ h+k } \times I_4$, $e^{h-k }\times I_4$, $-e^{h+k}\times I_4$, $-e^{- h-k}\times I_4$, $e^{- h+k}\times I_4$, $-e^{- h-k}\times I_4$ and $e^{h-k}\times I_4$ ($I_4$ is the identity). Let us exemplify such transitions thanks to the path of Fig.~\ref{spin_path_2}. The first transition at $s_1=t_1^2$ corresponds to the transition AA$\to$AB. Its amplitude is given by the term of the first column and the second raw of the top left matrix $W$. The next transition at $s_2=t_1^1$ corresponds to the transition AB$\to$CB. Its amplitude is given by the term of the second column and the second raw of the matrix $D_{\textrm{A} \to \textrm{C}}$.\\

 Finally, the dynamics of the 16 dimensional spin reduced density matrix is governed by an effective SSE with Hamiltonian $H_1$:
\begin{equation} 
p_1(t)=\overline{\langle \Phi_f | \Phi (t) \rangle},
\label{eq:p1:two_spins}
\end{equation}
where $\langle \Phi_f |=(e^{-2k(s_{2N})},0,0,0,0,0,0,0,0,0,0,0,0,0,0,0)$ and  $| \Phi \rangle$ is the solution of the stochastic Schr\"{o}dinger equation 

\begin{equation} 
i \partial_t | \Phi \rangle = H_1 (t) | \Phi \rangle
\label{eq:SSE:two_spins}
\end{equation}
with initial condition 
\begin{equation}
|\Phi_i \rangle=(e^{2k(t_0)},0,0,0,0,0,0,0,0,0,0,0,0,0,0,0)^T.
\end{equation}

Similarly to the one-spin case, simplifications occur in the scaling regime, as shown in Appendix E. In this Appendix, we also investigate numerical convergence issues, as well as other initial and final conditions, which lead to a different choice for the vectors $|\Phi_i \rangle$ and $|\Phi_f \rangle$.


\subsection{Nonequilibrium dynamics and quantum phase transition in the dimer model} 

Here, we apply the SSE methodology in order to tackle the non-equilibrium spin dynamics in the presence of strong dissipative interactions in the case of two spins. 

We define the triplet subspace spanned by the three states $\{|T_-\rangle=|-_z, -_z\rangle,|T_0\rangle=1/\sqrt{2}\left[|+_z, -_z\rangle+|-_z,+_z\rangle\right],|T_+\rangle=|+_z,+_z\rangle\}$, while the singlet state is $|S\rangle=1/\sqrt{2}\left[|+_z,-_z\rangle-|-_z,+_z\rangle\right]$ and remains isolated in the dynamics. This problem is well-known to exhibit a dissipative quantum phase transition\cite{Garst_Vojta,sougato,Peter_two_spins,Winter_Rieger} where the bath entirely polarizes the two spins either in the $|T_+\rangle$ or $|T_-\rangle$ state, by analogy to a ferromagnetic phase. The transition line can be located thanks to the evolution of the entanglement entropy with respect to $\alpha$ (see Fig.~5 of Ref. \onlinecite{Peter_two_spins}) or to the evolution of the connected correlation function $C=\langle \sigma^z_1 \sigma^z_2 \rangle-\langle \sigma^z_1 \rangle \langle \sigma^z_2 \rangle$ (see Fig.~10 of Ref. \onlinecite{Winter_Rieger}).\\

\begin{figure}[t!]
\center
\includegraphics[scale=0.5]{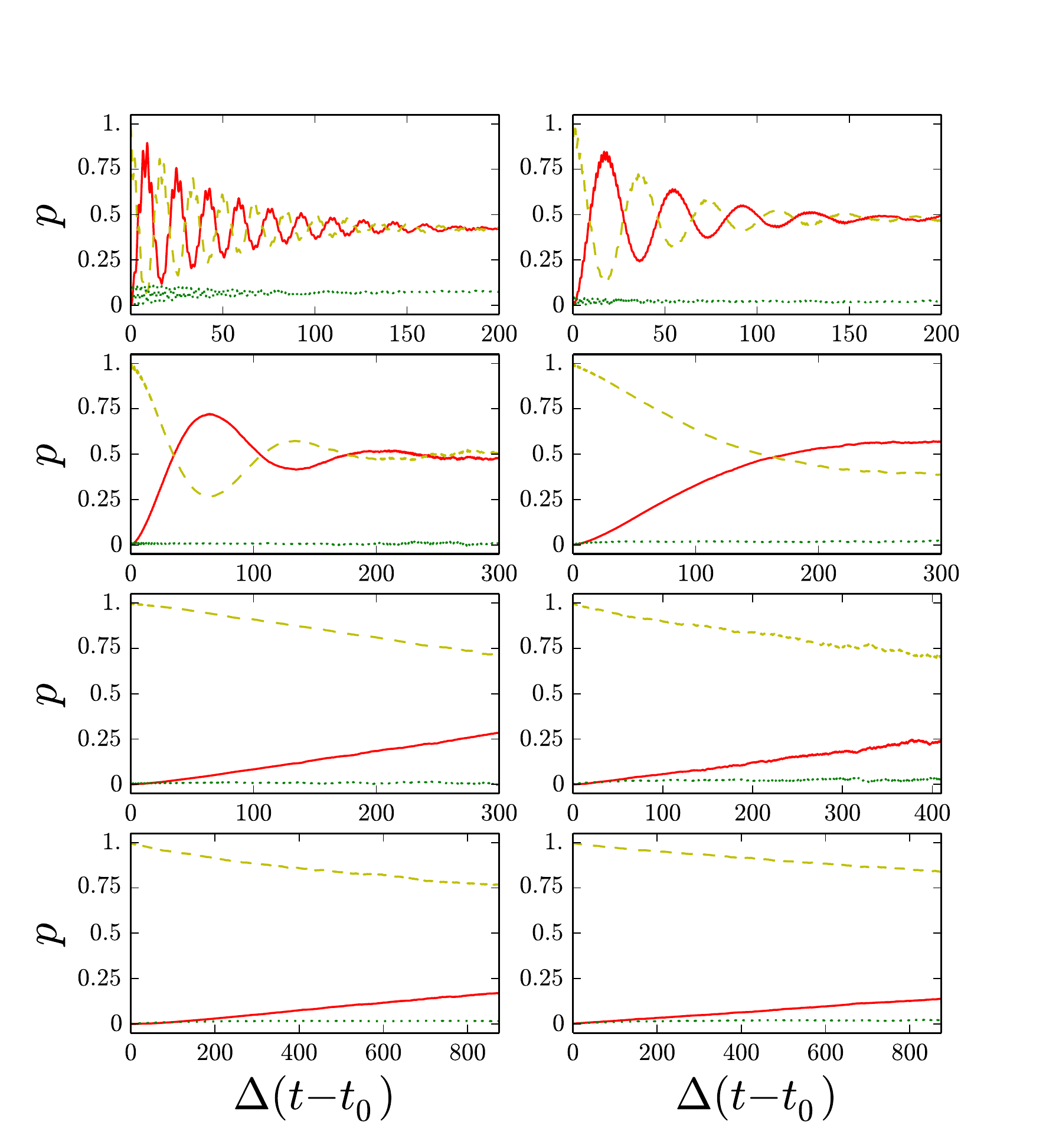}  
\caption{(Color online) Dynamics of the dimer model in the unpolarized phase: the dashed yellow line represents $p_{|T_+\rangle}$, the full red line represents  $p_{|T_-\rangle}$ and the dotted green line represents $p_{|T_0\rangle}$. From the top left to the bottom right, we have $\alpha=0.01$, $\alpha=0.02$, $\alpha=0.04$, $\alpha=0.06$, $\alpha=0.08$, $\alpha=0.1$, $\alpha=0.12$ and , $\alpha=0.14$. The system starts in the state $|T_+\rangle$ for all the plots. We have taken $\omega_c/\Delta=100$ and $K=0$ for all plots.}
\label{free_dynamics_21_10_2015}
\end{figure}

With our approach, we are able to address the non-equilibrium dynamics of the system both in the unpolarized and in the polarized phase, and thus reproduce the quantum phase transition. We consider that the system initially starts from the state $|T_+\rangle$ at the time $t_0$, when spin and bath are brought into contact. We show in Fig.~\ref{free_dynamics_21_10_2015} the time evolution of $p_{|T_0\rangle}$, $p_{|T_+\rangle}$ and $p_{|T_-\rangle}$, which are the occupancies of the states $|T_0\rangle$ , $|T_+\rangle$ and $|T_-\rangle$. 

The different panels correspond to different values of $\alpha$ from $\alpha=0.01$ (top left) to $\alpha=0.14$ (bottom right). All these values corrrespond to the unpolarized phase in the range of used parameters ($K=0$ and $\omega_c=100$).

 \begin{figure}[t!]
\center
\includegraphics[scale=0.34]{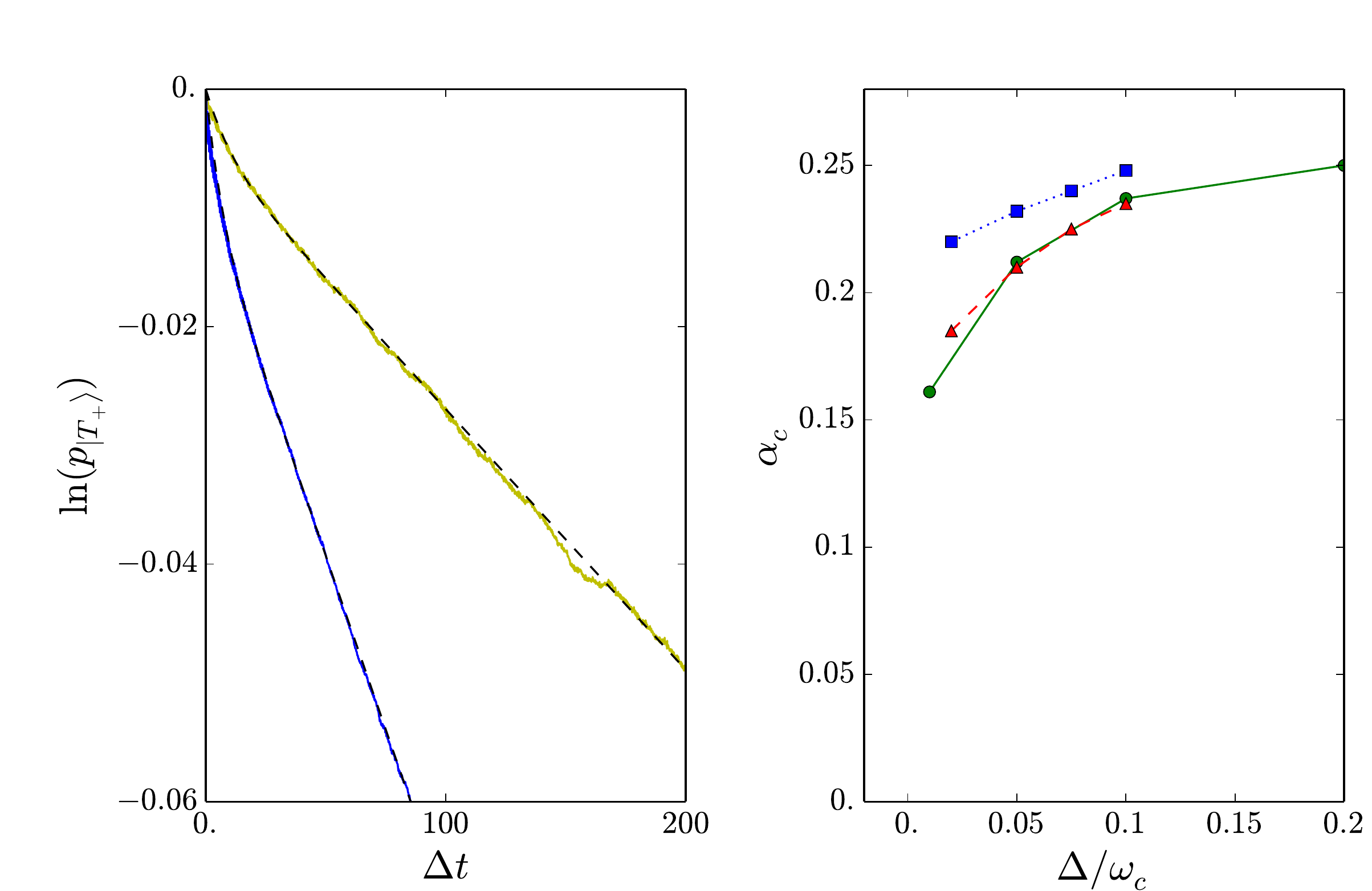}  
\caption{(Color online) Left panel: evolution of $\ln(p_{|T_+\rangle})$ at $\alpha=0.14$, for $\Delta/\omega_c=0.01$ (yellow line-top) and $\Delta/\omega_c=0.05$ (blue line-bottom), and bi-exponential fit (dashed black line). Right panel: Critical line with respect to $\Delta/\omega_c$ at $K=0$ (green dots and full green line) and comparison with the results obtained in Ref. \onlinecite{Peter_two_spins} (TDNRG) (red triangles and red dashed line) and Ref. \onlinecite{Winter_Rieger} (QMC) (blue squares and dotted blue line)}.
\label{decay_phase_diagram}
\end{figure}

\begin{figure}[t!]
\center
\includegraphics[scale=0.34]{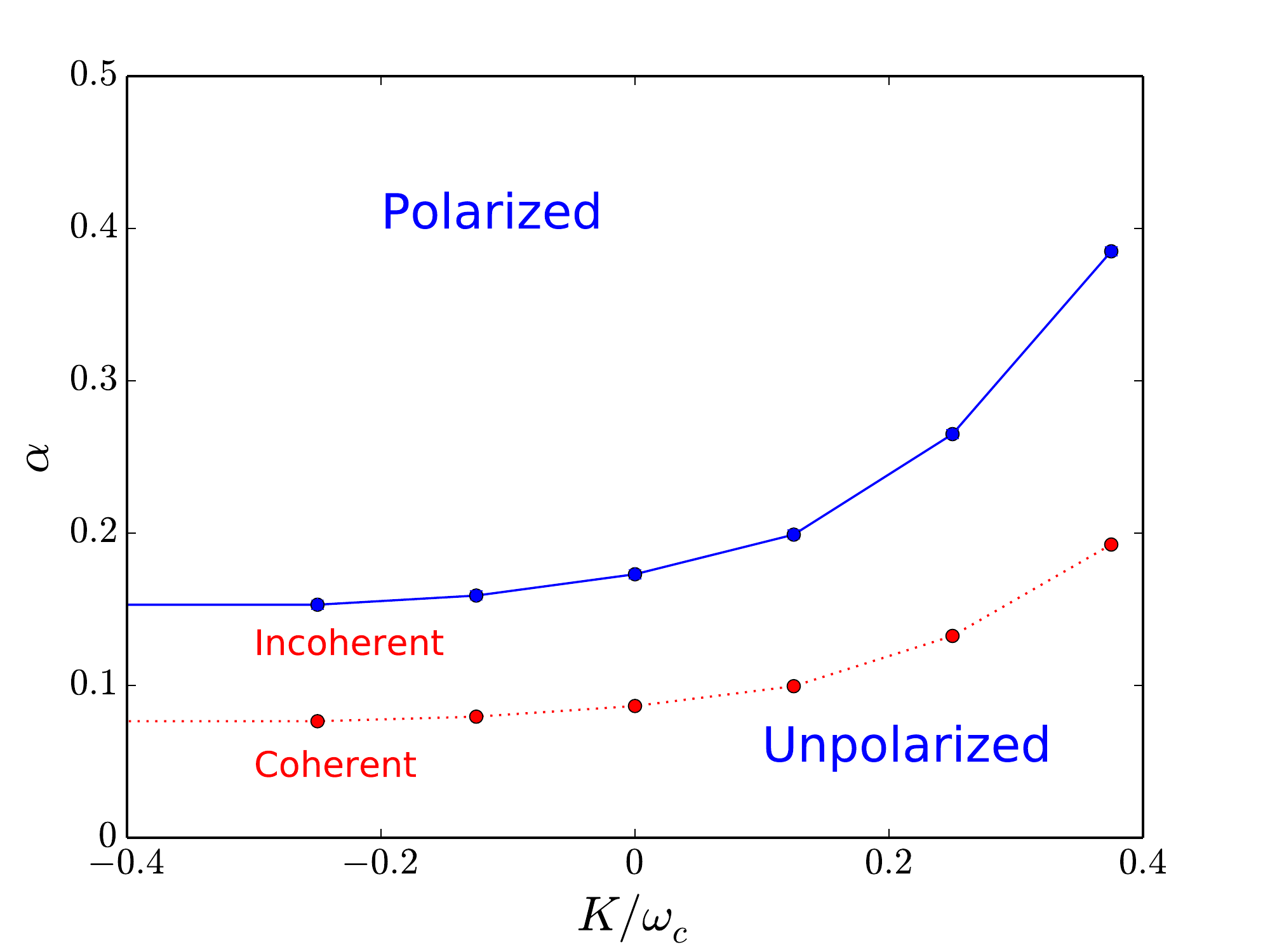}  
\caption{(Color online) Critical line with respect to $K$ for $\Delta/\omega_c=0.01$ (blue points and full blue line). Above the line, the system relaxes to a polarized steady-state. The red dots and the dotted red line show the location of the crossover line from coherent to incoherent behaviour for the spin oscillations.}
\label{phase_diagram}
\end{figure}


We first note in Fig.~\ref{free_dynamics_21_10_2015} a progressive suppression of the Rabi oscillations between the two states $|T_+\rangle$ and $|T_-\rangle$ when increasing the parameter $\alpha$. This behavior is similar to the one observed in the case of the single spin-boson model, where the crossover from coherent oscillations to an incoherent dynamics occurs at $\alpha_c/2$. At high values of $\alpha$, the relaxation from the initial state $|T_+\rangle$  becomes slower due to the strong ferromagnetic interaction, and it is numerically harder to investigate the dynamics in the zone $\alpha \geq 0.1$, due to the time scales involved (other initial states lead to an easier numerical investigation, allowing to determine accurately the equilibrium density matrix at long times). In the zone $\alpha_c/2<\alpha<\alpha_c$, we find a monotonic relaxation towards the equilibrium. In this zone, for the case of one spin, conformal field theory has predicted that several timescales are involved in the dynamics, leading to a multi-exponential decay\cite{Lesage_Saleur} (which has not been seen in NRG \cite{KLHQPT}). A bi-exponential decay was found in this case thanks to a multilayer multiconfiguration time-dependent Hartree method\cite{Wang_Thoss}. Here, for two spins and at small to intermediate times, we obtain results which are also consistent with a bi-exponential relaxation, as shown on the left panel of Fig.~\ref{decay_phase_diagram}. Other studies have predicted more complicated forms for the relaxation, without any pure exponential decay (see for example the results of Ref. \onlinecite{Kashuba_Schoeller} obtained with renormalization group methods). 

We are then able to locate the phase transition from the divergence of the associated time scale. The transition line is shown on the right panel of Fig.~\ref{decay_phase_diagram}, together with the previous results obtained with a time dependent Numerical Renormalization Group (TDNRG) method\cite{Peter_two_spins}, or with a Quantum Monte-Carlo (QMC) method\cite{Winter_Rieger}. This plot corresponds to a vanishing direct Ising interaction $K=0$, and different values of $\omega_c$. The phase diagram of the system with respect to the parameter $K$ is shown in Fig.~\ref{phase_diagram}. The full blue line shows the phase transition line between the polarized and the unpolarized phase, while the dotted red line shows the crossover line from coherent to incoherent Rabi oscillations in the dynamics \cite{Peter_two_spins}.\\






Next, we show results concerning the dynamics in the polarized phase ($\alpha>\alpha_c$), corresponding to a quantum quench across the critical line, from $\alpha=0$ to $\alpha>\alpha_c$. Some theoretical studies have focused on this question in spins \citep{essler,gambasi,delcampo} or bosonic systems \citep{roux_kollath,sciolla_biroli,rancon}.  For example, at $K=0$ and $\alpha=0$, the initial state of the system is given by $|\psi\rangle=|-_x\rangle \otimes |-_x\rangle=1/2(|T_+\rangle+|T_-\rangle)-1/\sqrt{2} |T_0\rangle$. The associated spin density matrix is

\begin{equation}
\rho_{S} (t_{0})=\frac{1}{4}\left(\begin{array}{cccc}
1 & -1 &-1 & 1   \\
-1 & 1 &1 & -1   \\
-1 & 1 &1 & -1   \\
1 & -1 &-1 & 1  \end{array} \right).
\label{eq:density_matrix_two_spins}
\end{equation}

 After a sudden change of the parameter $\alpha$, the system is in a nonequilibrium state. We compute the spin dynamics for different values of $\alpha>\alpha_c$ and for different values of $\Delta/\omega_c$. We find numerically that the system evolves towards the final density matrix 
 
 \begin{figure}[t!]
\center
\includegraphics[scale=0.36]{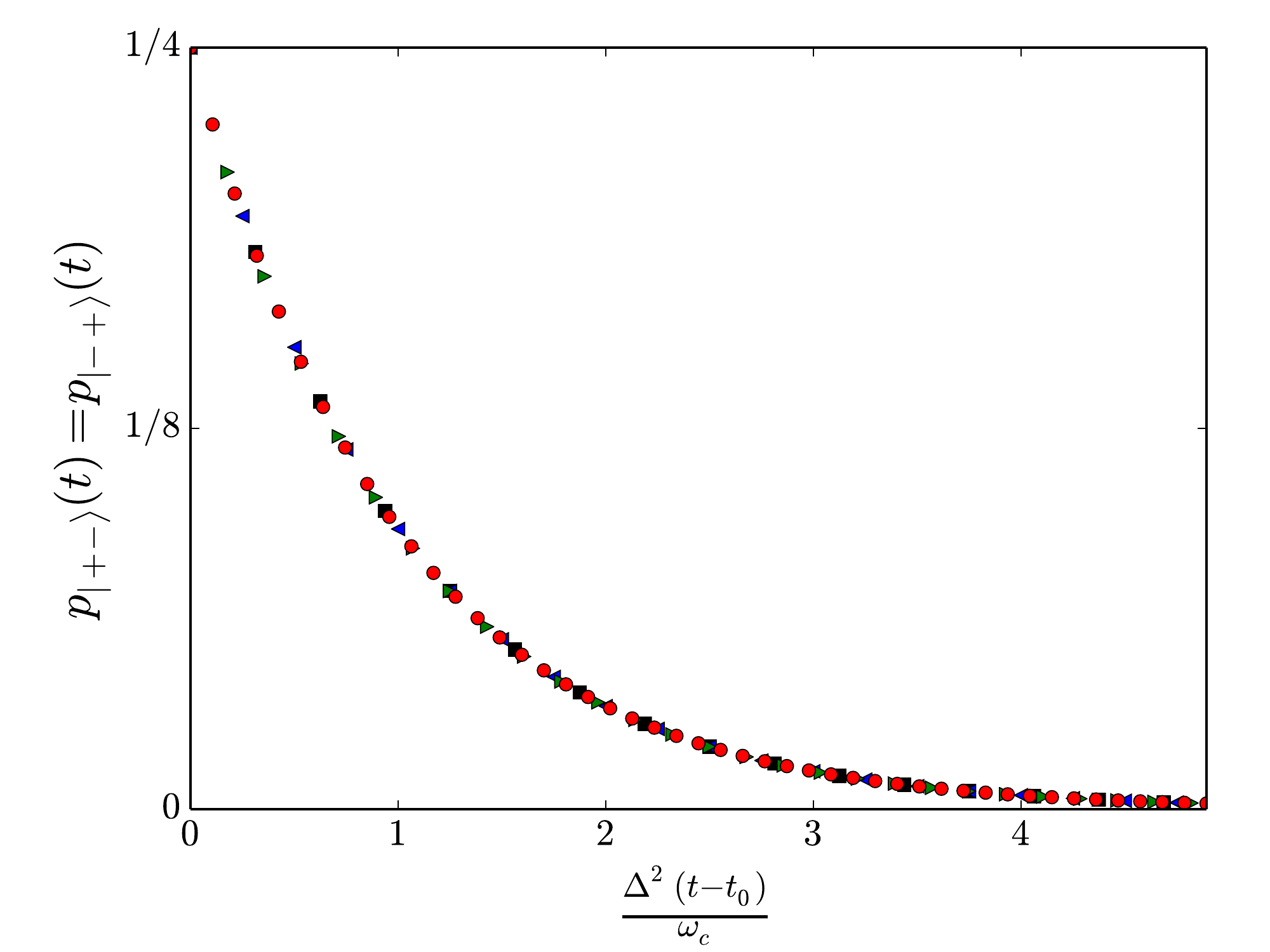}  
\caption{(Color online) Universal dynamics of the dimer model in the polarized phase. The system starts in the nonequilibrium state described by the density matrix of Eq. (\ref{eq:density_matrix_two_spins}), and relax towards a statistic superposition of $|T_+\rangle$ and $|T_-\rangle$. The parameters are $\alpha=0.2$, $\omega_c/\Delta=100$ (red points); $\alpha=0.25$, $\omega_c/\Delta=50$ (right pointing green triangles); $\alpha=0.22$, $\omega_c/\Delta=80$ (left pointing blue triangles); $\alpha=0.3$, $\omega_c/\Delta=20$ (black squares). Taking $K \neq 0$ gives the same exponential relaxation.}
\label{fig_localized}
\end{figure} 
 
 \begin{equation}
\lim_{t \rightarrow \infty} \rho_s(t) = \frac{1}{2}\left(\begin{array}{cccc}
1 & 0 &0 & 0   \\
0 & 0 &0 & 0   \\
0 & 0 &0 & 0   \\
0 & 0 &0 & 1  \end{array} \right),
\label{eq:density_matrix_two_spins_final}
\end{equation}
corresponding to a statistical superposition of the states $|T_+\rangle$ and $|T_-\rangle$ (up to an error of around $10^{-2}$). We find moreover that the spin dynamics is universal in the polarized phase, in the sense that it does not depend on $\alpha$ and $K$. More precisely, we find that 
 \begin{equation}
p_{|+-\rangle}(t)=p_{|-+\rangle}(t)=p_0 \exp\left[-\frac{\Delta^2 (t-t_0)}{\omega_c} \right],
\label{Eq:relaxation}
 \end{equation}
 as shown in Fig.~\ref{fig_localized}, for a quench from $\alpha=0$ to $\alpha>\alpha_c$. $p_{|+-\rangle}(t)$ ($p_{|-+\rangle}(t)$) is the probability to find the system in the state $|+_z,-_z\rangle$ ($|-_z,+_z\rangle$) at time $t$, given by the diagonal term of the density matrix $[\rho_S]_{22}$ ($[\rho_S]_{33}$). This simple form of the damping, and its independance with respect to $K$ or $\alpha$, can be accounted for by a very fast relaxation towards the spin ground state, without the emission of photons. The strong bath-induced Ising interaction and the orthogonality between the polarized state lead to a rapid evolution independent of the other external parameters.\\
 
  \begin{figure}[t!]
\center
\includegraphics[scale=0.42]{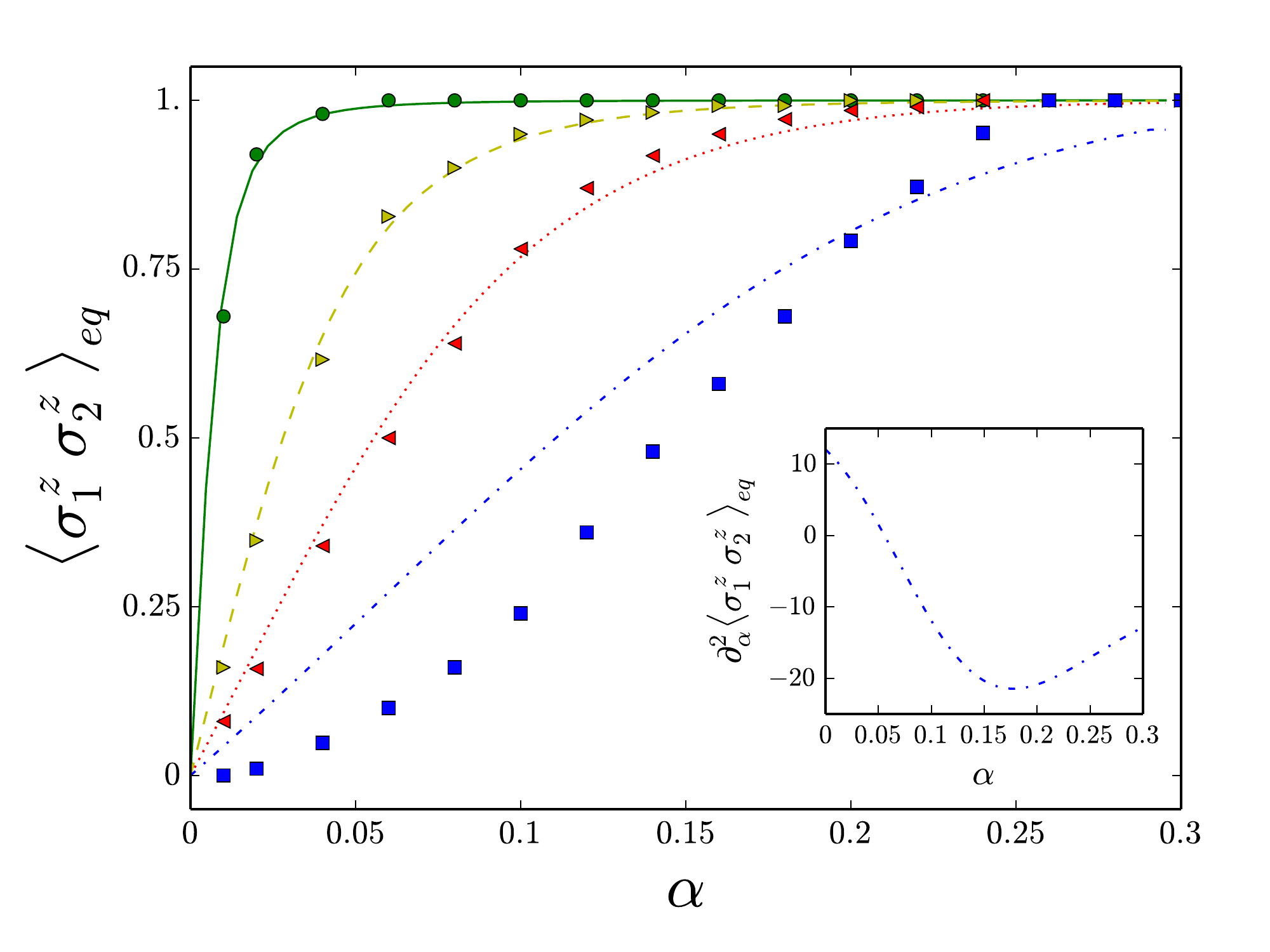}  
\caption{(Color online) Equilibrium value of $\langle \sigma^z_1 \sigma^z_2  \rangle$ as a function of $\alpha$ for $\Delta/\omega_c=0.01$ (green circles), $\Delta/\omega_c=0.05$ (yellow right-pointing triangles), $\Delta/\omega_c=0.1$ (red left-pointing triangles) and $\Delta/\omega_c=0.2$ (blue squares). We have $K=0$. The lines correspond to the value predicted by a toy-model of two interacting spins with tunneling element $\tilde{\Delta}_r$ and Ising interaction $\tilde{K}_r$ obtained thanks to a variational procedure. Parameters are $\Delta/\omega_c=0.01$ (full green line), $\Delta/\omega_c=0.05$ (yellow dashed line), $\Delta/\omega_c=0.1$ (red dotted line) and $\Delta/\omega_c=0.2$ (blue mixed line). The inset shows the evolution of $\partial^2_{\alpha} \langle \sigma^z_1 \sigma^z_2  \rangle $ with $\alpha$ for $\Delta/\omega_c=0.2$. The sign of this quantity changes when increasing $\alpha$.}
\label{Correlation_function_two_spins}
\end{figure}

We also remark that, in the unpolarized phase, the value of $\langle \sigma^z_1\sigma^z_2 \rangle_{eq}$ is non-zero due to the strong ferromagnetic interaction mediated by the bath. We compute this quantity as the limit of $tr_B\left[\rho_S (t) \sigma^z_1 \sigma^z_2\right]$ at long times, and plot its evolution with respect to $\alpha$ for different values of $\omega_c$ in Fig.~\ref{Correlation_function_two_spins}. At very small $\Delta/\omega_c$ we have roughly $\langle \sigma^z_1\sigma^z_2 \rangle_{eq}=\alpha \omega_c /\sqrt{(\alpha \omega_c)^2+\Delta_r^2}$, which would be the equilibrium value of this quantity in a two-spins Ising model governed by the Hamiltonian 

\begin{align}
H_I=\frac{\Delta_r}{2}\left( \sigma_1^x+\sigma_2^x \right)-K_r \sigma_1^z \sigma_2^z,
\label{toy_model_ising_two_spins}
\end{align}
where $\Delta_r=\Delta (\Delta/\omega_c)^{\alpha/(1-\alpha)}$ is the renormalized tunneling element obtained by an adiabatic renormalization procedure\cite{leggett,weiss} (see the Introduction).

There are notable deviations with respect to this toy-model, especially when $\Delta/\omega_c$ becomes larger ($\Delta/\omega_c \geq 0.02$). In this case, the adiabatic renormalization procedure is no longer valid, as the bath and spin degrees of freedom evolution time scales are not well separated. The assumption of fully polarized bath states associated to one given spin polarization no longer holds and we need to refine the analysis, for example by using a variational technique on the ground state wavefunction following the ideas of Refs. \onlinecite{Silbey_Harris,sougato}. We write the Hamiltonian of the system in a displaced oscillator basis defined by the four states $\{|B_{++}\rangle \otimes |+_z,+_z\rangle,|B_{0}\rangle \otimes|+_z,-_z\rangle,|B_{0}\rangle \otimes|-_z,+_z\rangle,|B_{--}\rangle \otimes|-_z,-_z\rangle \}$, with

\begin{align}
B_{++}&=\prod_{k} \exp\left[-\frac{f_k}{\omega_k}\left( b^{\dagger}_k- b_k\right) \right]|B_{0}\rangle   \\
B_{--}&=\prod_{k} \exp\left[\frac{f_k}{\omega_k}\left(b^{\dagger}_k- b_k\right) \right] |B_{0}\rangle,
\end{align}
where $|B_{0}\rangle$ is the ground state of the bosonic bath taken in isolation at zero temperature. $f_k$ are variational parameters with $f_k \neq \lambda_k$ at a general level. With this ansatz we do not specify the amplitude with which a given mode is displaced \textit{ab initio}, but these coefficients are found by minimizing the free energy of the total system. The displacement from the equilibrium position of a given oscillator may then depend on other parameters. Following Ref. \onlinecite{sougato}, we find self-consistent equations for the bath-induced Ising interaction $\tilde{K}_r$ and the renormalized tunneling element $\tilde{\Delta}_r$,

\begin{align}
\tilde{\Delta}_r&=\Delta \exp\left[-\alpha\int_{0}^{\infty} d\omega \frac{G(\omega)^2}{\omega} e^{-\omega/\omega_c} \right],\\
\tilde{K}_r&=\alpha \int_{0}^{\infty} d\omega G(\omega)[2-G(\omega)]  e^{-\omega/\omega_c},\\
G(\omega)&=\frac{\sqrt{\tilde{K}_r^2+\tilde{\Delta}_r^2}+\tilde{K}_r}{\sqrt{\tilde{K}_r^2+\tilde{\Delta}_r^2}+\tilde{K}_r+\frac{\tilde{\Delta}_r^2}{\omega}}.
\end{align}

We plot the corresponding evolution of $\langle \sigma^z_1  \sigma^z_2 \rangle_{eq}=\tilde{K}_r /\sqrt{(\tilde{K}_r)^2+\tilde{\Delta}_r^2}$ with respect to $\alpha$ for different values of $\omega_c$ in Fig.~\ref{Correlation_function_two_spins}. We find a good agreement with the exact results given by the SSE method as long as $\Delta/\omega_c$ remains small ($\Delta/\omega_c \leq 1$). We notably recover a change of the concavity of $\langle \sigma^z_1  \sigma^z_2 \rangle_{eq}$ with respect to $\alpha$, as shown in the inset of Fig. \ref{Correlation_function_two_spins} where we plot the evolution of the second derivative of $\langle \sigma^z_1  \sigma^z_2 \rangle_{eq}$ for $\Delta/\omega_c=0.2$. This feature cannot be recovered by the adiabatic renormalization procedure, but we see that this effect is far more pronounced in the results of the SSE than in the variational treatment. The dynamical adjustment of both the bath and spin degrees of freedom can thus explain some features of the results obtained numerically, especially at small $\Delta/\omega_c \leq 0.1$ but this variational approach fail at quantitatively describing the regime of strong coupling and the dissipative quantum phase transition. From the analytical point of view, we also note some efforts with multi-polaron approaches\cite{Florens}. As seen in Fig.~\ref{Correlation_function_two_spins}, the main effect at large $\omega_c/\Delta$  is to induce a large ferromagnetic interaction. We will use this feature below in the synchronization and LZ interferometry phenomena.


\subsection{Synchronization} 

Synchronization phenomena occur spontaneously in a wide range of physical systems\cite{synchronization}. Here we quantitatively describe synchronization mechanisms between two spins 1/2 starting from the polarized state 
$|+_z,+_z\rangle$, without drive. In this two-spin problem coupled to a ohmic bath, some results were also obtained using the NRG \cite{Peter_two_spins}. A comparison between classical and quantum regimes for this kind of problems without dissipation was recently done in Ref. \onlinecite{synchronization_fuchs}.

\begin{figure}[t!]
\center
\includegraphics[scale=0.4]{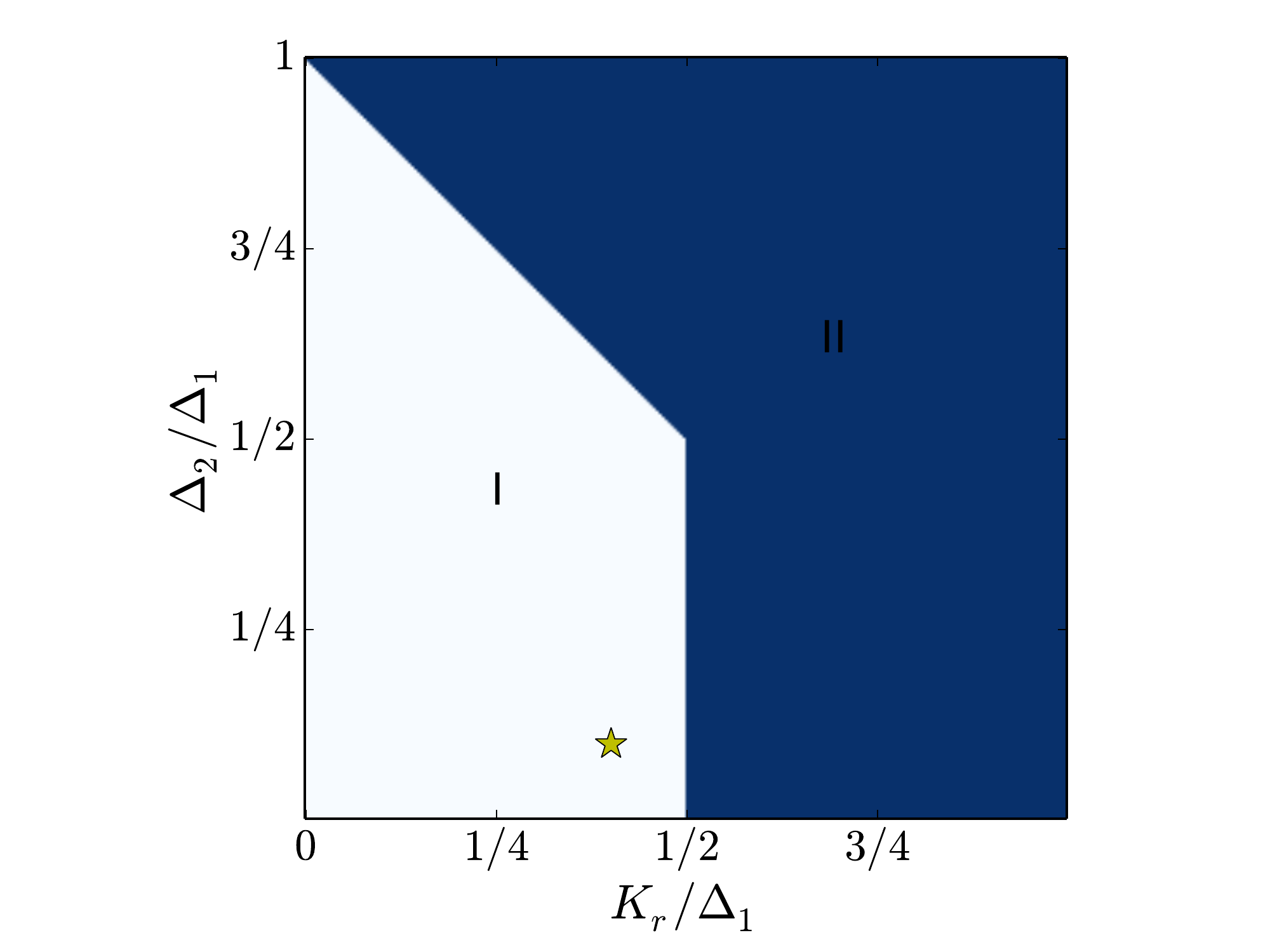}
\caption{(Color online) Synchronization phase diagram in the case of direct Ising interaction $K_r=K$. Region I (in white) corresponds to the unsynchronized regime : $\langle \sigma^z_1 \sigma^z_2 \rangle$ vanishes periodically. The yellow star shows the point for which we compare direct and bath-induced interaction (see text). Region II (in blue) corresponds to the synchronized regime : $\langle \sigma^z_1 \sigma^z_2 \rangle>0$ at all times.
}
\label{Synchronization_phase_diagram_1}
\end{figure}

We consider the dynamics of two interacting spins with different bare oscillation frequencies $\Delta_1$ and $\Delta_2$ (with $\Delta_1>\Delta_2>0$), starting from the same initial state. We quantify the synchronization due to the interaction, thanks to spin-spin correlations in time. We will compare the case of direct versus bath-induced interaction. We denote by $K_r$ the effective strentgh of the interaction between the spins. In the case of a coupling through the bath we identify $K_r=\alpha \omega_c$ while we have $K_r=K$ in the case of a direct Ising interaction. Some efforts were done to study this effect in Ref.~\onlinecite{Peter_two_spins}. 

Let us first consider the case of direct Ising interaction $K$. A quantitative description of this type of synchronization can be done by studying the time-evolution of $\langle \sigma^z_1 \sigma^z_2 \rangle$. The system starts in the state $|+_z,+_z\rangle$, so that  $\langle \sigma^z_1 \sigma^z_2 \rangle (t_0)=1$ at the initial time. We define the synchronized regime as the region in the parameters space for which $\langle \sigma^z_1 \sigma^z_2 \rangle$ stays positive at all times. We show in Fig~\ref{Synchronization_phase_diagram_1} the synchronization phase diagram with respect to $\Delta_2/\Delta_1$ and $K_r/\Delta_1=K/\Delta_1$. In the region I (in white), the two spins are not synchronized and the correlation function $\langle \sigma^z_1 \sigma^z_2 \rangle$ changes sign periodically. In the other region (region II in blue in Fig.~\ref{Synchronization_phase_diagram_1}) $\langle \sigma^z_1 \sigma^z_2 \rangle$ always stays positive. For $K_r/\Delta_1>1/2$ the Ising interaction dominates and the dynamics is synchronized for all values of $\Delta_2$. When $\Delta_2$ approaches $\Delta_1$, the two spins have comparable oscillating frequencies and the synchronization is then easier.

\begin{figure}[t!]
\center
\includegraphics[scale=0.2]{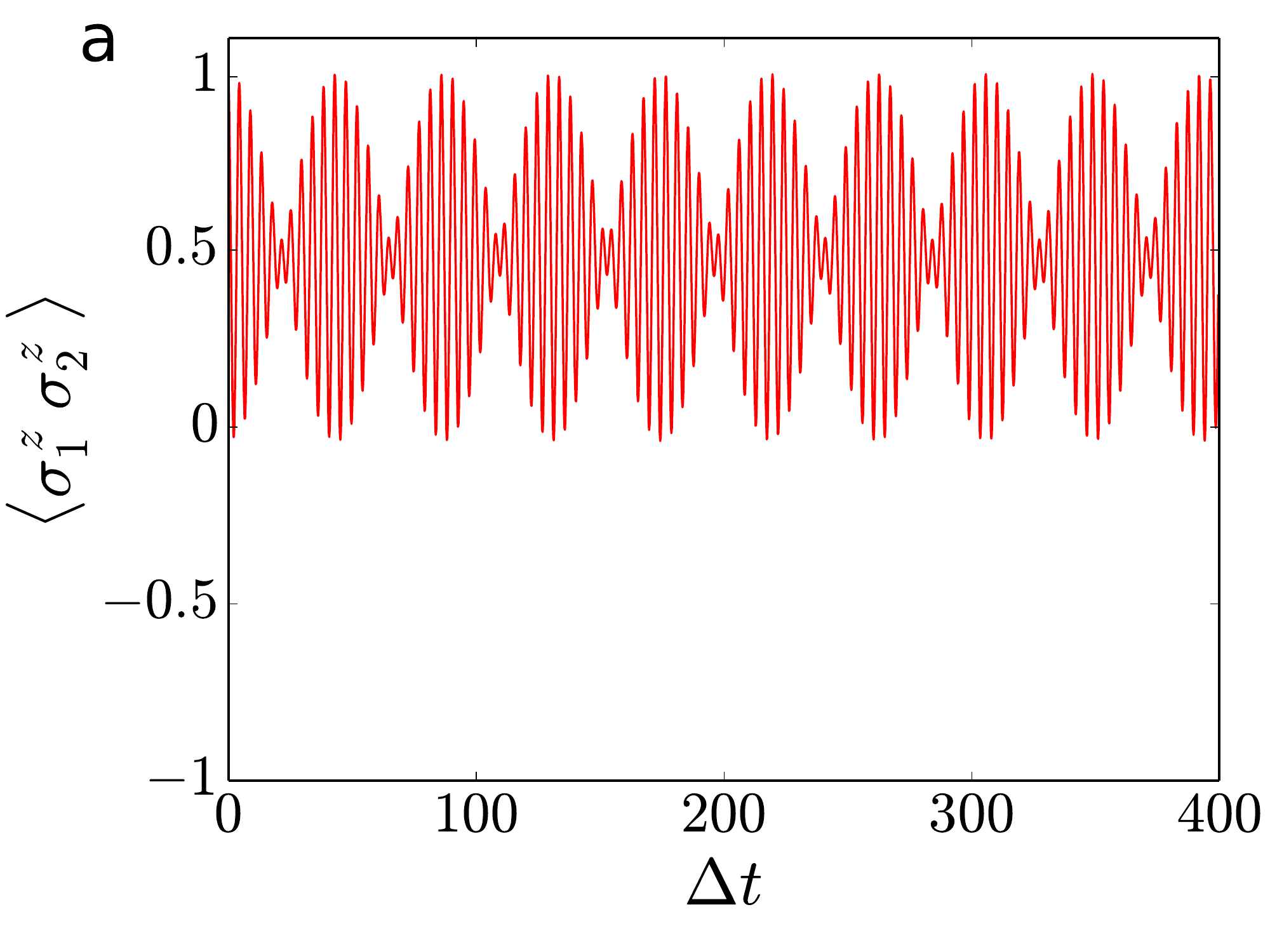}  \includegraphics[scale=0.2]{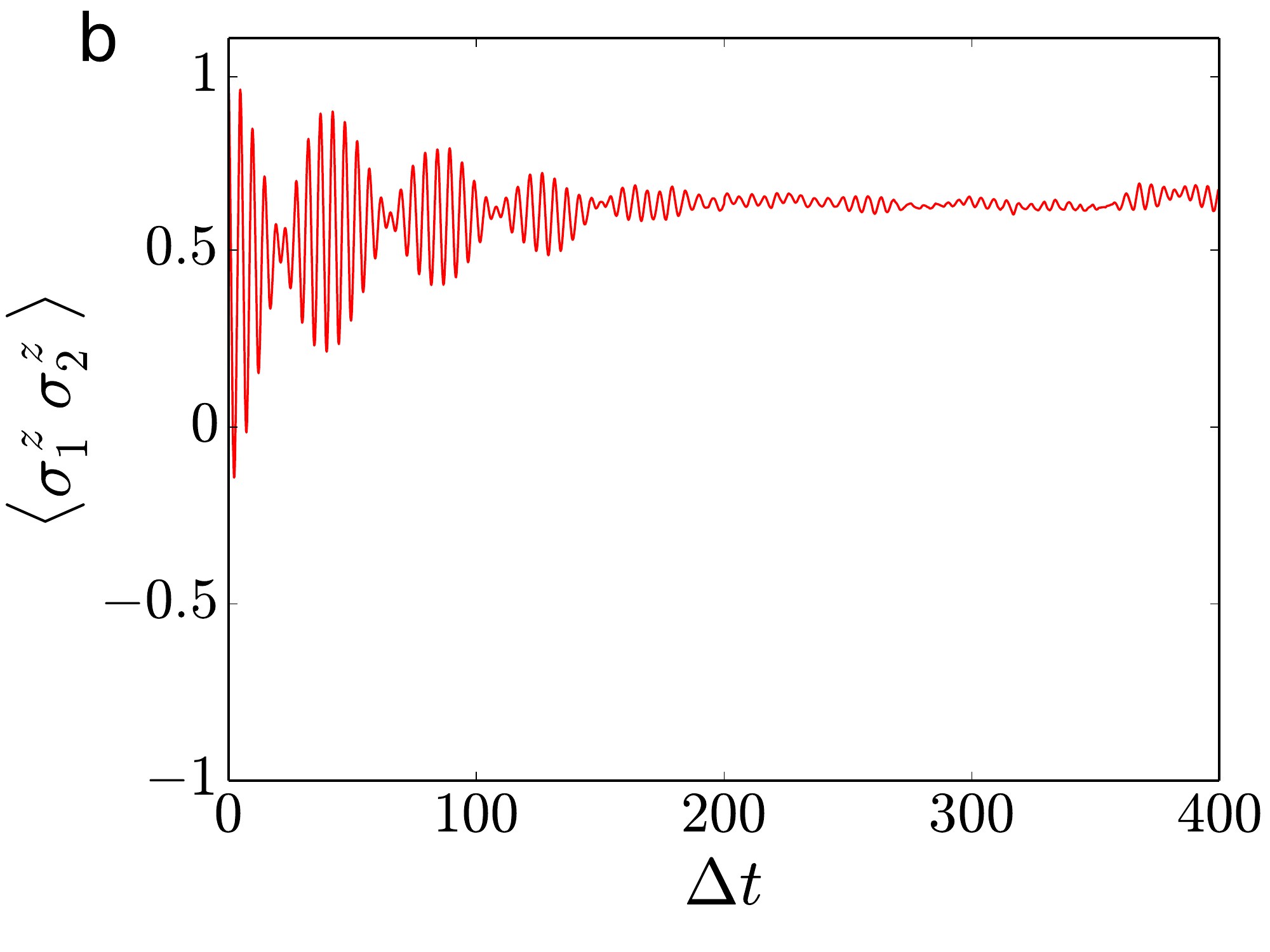}  \\
\includegraphics[scale=0.2]{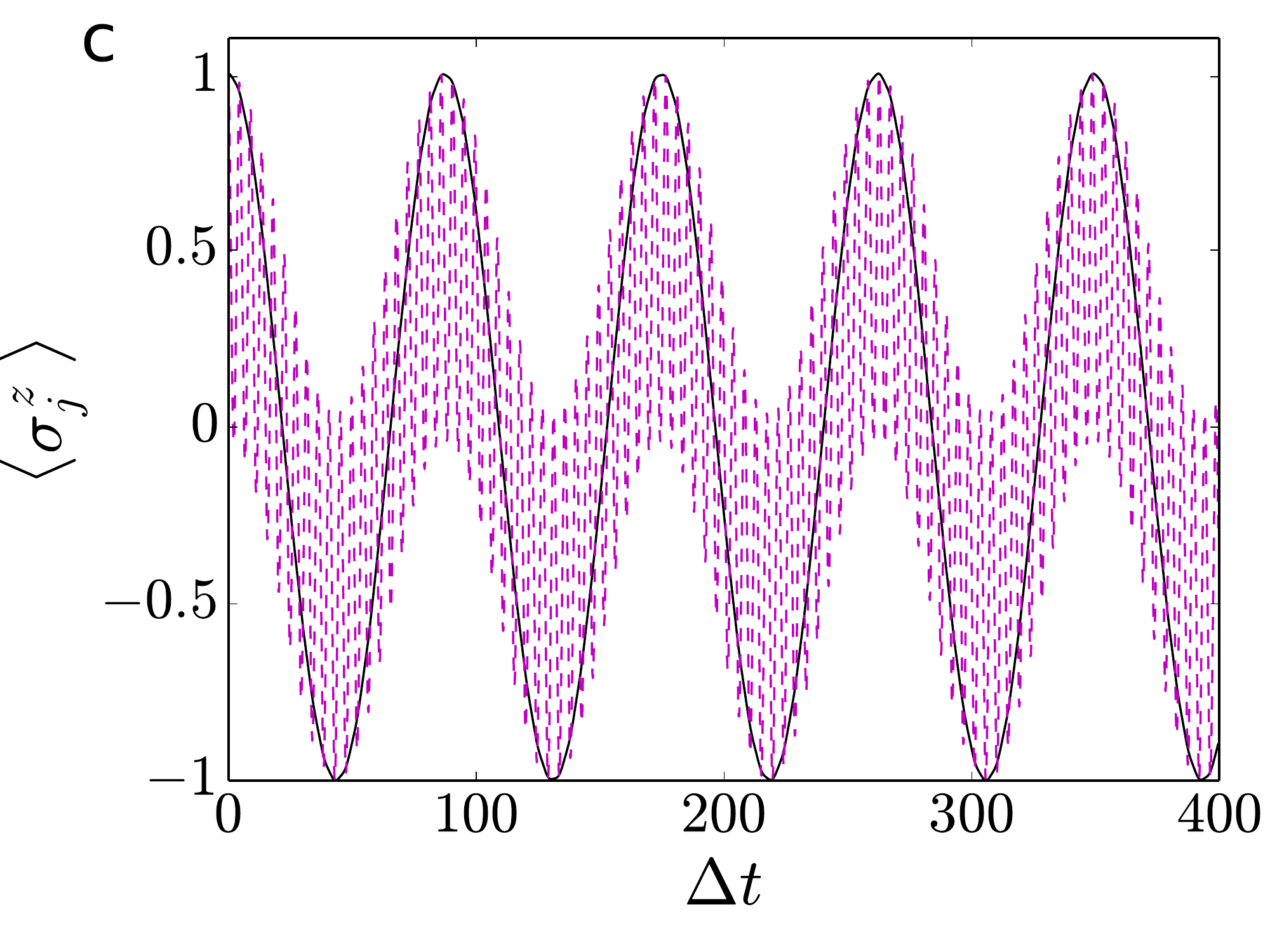}  \includegraphics[scale=0.2]{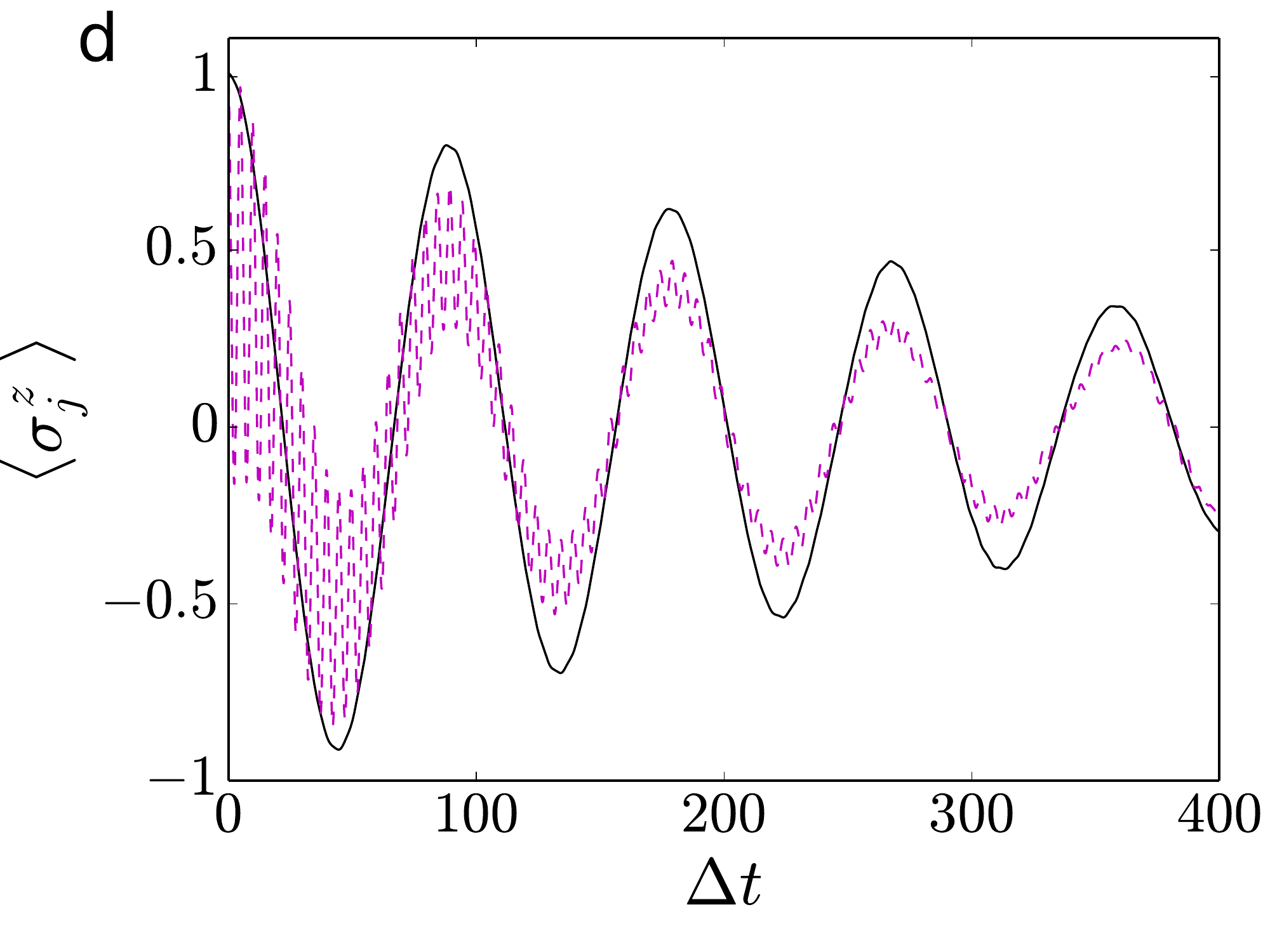}  
\caption{(Color online) Panels a and b: time evolution of $\langle \sigma^z_1 \sigma^z_2 \rangle$ for a direct Ising interaction interaction $K_r=K$ (panel a) and for a bath-induced interaction  $K_r=\alpha \omega_c$ (panel b). Panels c and d: time evolution of $\langle \sigma^z_1\rangle $ and $\langle \sigma^z_2 \rangle$ for a direct Ising-like interaction $K_r=K$ (panel c) and a bath-induced interaction $K_r=\alpha \omega_c$ (pannel d). We have $K_r/\Delta_1=0.4$, $\Delta_2/\Delta_1=0.1$ and $\omega_c=20 \Delta_1$. 
}
\label{Synchronization_two_spins_1}
\end{figure}

The dissipative case, for which the interaction originates from the interaction with the bath, shows a similar phase diagram. There are however notable differences in the unsynchronized regime close to the transition line. In this region, the interaction with the bath leads to an effective synchronization after a short time unsynchronized dynamics. To exemplify this effect, we focus on the spin dynamics at $K_r/\Delta_1=0.4$ and $\Delta_2/\Delta_1=0.1$ in both cases. These parameters correspond to the yellow star in Fig.~\ref{Synchronization_phase_diagram_1}. The evolution of $\langle \sigma^z_j\rangle $ and $\langle \sigma^z_1 \sigma^z_2\rangle $ is shown in Fig.~\ref{Synchronization_two_spins_1} in both cases. We remark that in the case of direct Ising coupling (panel a), there is no synchronization transition as $\langle \sigma^z_1 \sigma^z_2\rangle $ changes sign periodically. By contrast, we remark that $\langle \sigma^z_1  \sigma^z_2 \rangle$ only vanishes a finite number of times (see panel b). After this short time behaviour, the system enters a synchronized regime for which $\langle \sigma^z_1  \sigma^z_2 \rangle$ no longer vanishes and tends to a non-zero equilibrium value corresponding to a polarized equilibrium state. \\

This synchronization effect is the sole consequence of the Ising-like interaction between spins. We found that dissipation processes enhance synchronization, as they favor the evolution towards more stable polarized states. The cases of Markovian or Non-Markovian bath may lead to the comparable enhancement. We note recent experiments in ultra-cold atoms exemplifying the synchronization phenomena between bosons and fermions \cite{synchronization_salomon}.\\

\subsection{Landau-Zener-Stueckelberg-Majorana interferometry} 

In this Section, we investigate the non-equilibrium behavior of the dimer system under an additional linear driving term $\epsilon (t)/2 \sum_{j=1}^2 \sigma_j^z$.\\

We focus on a single linear passage, known as Landau-Zener problem. It corresponds to $\epsilon (t) =\epsilon_0+v(t-t_0)$, $(v>0)$. We choose $\epsilon_0<0$ with $|\epsilon_0|/\Delta\gg 1$ so that the initial state $|T_+\rangle$ corresponds to the ground state at the initial time $t_0$. Landau \cite{Landau}, Zener \cite{Zener}, Stueckelberg \cite{Stueckelberg} and Majorana \cite{Majorana} provided an analytical description of this problem in the case of an isolated two-level system subject to a linear sweep ($K=0$ and $\alpha=0$). The survival probability $p_{lz}$ that the spin remains in its initial state after the sweep, is fully determined by the velocity of the sweep $v$, and we have $p_{lz}=\exp [-\pi \Delta^2/2v]$. It was shown in Refs. \onlinecite{kayanuma_1,kayanuma_2} that the presence of a gaussian dissipative bath does not affect the transition probability in the case of the Landau-Zener sweep for one single spin, as long as the coupling is along the z-direction. It is no longer true for two spins and the presence of the bath affects the final state.

For a symmetric drive only the triplet states are coupled to the bath, and three levels participate to the dynamics. The system then constitutes a $SU(3)$ Landau-Zener-Stueckelberg-Majorana interferometer\citep{su3}.

 \begin{figure}[t!]
\center
\includegraphics[scale=0.4]{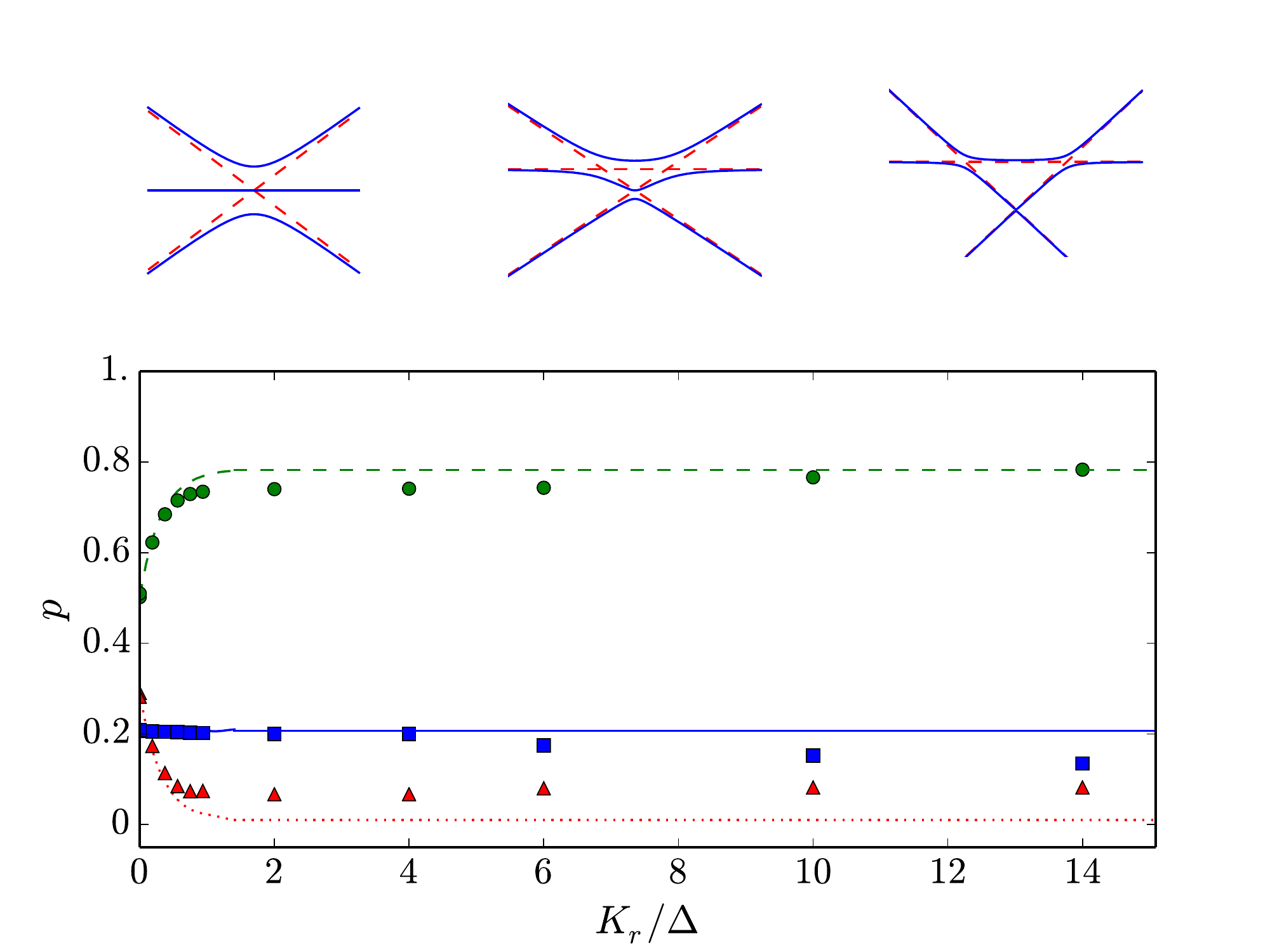}  
\caption{(Color online) Top: evolution of the energy levels with respect to the drive $\epsilon$, for different values of the direct Ising coupling at $\alpha=0$. Main figure: evolution of the final transition probabilities after a linear sweep of velocity $v=2\Delta^2$ as a function of $K_r/\Delta=(K+\alpha\omega_c)/\Delta$.  The lines correspond to a direct Ising interaction $K_r=K$ and $\alpha=0$, and the markers correspond to a bath-induced coupling $K_r= \alpha \omega_c$ and $K=0$. Full blue line, and blue squares: $p_{|T_+\rangle}(t \to \infty)$. Dotted red line and red triangles: $p_{|T_-\rangle}(t \to \infty)$. Dashed green line and green points: $p_{|T_0\rangle}(t \to \infty)$. We take $\omega_c=100 \Delta$.}
\label{energy_levels_transition_proba}
\end{figure}

 In Fig.~\ref{energy_levels_transition_proba}, we plot the different probabilities $p_{|T\rangle}(t \to \infty)$, for $|T\rangle \in \{|T_-\rangle,|T_0\rangle,|T_+\rangle\}$ to end up in the state $|T\rangle$ at long times after a linear sweep of velocity $v=2\Delta^2$, as a function of $K_r/\Delta$. The lines correspond to the case where $\alpha=0$, so that the interaction between the two spins is only due to the direct Ising interaction $K_r=K$. We remark that the value of $p_{|T_+\rangle}(t \to \infty)$ is not affected by the Ising interaction.  $p_{|T_-\rangle}(t \to \infty)$ goes to zero when the Ising interaction increases, while $p_{|T_0\rangle}(t \to \infty)$ simultaneously increases. This can be easily understood by the structure of the energy levels for the different values of $K$. On the upper part of Fig.~\ref{energy_levels_transition_proba}, we draw the energy levels of the triplet states as a function of $\epsilon$, for different values of the Ising coupling $K$, increasing from left to right. The system always starts at time $t=t_0$ on the lower branch at negative bias $\epsilon_0$. At the velocity considered here, we go from the regime of independent crossings (the two spins behave independently when $K=0$, see left drawing) to the regime of one single crossing between $|T_+\rangle$ and $|T_0\rangle$ while we increase the value of $K$. When $K/\Delta \gg 1$, the lowest anticrossing can be ignored and the probability to end up in the state $|T_-\rangle$ then vanishes as the first gap closes (see the right drawing).\\
 

The markers in Fig.~\ref{energy_levels_transition_proba} correspond to the same protocol for $K=0$ and $\alpha$ not zero. As can be seen, the dominant effect of the bath at high $\omega_c$ is to induce a ferromagnetic Ising-like interaction. Here however, the probability to end up in the state $|T_-\rangle$ does not vanish when increasing the value of $\alpha$. This is due to transitions from $|T_0\rangle$ to $|T_-\rangle$ associated to emissions of a bosonic excitations after the crossing of the critical point.  For very rapid transitions, losses become negligible and the fidelity is higher. \\

Multiple consecutive and rapid passages may result in constructive or destructive interferences, depending on the phases acquired during the adiabatic and the non-adiabatic evolutions\cite{LZSM_shevshenko}, allowing to propose an entanglement generation protocol by tuning the external drive, which is of great importance for quantum information purposes. \\
 

\section{Array}

 For greater values of $M$, the problem becomes rapidly untractable numerically, as the density matrix of the spin system becomes too large. We will then extend the method at a mean field level in the case of the array ($M\to \infty$) in the subsection A. In the subsection B, we investigate Landau-Zener sweeps for the array and interpret the results with a Kibble-Zurek type argument. Recent developments linked non non-equilibrium physics in these lattice systems involve Matrix Product States \cite{Garrahan,Marco,Sanchez}; stochastic mean-field methods also allow to describe non-equilibrium light-matter systems\cite{Keeling}.\\
 
\subsection{Mean-field approximation in the limit $M\to \infty$}

 We proceed as in the one-spin and two-spin cases and follow the steps exposed in Sec. II. We start with all the spins initially in the state $|+_z\rangle$ so that $\rho_{S} (t_{0})=\prod_{j=1}^M | +_{z,j} \rangle\langle +_{z,j} | $.  At a given time $t$, the elements of the spin reduced density matrix read

\begin{align}
\langle \bm{\sigma}_f | \rho_S (t) | \bm{\sigma}_f'\rangle&=\sum_{n}   \langle  u_{n}, \bm{\sigma}_f | U(t) \rho(t_0) U^{\dagger}(t) |  u_{n},\bm{\sigma}_f '\rangle,
\label{eq:densitymatrix}
\end{align}
where we define the $M$-dimensional spin vector $|\bm{\sigma}\rangle=|\sigma_1,\sigma_2,..,\sigma_M\rangle$. The time-evolution of the spin reduced density matrix can be then re-expressed as, 

\begin{equation}
\langle \bm{\sigma}_f | \rho_S (t) | \bm{\sigma}_f'\rangle=  \int D\bm{\sigma} D\bm{\sigma}' \exp\left\{i\left[ S_{\bm{\sigma}}-S_{\bm{\sigma}'}\right]\right\} \mathcal{F}_{[\bm{\sigma}, \bm{\sigma}']}.
\label{eq:densitymatrixelement_N_spins}
\end{equation}

The integration runs over all $M$-dimensional constant by part spin paths $\bm{\sigma}$ and $\bm{\sigma}'$ such that $|\bm{\sigma}(t_0)\rangle=|\bm{\sigma}'(t_0)\rangle=|\bm{+}\rangle=\prod_{j=1}^M | +_{z,j} \rangle$, $|\bm{\sigma}(t)\rangle=|\bm{\sigma_f}\rangle$ and $|\bm{\sigma}'(t)\rangle=|\bm{\sigma_f'}\rangle$. $ S_{\bm{\sigma}}$ denotes the free action to follow one given $M$-dimensional spin path without the environment. This free action contains the transverse field terms, and the Ising interaction terms. The effect of the environment is fully contained in the influence functional $\mathcal{F}_{[\bm{\sigma}, \bm{\sigma}']}$, which reads in this case:

\begin{widetext}
\begin{equation}
\mathcal{F}[\bm{\sigma},\bm{\sigma'}]=e^{-\frac{1}{\pi} \int_{t_0}^t ds \int_{t_0}^s ds' \sum_{i,j}\left\{-i \mathcal{L}_1(s-s',x_i-x_j)\frac{ \sigma_i (s)-\sigma_i '(s) }{2} \frac{ \sigma_j (s')+\sigma_j '(s') }{2} +\mathcal{L}_2(s-s',x_i-x_j)\frac{\sigma_i (s)-\sigma_i '(s) }{2} \frac{ \sigma_j (s')-\sigma_j '(s')}{2}\right\}}\times \mathcal{G}[\bm{\sigma},\bm{\sigma'}],
\label{eq:influence_N_spins}
\end{equation}
\end{widetext}

where we have:
\begin{align} &\mathcal{L}_1(t,x)=\frac{1}{2}\left[L_1\left(t-\frac{x}{v_s}\right)+L_1\left(t+\frac{x}{v_s}\right) \right]   \notag \\
&\mathcal{L}_2(t,x)=\frac{1}{2}\left[L_2\left(t-\frac{x}{v_s}\right)+L_2\left(t+\frac{x}{v_s}\right) \right].
\label{Ls}
\end{align}

\begin{figure}[t!]
 \begin{minipage}[b]{0.95\linewidth}
\includegraphics[scale=0.43]{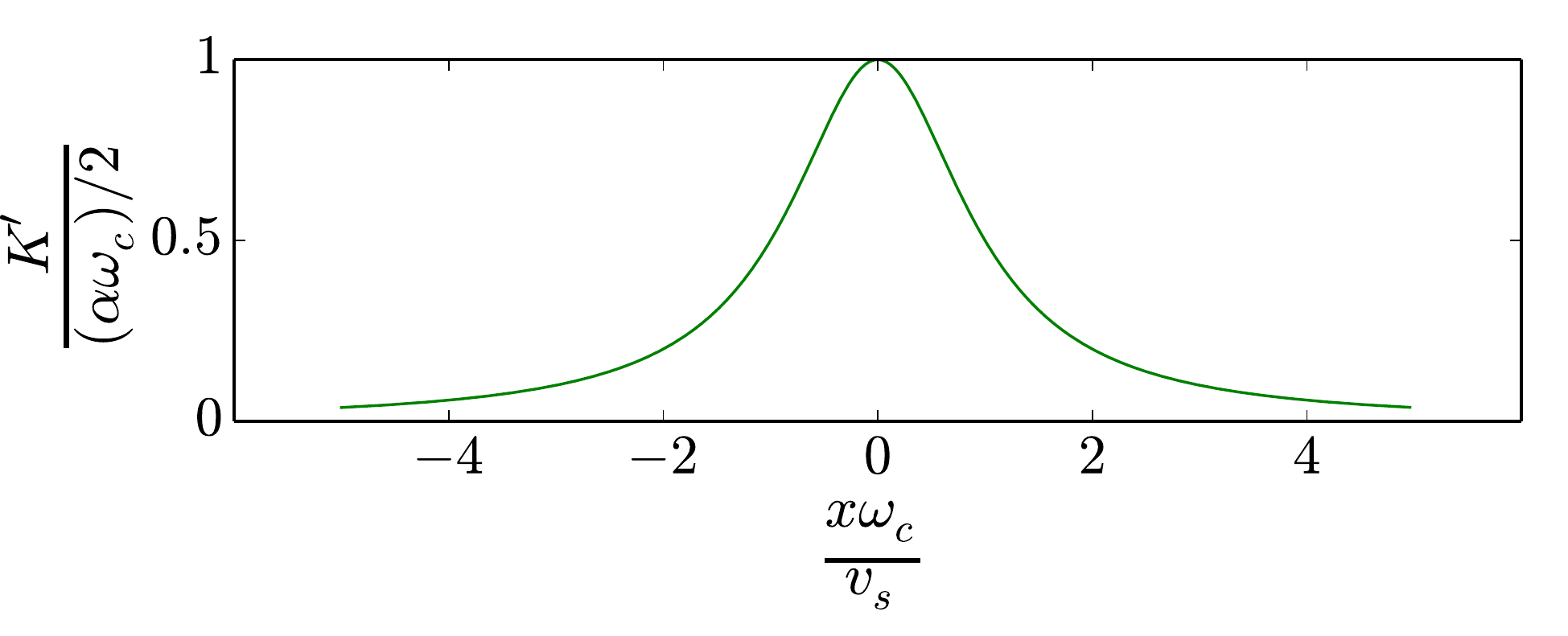} 
\end{minipage}\\
\begin{minipage}[b]{0.52\linewidth}
\includegraphics[scale=0.24]{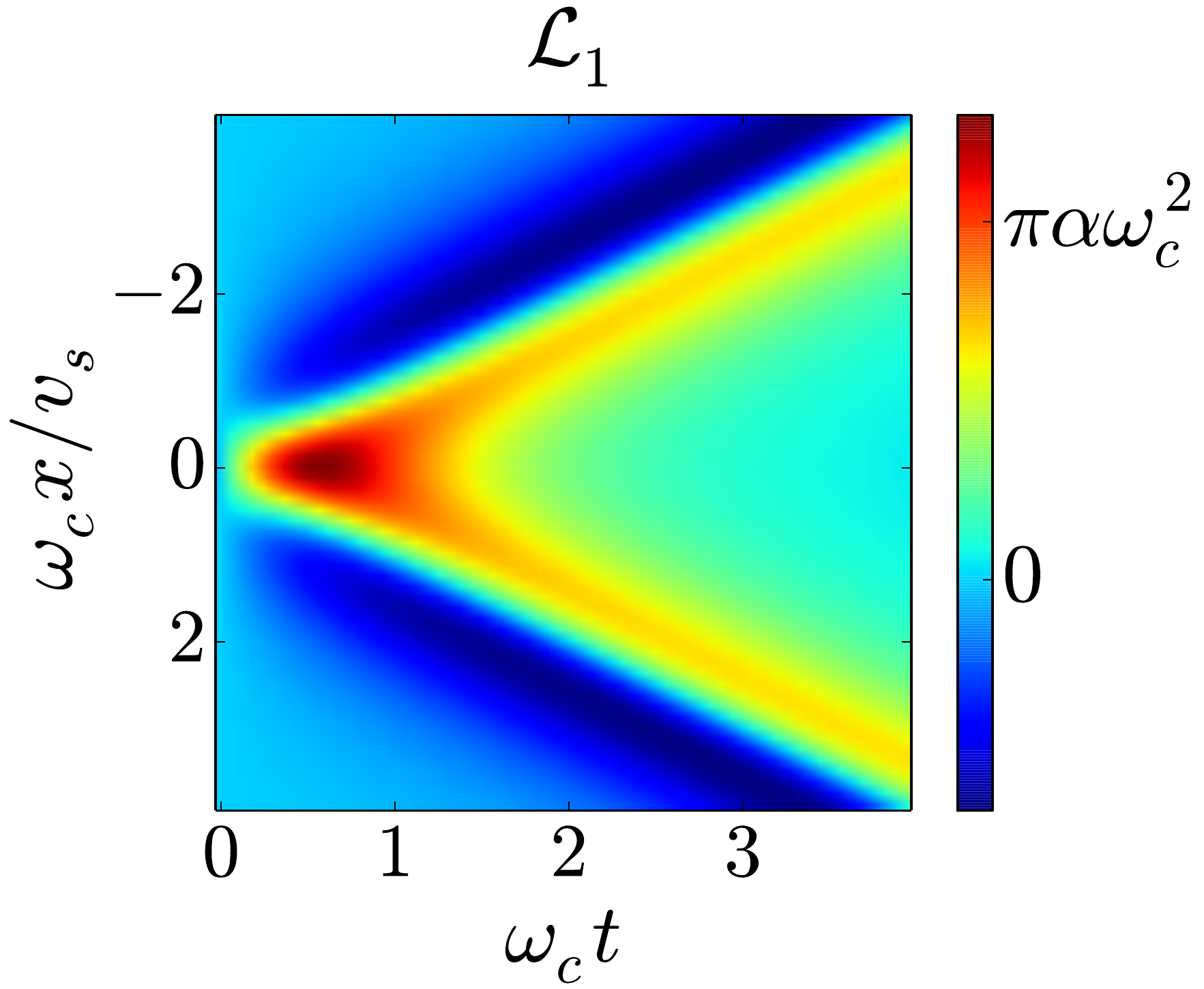} 
\end{minipage}
\begin{minipage}[b]{0.4\linewidth}
\includegraphics[scale=0.24]{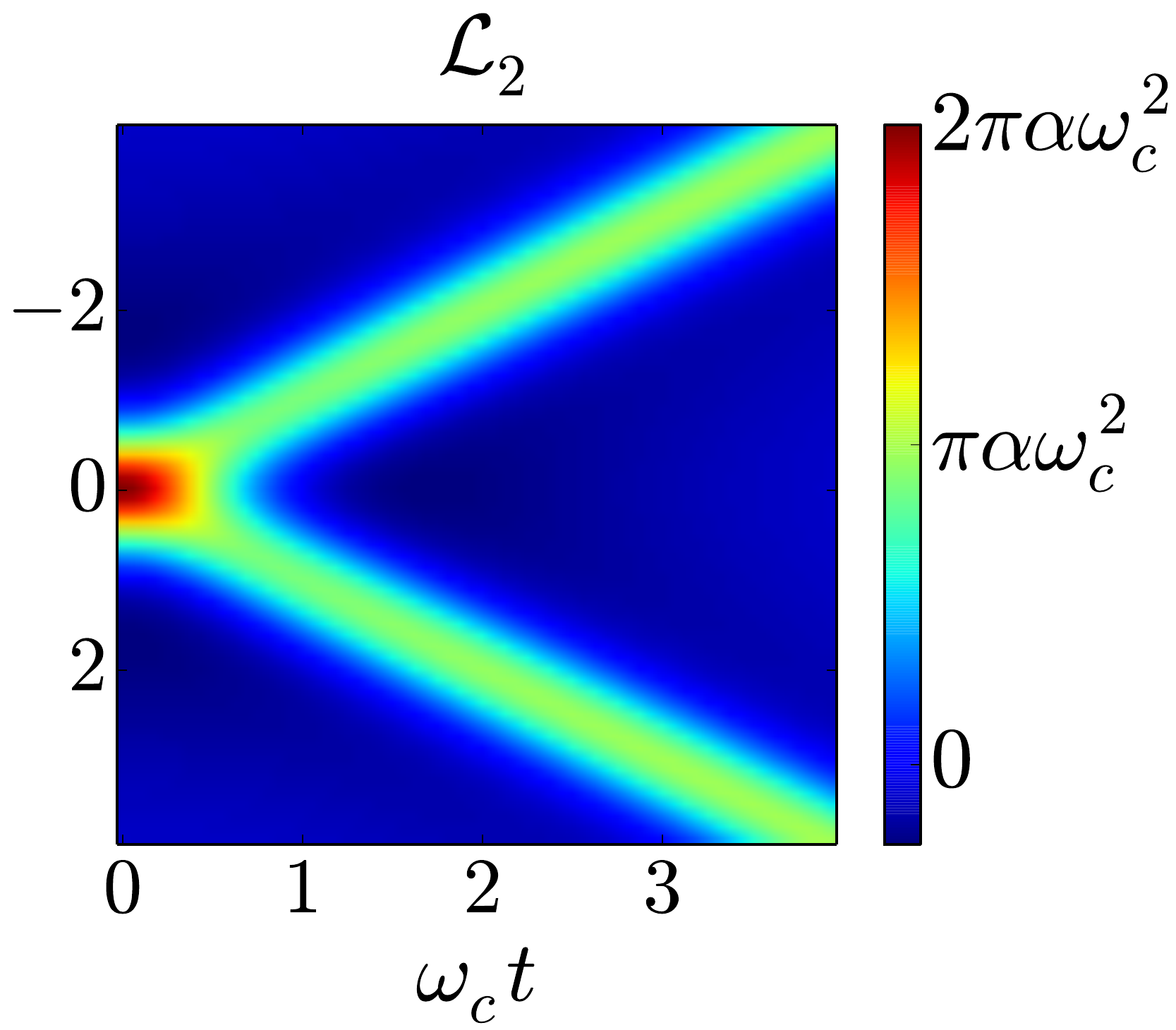}     
\end{minipage}
\caption{(Color online) Top: Evolution of the direct Ising interaction which is induced by the presence of the bath between two spins distant of $x$, as a function of  $x \omega_c/v_s$. Bottom: Space-time dependency of the coupling functions $\mathcal{L}_1$ (left) and $\mathcal{L}_2$ (right). The bath induces a long-range interaction between spins.}
\label{couplin_array}
\end{figure}

The bosonic environment couples the symmetric and anti-symmetric spin paths $\eta^p(t)=1/2[\sigma_p(t)+\sigma_p'(t)] $ and $ \xi^p(t)=1/2[\sigma_p(t)-\sigma_p'(t)]$ at different times and different lattice sites. In Fig.~\ref{couplin_array}, we plot the space and time coupling functions $\mathcal{L}_1$ (bottom left) and $\mathcal{L}_2$ (bottom right). We see that the bosons induce a long-range interaction between spins. The maximal effect between two spins separated by a distance $x$ occurs after a time $x/v_s$, due to the finite sound velocity $v_s$ of the excitations. \\

The last term of Eq. (\ref{eq:influence_N_spins}) reads
\begin{equation}
\mathcal{G}[\bm{\sigma},\bm{\sigma}']=e^{ i \frac{\mu}{2} \int_{t_0}^t ds \left[\sum_{j} \frac{\sigma_j (s)}{2} e^{i k x_j} \right]^2-\left[\sum_{j} \frac{\sigma_j ' (s)}{2} e^{i k x_j} \right]^2},
\label{eq:influence_N_spins_2}
\end{equation}
with  $\mu=2/\pi \int_0^{\infty} J(\omega)/\omega$. We recover that the bath is responsible for an indirect {\it ferromagnetic} Ising-like interaction between the spins $K'_{|j-p|}=1/(2\pi) \int_0^{\infty} J(\omega)/\omega \cos[(x_i-x_j)/v_s]$, whose expression is given in Eq. (\ref{renormalized_coupling}). We plot on the top panel of Fig.~\ref{couplin_array} the value of $K'_{|j-p|}$ with respect to $x\omega_c/v_s$, where $x=x_i-x_j$ is the distance between the two sites $i$ and $j$.\\

The bath is responsible for two distinct types of interactions. The first one is a retarded interaction mediated by the bosonic excitations, which travel at the speed $v_s$. The second one is an instantaneous interaction $K'$, of which we have given a physical interpretation thanks to the polaronic transformation in Eq.~(2). \\

Dealing with the spatial extent remains difficult, and we will treat the array problem at a mean-field level in the limit $M\to \infty$. The spins are coupled through three different terms: the instantaneous direct Ising interaction of strength $K$, the instantaneous interaction mediated by the bath in $\mathcal{G}$, and the retarded interaction mediated by the bath whose expression is given by the first term of the right hand side of Eq. (\ref{eq:influence_N_spins}). We will treat instantaneous spin-spin interactions at a mean field level in the thermodynamic limit $M\to \infty$. In the limit $\omega_c a/v_s \ll 1$, where $a$ is the lattice spacing, we see that the retarded interactions have no effect between \textit{different} spins at a mean field level, since we have $\int_{-\infty}^{\infty} dx \mathcal{L}_1 (s,x)=\int_{-\infty}^{\infty} dx \mathcal{L}_2 (s,x)=0 $. In the following, we will then neglect the retarded interaction between \textit{different} spins, and only conserve the retarded self-interaction. Finally the propagation integral can be factorized in a product of $M$ individual matrix elements, so that it is possible to write: 

\begin{align}
\langle \sigma_{p,f}   | \rho_{S,p} (t) | \sigma_{p,f}'\rangle= \int &D \sigma_p D \sigma'_p  A_p[\sigma_p] A_p[\sigma_p']^* \mathcal{F}_p[\sigma_p,\sigma_p'] \notag \\
&\times e^{-i K_r \int_{t_0}^t ds  \left[\sigma_p (s)-\sigma_p' (s)\right]  \langle \sigma^z_p (s) \rangle },
\label{propagation_array}
\end{align}  
where $\rho_{S,p}$ denotes the density matrix of spin $p$. $A_p[\sigma_p]$ denotes the amplitude to follow a given path for the spin $p$ in the sole presence of the transverse field. We have $K_r=K+2\sum_{j=1}^{\infty} K'_{j}$. The remaining term $\mathcal{F}_p[\sigma_p,\sigma_p']$ encapsulates the effect of the bosonic bath on the spin $p$,

\begin{align}
\mathcal{F}_p[\sigma_p,\sigma_p']=\exp\Big\{ \int_{t_0}^t ds \int_{t_0}^s ds' &\frac{i}{\pi} L_1(s-s') \xi^p(s)\eta^p(s') \notag\\
&-\frac{1}{\pi} L_2(s-s')\xi^p(s)\xi^p(s')  \Big\}.
\label{eq:influence_3}
\end{align}

 We will drop the $p$ index in the following, as all the sites are equivalent in the mean-field description. Following the same steps as in Sec. II, we focus on the computation of $p_2(t)=\langle +_z|\rho_S(t)|+_z\rangle$ and reach the same expression than for $p_0(t)$ (see Eq.~(\ref{eq:p(t)_1})), with 
 \begin{align}
 &\mathcal{F}_{n}= \mathcal{Q}_1 \mathcal{Q}_2 \mathcal{Q}_3. 
  \end{align}
The expressions (\ref{Q_1}) and (\ref{Q_2}) for the expressions of $\mathcal{Q}_1$ and $\mathcal{Q}_2$ are still valid, and we have the additionnal term
  \begin{align}
      &\mathcal{Q}_3 =\exp \left[ -2i K_r \sum_{j=1}^{2n} \Xi_j \int_{t_0}^{t_j} ds \langle \sigma^z (s) \rangle  \right]\label{Q_3} . 
 \end{align}
We then reach for $p_2(t)$ the same expression as the one obtained for $p_0(t)$ in Eq.~(\ref{scalar_prod_SSE}), with the same final vector and $|\phi\rangle$ solution of the SSE (\ref{SSE}), with the effective Hamiltonian given by (\ref{eq:spin_hamiltonian}) provided that we add to the stochastic field $h$ the field $h_I$ defined by $h_{I} (t)=-2iK_r\int_{t_0}^t ds \langle \sigma^z (s) \rangle $. We have then reached an auto-coherent equation, as $\langle \sigma^z(t) \rangle$ enters in the expression of $h_I (t)$. The numerical procedure requires a larger number of realizations of the field $h$ and $k$ compared to the one-spin case. For each realization, we solve the stochastic equation and $\langle \sigma^z(t) \rangle$ is obtained by averaging over the results. The effect of $\langle \sigma^z(t) \rangle$ in $h_I (t)$ is dynamically updated with the number of samplings. \\

We can use our method to compute the free spin dynamics in the limit of $M\to \infty$. We check that the bath causes a decay towards one of the two equilibrium states in the ferromagnetic phase as well as a renormalization of both the tunneling element and the Ising coupling. However, it does not affect the university class (second order with mean-field exponents for the paramagnetic-ferromagnetic transition) of the quantum phase transition as long as the direct Ising term $K$ is \textit{not} zero. This behavior can be understood thanks to a thermodynamic analysis of the action at low wave-vectors $q$ and low frequency $\omega$, which is dominated by the contribution of the long range Ising interaction, as shown in Appendix F.

\subsection{LZ transitions : Array}

We focus now on many-body Landau-Zener sweeps for the array, at a mean-field level. Let us underline that this protocol is different from the dynamical transition of the quantum Ising model in transverse field with nearest neighbours interactions studied in the litterature \cite{sengupta_powell_sachdev,dzarmaga} (and references therein), where the driving parameter is the transverse field and which can be studied elegantly in $k$ space. Here, we are interested in the dynamics of local spin variables at a mean field level. A rigorous description of the dynamics should involve all the energy levels of the system, and their respective avoided crossings. Our mean-field description greatly simplifies the problem and the interplay of all the levels is reduced to a single avoided crossing governed by the local self-consistent Hamiltonian,
\begin{align}
H_j=&\frac{\Delta}{2}\sigma^x_j+\left[\frac{\epsilon(t)}{2}-K_r \langle \sigma^z(t) \rangle \right] \sigma^z_j\notag \\
&+\sum_{k} \left[ \lambda_{k} e^{ik x_j} \left(b^{\dagger}_{-k}+b_k \right) \frac{\sigma_j^z}{2}+\omega_k b^{\dagger}_k b_k \right].
\label{N_spins}
\end{align}

 \begin{figure}[t!]
\center
\includegraphics[scale=0.4]{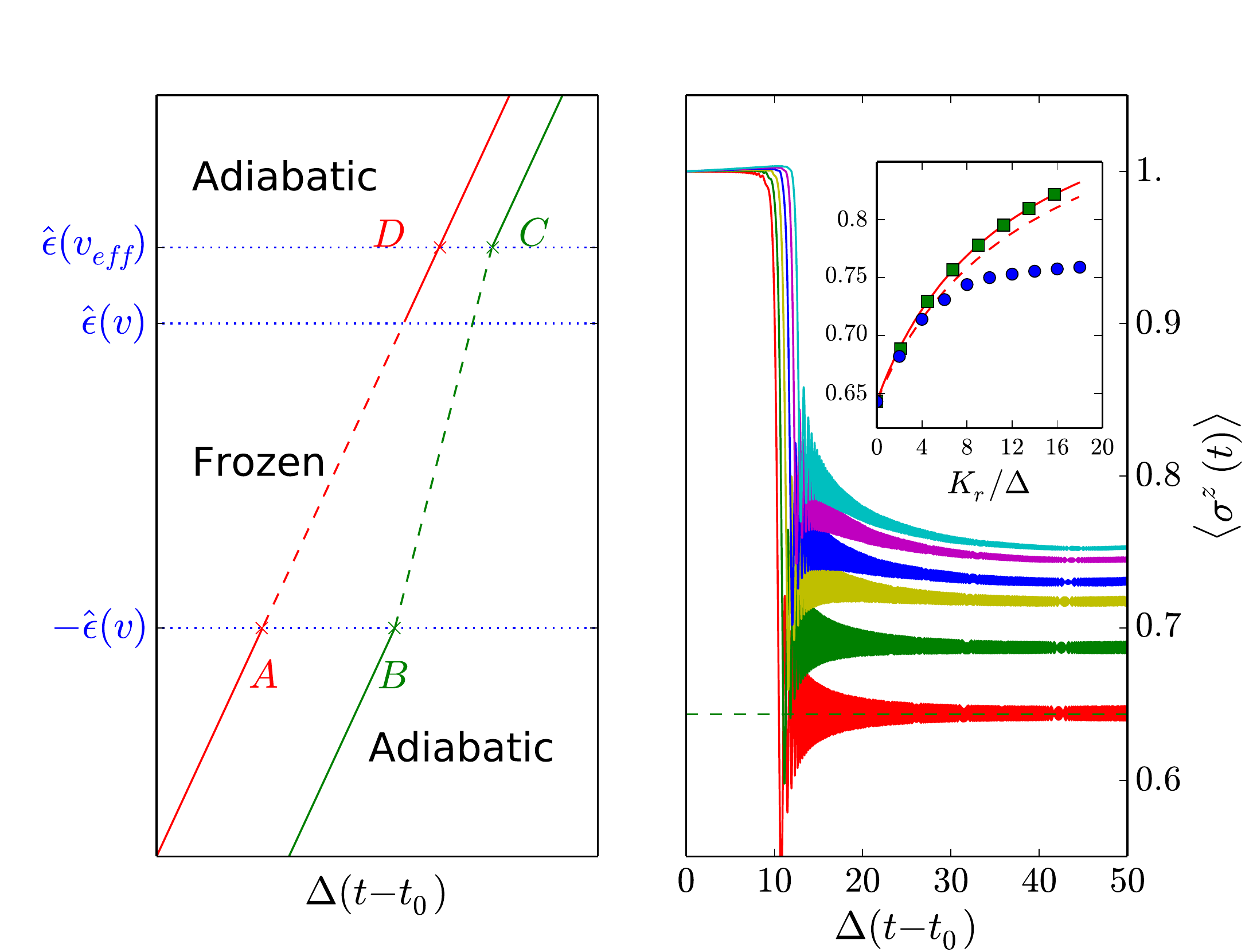}  
\caption{(Color online) Left: schematic interpretation of the Landau-Zener sweep for the array in the framework of the Kibble-Zurek mechanism. The line $(AD)$ shows the evolution of the bare bias field with respect to time, while the broken line connecting points $B$ and $C$ represents the effective bias field. The lines are full during the adiabatic stages, and dashed during the frozen (non-adiabatic) period. Right: Fast sweep ($v/\Delta^2=8$) in the array, for different values of $\alpha$ corresponding to $K_r=0$ (red curve), $K_r=2$ (green curve), $K_r=4$ (yellow curve), $K_r=6$ (blue curve), $K_r=8$ (magenta curve) and $K_r=10$ (cyan curve). We have $K=0$ and $\omega_c=100 \Delta$. The dashed line is the usual theoretical prediction for one spin, following Landau, Zener, Stueckelberg and Majorana; see Sec. III D. Inset: the blue points show the values of $\langle \sigma^z(t\to \infty) \rangle$ with respect to $K_r/\Delta$, corresponding to the parameters of the main plot (interaction mediated by the bath $K_r=\alpha \omega_c$ ($K=0$)). Green squares correspond to a direct Ising interaction $K_r=K$.  The full (dashed) red line shows the expectation value of $\langle \sigma^z(t\to \infty) \rangle$ with respect to $K_r/\Delta$ ($K_r/\Delta_r$) deduced from the Kibble-Zurek mechanism.}
\label{fig_KBZ}
\end{figure}

The presence of the Ising interaction or the presence of the bath both lead to a change in the final value of $\langle \sigma^z (t \to \infty) \rangle$. The origin is the same in both cases at weak coupling and large $\omega_c \gg \Delta$, as the dominant effect of the bath is to induce a ferromagnetic Ising-like interaction $K_r$. In the following we use a Kibble-Zurek argument \cite{Kibble,Zurek} in order to describe quantitatively this effect. The single site fast Landau-Zener transition can indeed be described thanks to the Kibble-Zurek mechanism, which predicts the production of topological defects in non-equilibrium phase transitions\cite{dzarmaga,damski,review_KBZ}. This description splits the dynamics into three consecutive stages: it is supposed to be adiabatic in the first place, then evolves in a non-adiabatic way near the transition point, and finally becomes adiabatic again. The impossibility of the order parameter to follow the change applied on the system provokes this non-adiabatic stage, where the dynamics is said to be ``frozen". It is convenient to introduce the characteristic energy scale \cite{damski} 
\begin{equation}
\hat{\epsilon}=\Delta/\sqrt{2}\left\{ \left[1+16v^2/(\pi^2\Delta^4)\right]^{1/2}-1\right\}^{1/2}, 
\end{equation}
which sets the limit between adiabatic and frozen stages (see left pannel of Fig.~\ref{fig_KBZ}).\\

We first focus on the case where $\alpha=0$ and the direct Ising interaction $K$ is not zero, so that $K_r=K$. The effective field felt by one site is the sum of the bias field $\epsilon (t)$ and the Ising interaction, and will be denoted $\epsilon_{eff} (t)$.  The dynamics always enters in the frozen stage with $\langle \sigma^z \rangle \simeq 1$, so that we have $\epsilon_{eff} (t)=\epsilon (t)-K_r$ during the first adiabatic stage. At the end of the frozen stage, the spin expectation value has changed, and the effective field becomes $\epsilon_{eff} (t)=\epsilon (t)-K_r \langle \sigma^z (t) \rangle$. This leads to a change of the effective speed at which the frozen zone is crossed through, and ultimately of the transition probability. This can be seen on the left pannel of Fig.~\ref{fig_KBZ}, where we show the evolution of both the bare and the effective bias fields with respect to time. We can estimate the renormalization of the effective speed self-consistently thanks to basic geometrical considerations in the trapezoid $(ABCD)$ of the Fig.~\ref{fig_KBZ} (left panel). The effective crossing speed is given by
\begin{equation}
v_{eff}=\frac{\hat{\epsilon}(v)+\hat{\epsilon}(v_{eff})}{t_C(v_{eff})-t_B}.
\label{veff_1}
\end{equation}
The denominator can be simplified by writing that $t_C(v_{eff})-t_B=\left[t_C(v_{eff})-t_D\right]+(t_D-t_A)-(t_B-t_A)$. We know that $(t_D-t_A)=\left[\hat{\epsilon}(v)+\hat{\epsilon}(v_{eff})\right]/v$, and $\left[t_C(v_{eff})-t_D\right]-(t_B-t_A)$  can be expressed as $-K_r\left[1-\langle \sigma^z (t_C,v_{eff}) \rangle\right]$. Next we suppose that we can approximate $\langle \sigma^z (t_C,v_{eff}) \rangle$ by $\langle \sigma^z (t \to \infty) \rangle$. Altogether, we get 
\begin{equation}
\frac{v_{eff}}{v}=\frac{\hat{\epsilon}(v)+\hat{\epsilon}(v_{eff})}{\hat{\epsilon}(v)+\hat{\epsilon}(v_{eff})-2K_r\left[1-p_{lz}(v_{eff})\right]}.
\label{veff_2}
\end{equation}

It allows us to know the variation of the effective speed $v_{eff}$ at which the transition is crossed with respect to the Ising interaction $K_r$. The spin expectation value $\langle \sigma^z (t \to \infty) \rangle$ is then estimated thanks to the Landau-Zener formula, and its evolution with respect to $K_r=K$ is shown by the red curve in the inset of the right part of Fig.~\ref{fig_KBZ}. The estimation matches well the results obtained numerically (green squares).\\

Now we take $K=0$ and $\alpha$ not zero so that $K_r=\alpha \omega_c$. We plot on the right panel of Fig.~\ref{fig_KBZ} the dynamics obtained with the SSE. We see in the inset that, at small $\alpha$, the estimation of the final value of the spin variable thanks to Eq. (\ref{veff_2}) is correct. However, it breaks down when the dissipation strength is increased because the assumption $\langle \sigma^z (t_C,v_{eff}) \rangle \simeq\langle \sigma^z (t \to \infty) \rangle$ used to derive $v_{eff}$ is no longer correct. Relaxation processes occur after the crossing of the frozen zone which lower $\langle \sigma^z(t \to \infty)\rangle$. This can be seen on the behaviour of the curves obtained at large values of $\alpha$ (the cyan curve for example), where the spin expectation value continues to go down during a rather long time after the crossing. The dotted red curve takes into account the renormalization of the tunneling frequency $\Delta_r$ due to the presence of the bath.\\

We have studied dissipative sweep protocols in the case of the dimer model and for an infinite array. The coupling to a dissipative environment is responsible for both an Ising-type interaction and relaxation processes. In the regime where $\Delta/\omega_c \ll 1$ the predominance of the bath-induced Ising like interaction on relaxation mechanisms renders possible a quantitative prediction of the dynamics. In the case of two spins, it was indeed possible to understand the evolution of the energy levels and take into account the three avoided crossings. Increasing the number of spins would lead to a larger number of level crossings and a more complex energy level structure, as one should take into account the side-by-side avoided crossings of all the energy levels (except the eventual singlet which remains isolated). In the case of a large number of spins with long-range interactions, we recover a local spin-1/2 view on the dynamics which can be understood with a single-crossing view, by analogy with mean-field methods. In this case, the self-consistency equation comes from a Kibble-Zurek type criterion with adiabaticity considerations.
 
\section{Conclusion}
 
To summarize, we have developed a stochastic approach to address the spin dynamics in dissipative quantum spin arrays. Using complex gaussian random fields, we have carefully studied the applicability of the method in 
all the applications. We focused first on the quantum phase transition displayed by two spins in contact with the same ohmic bath, and studied quenched dynamics both in the unpolarized and in the polarized phase. We also investigated quantitatively bath-induced synchronization phenomena occurring in this system. Then we considered non-equilibrium sweep protocols, both in the case of two spins and for the quantum Ising chain with long-range forces. In this latter case, the dynamics can be understood thanks to a simple Kibble-Zurek type argument. Our results can be tested in ultra-cold atom systems \cite{recati_fedichev,orth_stanic_lehur}. The method could also be applied to the sub-ohmic spin-boson model \cite{Bulla,Doucet}, Jaynes-Cummings or Rabi arrays \cite{Rabi_article,review}, for topological problems with Dirac points \cite{Montambaux}, and for fermionic environments \cite{Schiro,Millis,DemlerKondo}, as in Kondo lattices \cite{Si}. 

We thank Camille Aron, Lo\"{i}c Herviou, Walter Hofstetter, Christophe Mora, Peter P. Orth, Zoran Ristivojevic, Guillaume Roux, Marco Schiro for discussions. This work has been supported by PALM Labex, project Quantum-Dyna (ANR-10-LABX-003).  


\appendix
  \onecolumngrid
\section{Feynman-Vernon influence functional}    

Here we derive the expression (7) given in the main text, using the method of Ref. \onlinecite{Brandes_course}. In order to simplify the derivation, we first consider that one single bosonic mode is coupled to the spin, and we have the Hamiltonian
\begin{equation}
H=\frac{\Delta}{2} \sigma^x+\frac{\lambda}{2} (b+b^{\dagger}) \sigma^z + \omega b^{\dagger} b.
\end{equation}
The general case of several modes will be deduced from this simpler case at the end of this appendix.  Let us call $H_S=\Delta/2 \sigma^x$ the spin-part and $H_B=\hbar \omega b^{\dagger}b+\frac{\lambda}{2} \sigma^z (b+b^{\dagger})$ the interaction part. From Eq.~(4) of the main text and after the introduction of the identity both on the left and on the right of the term $\rho (t_0)$, we get 

\begin{align}
\langle \sigma_f |\rho_S (t)|\sigma_f' \rangle=\sum_{n,m,p,k,k'}& \Big\{ \langle u_n,\sigma_f |U(t)|u_m,\sigma_k \rangle  \langle u_m,\sigma_k |\rho(t_0)|u_p,\sigma_{k'} \rangle \langle u_p,\sigma_{k'} |U(t)|u_m,\sigma_f' \rangle \Big\}.
\end{align}

Next, we use the factorising initial condition ${\rho(t_0)=\rho_B(t_0) \otimes |+_z\rangle \langle +_z|}$ and reach

\begin{align}
\langle \sigma_f |\rho_S (t)|\sigma_f' \rangle=\sum_{n,m,p}& \Big\{  \langle u_m |\rho_B (t_0)|u_p \rangle \langle u_n,\sigma_f |U(t)|u_m,+_z \rangle\langle u_p,+_z |U(t)|u_m,\sigma_f' \rangle \Big\}.
\end{align}

The last two terms can be expressed thanks to a path integral. The resulting action can be divided into two parts $S_S$ and  $S_B$, the first one resulting from the spin Hamiltonian alone and the second one resulting from the remaining part. Factorising the spin part, we reach the equation (6) of the main text, where we get
\begin{equation}
F[\sigma, \sigma']=tr_B \left\{\rho_{B}(t_0) U_{B}[\sigma] (t) U_{B}^{\dagger}[\sigma'](t) \right\} ,
\end{equation}
where $U_B[\sigma]$ being the time evolution operator related to $H_B$ where $\sigma$ is a classical time-dependent spin-path. In order to evaluate this functional we need to derive the expression of the bath evolution operator. To do so, we switch to the interaction picture (where $V=\lambda/2(a+a^{\dagger})\sigma$ is the interaction term) and define $\tilde{U}_{B}[\sigma](t)$ the corresponding time evolution operator. We have :
\begin{equation}
i \hbar \partial_t \tilde{U}_{B}[\sigma](t)=\tilde{V}(t) \tilde{U}_{B}[\sigma](t) 
\end{equation}
Defining $\hat{X}=\frac{b+b^{\dagger}}{\sqrt2}$, and $\hat{P}=\frac{b-b^{\dagger}}{i \sqrt2}$, the commutation relations gives:\newline

$\left\{
\begin{array}{l}
  e^{-i\omega b^{\dagger}b t} \hat{X} e^{i\omega b^{\dagger}b t}=\hat{X}+\omega t ~e^{-i\omega b^{\dagger}b t}  \hat{P} e^{i\omega b^{\dagger}b t} \\
  e^{-i\omega b^{\dagger}b t} \hat{P} e^{i\omega b^{\dagger}b t}=\hat{P}-\omega t ~ e^{-i\omega b^{\dagger}b t}  \hat{X} e^{i\omega b^{\dagger}b t}
\end{array}
\right.$\newline

which results in:
\begin{equation}
\tilde{V}(t)=\frac{\lambda}{2} \sigma^x (t) [(b+b^{\dagger}) \cos \omega t+\frac{b-b^{\dagger}}{i} \sin \omega t ].
\end{equation}

As the evolution operator $\tilde{U}_{B}[\sigma](t)$ is unitary, we suppose that we can write it as ${e^{-i \alpha (t)} ~e^{-i \beta(t)(b+b^{\dagger}) } ~e^{-i \gamma (t) \frac{(b-b^{\dagger})}{i}}}$. The Schr\"{o}dinger equation gives us the expression of $\alpha$, $\beta$, $\gamma$:\newline

$\left\{
\begin{array}{l}
\beta(t)= \int_{t_0}^t ds \frac{\lambda}{2} \sigma (s) \cos \omega s   \\
 \gamma(t)=\int_{t_0}^t ds  \frac{\lambda}{2}  \sigma (s) \sin \omega s    \\
 \alpha(t)=-\int_{t_0}^t ds \int_0^s ds' \left(\frac{\lambda}{2}\right)^2 \sigma (s) \sigma (s') \cos \omega s'  \sin \omega s 
\end{array}
\right.$\\

Then, we have :

\begin{align}
 F[\sigma, \sigma']=e^{i \left[\alpha'\left(t\right)-\alpha(t) \right]}\int dX \langle X|\rho_B&(0) ~e^{i\gamma'(t)\hat{P}} ~e^{i\beta'(t)\hat{X}} e^{-i\beta(t)\hat{X}} ~e^{-i\gamma(t)\hat{P}}  | X \rangle,
\end{align}
where the states $|X\rangle$ represent a complete set of position eigenstates. It simplifies into
\begin{align}
F[\sigma, \sigma']=e^{i\left[\alpha'(t)-\alpha(t)\right]}\int dX &\langle X|\rho_B(0) | X+\gamma(t)-\gamma'(t) \rangle e^{i\left[\beta'(t)-\beta(t)\right]\left[X+\gamma(t)\right]}.
\end{align}

In order to evaluate the element ${\langle X|\rho_B(t_0) | X+\gamma(t)-\gamma'(t) \rangle}$, we assume a thermal equilibrium at inverse temperature $\beta$ for the operator $\rho_B(t_0)$:

\begin{align}
\langle X_1|\rho_B(0) | X_2\rangle=\frac{1}{Z} &\left(\frac{1}{2\pi \sinh \beta \omega_0 }\right)^{\frac{1}{2}} e^{-\frac{1}{2 \sinh \beta \omega_0 }\left[\left(X_1^2+X_2^2\right) \cosh \beta \omega_0 -2X_1X_2\right] }. 
\end{align}

Using the properties of Gaussian integrals, as well as the identity $\frac{\cosh \beta \omega_0 -1}{ \sinh \beta \omega_0 }=\tanh \beta \omega_0 /2$, we get:

\begin{align}
F[\sigma, \sigma']=&e^{i\big[\alpha'(t)-\alpha(t)\big]+i\big[\beta'(t)-\beta(t)\big]\big[\gamma(t)+\gamma'(t)\big]  } e^{-\frac{1}{4} \coth \beta \omega_0 / 2\big[(\beta'(t)-\beta(t))^2+(\gamma(t)-\gamma'(t))^2\big] }.
\end{align}
Hence re-inserting the expressions of $\alpha$, $\beta$ and $\gamma$ and after trigonometric calculations and using the symmetry of the integrand we finally  recover Eq.~(7) of the main text, with $L_1(t)=\pi \lambda^2 \sin \omega_0 t$ and $L_2(t)=\pi \lambda^2 \cos \omega_0 t \coth \beta \omega_0/2$. The generalization to an infinite number of modes is straightforward.
  
\section{Blip-sojourn development and derivation of Eq. (\ref{Q_1}) and Eq. (\ref{Q_2})}
Given a path (see Fig.~\ref{spin_path_1} of the main text for example), we can evaluate Eq. (\ref{eq:influence}) of the main text. First we evaluate the contribution given by $L_1$.
\begin{align}
\frac{i}{\pi}\int_{t_0}^t ds \int_{t_0}^s ds' L_1(s-s') \xi(s) \eta(s')=&\frac{i}{\pi} \sum_{j>k=0}^n \xi_j \eta_k \int_{t_{2j-1}}^{t_{2j}} ds \int_{t_{2k}}^{t_{2k+1}} ds'L_1(s-s') \notag\\
=&\frac{i}{\pi} \sum_{j>k=0}^n \xi_j \eta_k [ Q_1(t_{2j-1}-t_{2k})+Q_1(t_{2j}-t_{2k+1})-Q_1(t_{2j}-t_{2k}) -Q_1(t_{2j-1}-t_{2k+1})] \notag\\
=&\frac{i}{\pi} \sum_{j>k=0}^{2n} \Xi_j \Upsilon_k  Q_1(t_j-t_k),
\end{align}
with $Q_1$ the opposite of the second integral of $L_1$ with $Q_1(0)=0$. Then we evaluate the contribution given by $L_2$.
\begin{align}
-\frac{1}{\pi}\int_{t_0}^t ds \int_{t_0}^s ds' L_2(s-s') \xi(s) \xi(s')=&-\frac{1}{\pi} \sum_{j>k=0}^n \xi_j \xi_k \int_{t_{2j-1}}^{t_{2j}} ds \int_{t_{2k-1}}^{t_{2k}} ds'L_2(s-s') -\frac{1}{\pi} \xi_j \xi_j \int_{t_{2j-1}}^{t_{2j}} ds \int_{t_{2j-1}}^{s} ds'L_2(s-s') \notag\\
=&\frac{1}{\pi} \sum_{j>k=0}^n \xi_j \xi_k [ Q_2(t_{2j-1}-t_{2k-1})+Q_2(t_{2j}-t_{2k})-Q_2(t_{2j}-t_{2k-1}) -Q_2(t_{2j-1}-t_{2k})] \notag\\
&-\frac{1}{\pi} \sum_{j} \xi_j \xi_j Q_2(t_{2j}-t_{2k-1})\notag\\
=&\frac{1}{\pi} \sum_{j>k=0}^{2n} \Xi_j \Xi_k  Q_2(t_j-t_k),
\end{align}
with $Q_2$ the second integral of $L_2$ with $Q_2(0)=0$. We recover Eq. (\ref{10}), (\ref{Q_1}) and (\ref{Q_2}) of the main text, where $\mathcal{Q}_1$ contains the coupling of the blips to all the previous sojourns, and $\mathcal{Q}_2$ contains the coupling of the blips to all the previous blips (including self-interaction). 
  
\section{Sampling of the stochastic variables and numerical convergence}  
In order to sample the variables $h$ and $k$ which verify the correlations of Eq. (\ref{height_1}), (\ref{height_2}) and (\ref{height_3}), we use a Fourier series decomposition of the functions $Q_1$ and $Q_2$. To do so, we introduce the variable $\tau=t/t_f$ where $t_f$ is the final time of the experiment/simulation. Hence $\tau \mapsto Q_2 (\tau t_f)$ and $\tau \mapsto Q_1 (\tau t_f)\theta (\tau)$ are defined on $[-1,1]$. We extend their definitions by making them 2-periodic functions and it is then possible to expand them in Fourier series. In particular, we have:

\begin{align}
  &\frac{Q_2 \left[(\tau_j-\tau_k)t_f \right]}{\pi}=\frac{g_0}{2}+\sum_{m=1}^{\infty} \frac{g_m}{2}\left[\phi_m(\tau_j)\phi_m^*(\tau_k)+h.c. \right],\notag\\
   &\frac{Q_1 \left[(\tau_j-\tau_k)t_f \right]}{\pi}\theta(\tau_j-\tau_k)=\frac{f_0}{2}+\sum_{m=1}^{\infty} \frac{f^s_m}{2}\left[\phi_m(\tau_j)\phi_m^*(\tau_k)+h.c. \right]+\sum_{m=1}^{\infty} \frac{f^a_m}{2}\left[\phi_m(\tau_j)\phi_m^*(\tau_k)-h.c. \right],
\end{align}
  where $\phi_m:\tau \mapsto \exp(im\pi\tau)$, and we have for $m>1$, $\left\{g_m=\int_{-1}^1 d\tau \frac{Q_2 (\tau t_f)}{\pi} \cos m \pi \tau \right\}$, $\left\{f^s_m=\int_{-1}^1 d\tau  \frac{Q_1 \left(\tau t_f \right)}{\pi}\theta(\tau) \cos m \pi \tau \right\}$, and $\left\{f_m^{a}=\int_{-1}^1 d\tau  \frac{Q_1(\tau t_f)}{\pi}\theta(\tau) \sin m \pi \tau \right\}$. $g_0$ and $f_0$ are the constant Fourier coefficients. Then we define $h$ and $k$ as

  \begin{align}  
 h(\tau t_f) =\sum_{m=1}^{\infty} &\phi_m(\tau t_f)\Big[\left(\frac{g_m}{4}\right)^{\frac{1}{2}}(s_{1,m}+is_{2,m})+\left(\frac{f_m^s}{4}\right)^{\frac{1}{2}}(u_{1,m}+iu_{2,m})+\left(\frac{f^{a}_m}{4}\right)^{\frac{1}{2}}(v_{1,m}+iv_{2,m})      \Big] \notag \\
+&\phi_m^*(\tau t_f)\Big[\left(\frac{g_m}{4}\right)^{\frac{1}{2}}(s_{1,m}-is_{2,m})+\left(\frac{f_m^s}{4}\right)^{\frac{1}{2}}(u_{3,m}+iu_{4,m})+\left(\frac{f^{a}_m}{4}\right)^{\frac{1}{2}}(v_{3,m}+iv_{4,m})      \Big],\\
~\notag \\
 k(\tau t_f) =\sum_{m=1}^{\infty} &\phi_m(\tau t_f)\Big[\left(\frac{f_m^s}{4}\right)^{\frac{1}{2}}(u_{1,m}+iu_{2,m})+\left(\frac{f^{a}_m}{4}\right)^{\frac{1}{2}}(v_{1,m}+iv_{2,m})      \Big] \notag \\
+&\phi_m^*(\tau t_f)\Big[\left(\frac{f_m^s}{4}\right)^{\frac{1}{2}}(u_{3,m}+iu_{4,m})+\left(\frac{f^{a}_m}{4}\right)^{\frac{1}{2}}(v_{3,m}+iv_{4,m})      \Big],
\end{align}    

where $\{s_{i,m}\}$, $\{u_{i,m}\}$ and $\{v_{i,m}\}$ are standard normal variables. One can check that $h$ and $k$ verify the correlations given by Eqs. (\ref{height_1}), (\ref{height_2}) and (\ref{height_3}) of the main text. In general these fields are complex, and the presence of a non-zero real part may lead to an exponential slowing down of the convergence. We can check that $g_m<0$ for all $m$, such that the part of the field $h$ coming from $Q_2$ is purely imaginary. On the other hand, we always have terms of the form $u_1+i u_2$ when it comes to the decoupling of $Q_1$. It is then impossible to constrain the real part of the fields coming from the $Q_1$ decomposition. In the numerics, we use Fast Fourier Transform in order to increase the speed of the numerical procedure. We could have decomposed the fields on another basis of functions, but the choice of Fourier series decomposition seems natural, given the form of the functions $Q_1$ and $Q_2$ in Eq. (\ref{q1}) and Eq. (\ref{q2}).\\

\section{Expressions of $\mathcal{M}_1^p$ and $\mathcal{M}_2^p$}   

\begin{figure*}[h!]
  \includegraphics[width=4cm,height=5cm]{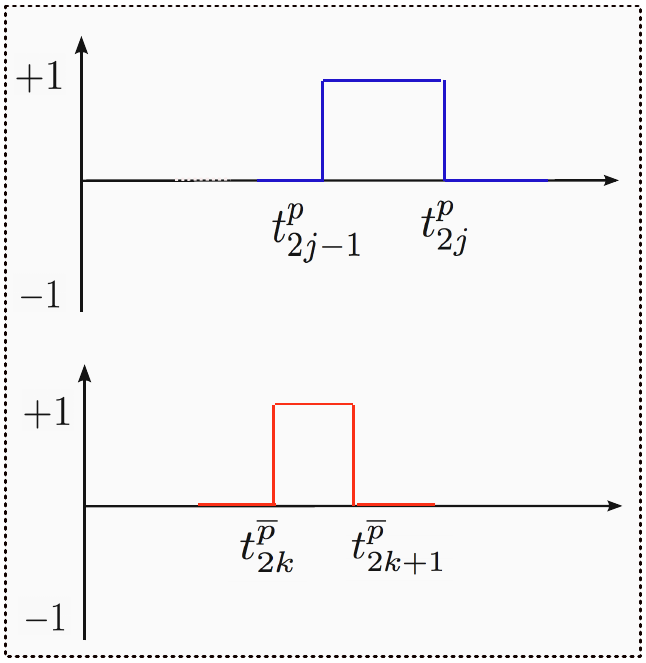}\includegraphics[width=4cm,height=5cm]{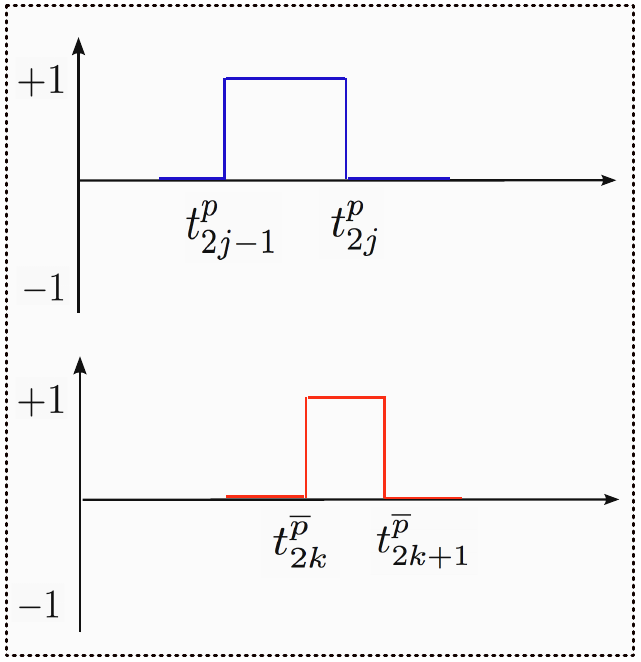}  \includegraphics[width=4cm,height=5cm]{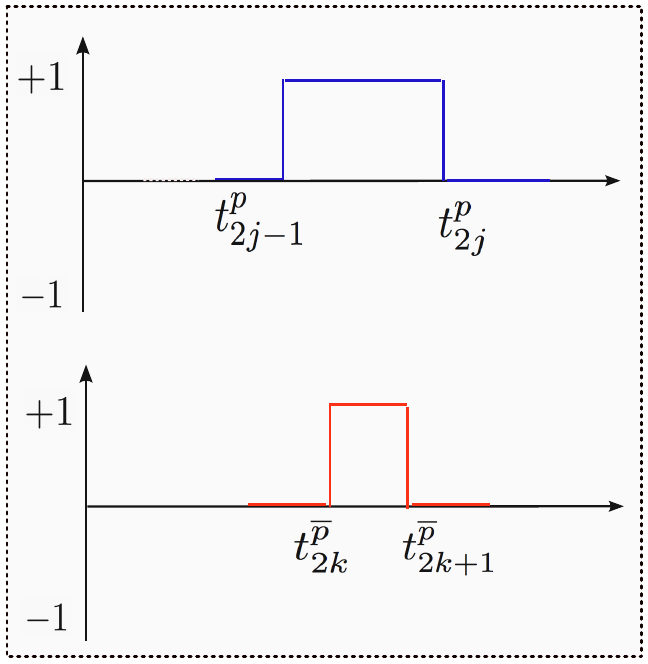}\includegraphics[width=4cm,height=5cm]{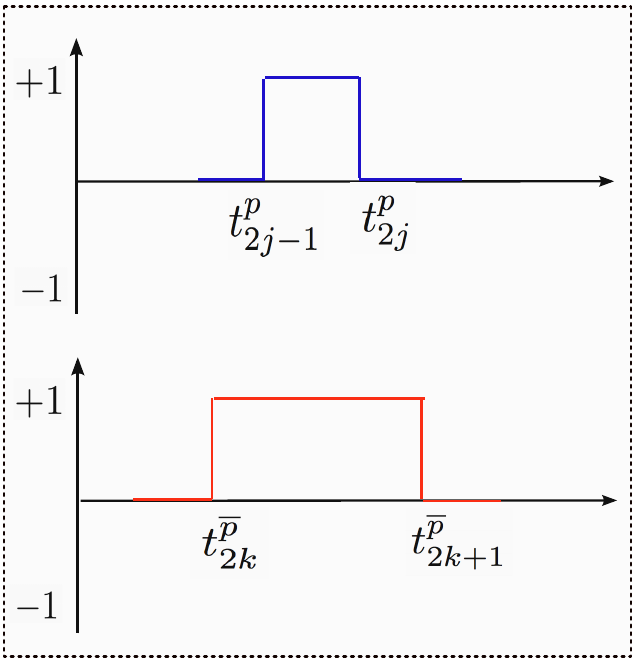}
  \caption{(Color online) Coupling of a blip of the spin $p$ with a simultaneous sojourn of the spin $\overline{p}$. There are four distinct configurations. }
  \label{blip_sojourn_appendix}
\end{figure*}

The derivation of $\mathcal{Q}_1^p$ and $\mathcal{Q}_2^p$ can be found in the Appendix B. In this case, the blip and sojourn variable cannot be simultaneously both non-zero. For $\mathcal{M}_1^p$ and $\mathcal{M}_2^p$, the situation is different as the state of the first spin does not constrain the state of the second one. More explicitly, for $\mathcal{M}_1^p$ for example, one of the spins may be in a blip state while the second one is in a sojourn state, as illustrated in Fig.~\ref{blip_sojourn_appendix}. In the following, we will compute the contribution of these particuliar blip-sojourn configurations.\\


The first case (left panel) yields,

\begin{align}
-\frac{i}{\pi} \int_{t_{2j-1}^p}^{t_{2j}^p} ds \int_{t_{2k}^{\overline{p}}}^{s} ds' \xi^p(s) &\eta^{\overline{p}}(s') L_1(s-s') =-\frac{i}{\pi} \Xi^p_{2j-1}\Upsilon^{\overline{p}}_{2k} \left[\int_{t_{2j-1}^p}^{t_{2k+1}^{\overline{p}}} ds \int_{t_{2k}^{\overline{p}}}^{s} ds' L_1(s-s')+ \int_{t_{2k+1}^{\overline{p}}}^{t_{2j}^p} ds \int_{t_{2k}^{\overline{p}}}^{t_{2k+1}^{\overline{p}}} ds' L_1(s-s') \right]  \notag \\
&=\frac{i}{\pi} \left[\Xi^p_{2j-1}\Upsilon^{\overline{p}}_{2k} Q_1(t^p_{2j-1}-t^{\overline{p}}_{2k})+\Xi^p_{2j}\Upsilon^{\overline{p}}_{2k} Q_1(t^p_{2j}-t^{\overline{p}}_{2k})+\Xi^p_{2j}\Upsilon^{\overline{p}}_{2k+1} Q_1(t^p_{2j}-t^{\overline{p}}_{2k+1}) \right].
\end{align}
The second configuration gives,
\begin{align}
-\frac{i}{\pi} \int_{t_{2j-1}^p}^{t_{2j}^p} ds \int_{t_{2k}^{\overline{p}}}^{s} ds' \xi^p(s) \eta^{\overline{p}}(s') L_1(s-s') &=-\frac{i}{\pi} \Xi^p_{2j-1}\Upsilon^{\overline{p}}_{2k} \left[\int_{t_{2k}^{\overline{p}}}^{t_{2j}^{p}} ds \int_{t_{2k}^{\overline{p}}}^{s} ds' L_1(s-s') \right]  \notag \\
&= \frac{i}{\pi} \left[\Xi^p_{2j}\Upsilon^{\overline{p}}_{2k} Q_1(t^p_{2j}-t^{\overline{p}}_{2k})\right].
\end{align}

The third configuration gives,
\begin{align}
-\frac{i}{\pi} \int_{t_{2j-1}^p}^{t_{2j}^p} ds \int_{t_{2k}^{\overline{p}}}^{s} ds' \xi^p(s) \eta^{\overline{p}}(s') L_1(s-s') &=-\frac{i}{\pi} \Xi^p_{2j-1}\Upsilon^{\overline{p}}_{2k} \left[\int_{t_{2k}^{\overline{p}}}^{t_{2k+1}^{\overline{p}}} ds \int_{t_{2k}^{\overline{p}}}^{s} ds' L_1(s-s')+\int_{t_{2k+1}^{\overline{p}}}^{t_{2j}^{p}} ds \int_{t_{2k}^{\overline{p}}}^{t_{2k+1}^{\overline{p}}} ds' L_1(s-s') \right]  \notag \\
&=\frac{i}{\pi} \left[\Xi^p_{2j}\Upsilon^{\overline{p}}_{2k} Q_1(t^p_{2j}-t^{\overline{p}}_{2k})+\Xi^p_{2j}\Upsilon^{\overline{p}}_{2k+1} Q_1(t^p_{2j}-t^{\overline{p}}_{2k+1})\right].
\end{align}

The  fourth configuration gives,
\begin{align}
-\frac{i}{\pi} \int_{t_{2j-1}^p}^{t_{2j}^p} ds \int_{t_{2k}^{\overline{p}}}^{s} ds' \xi^p(s) \eta^{\overline{p}}(s') L_1(s-s') &=-\frac{i}{\pi} \Xi^p_{2j-1}\Upsilon^{\overline{p}}_{2k} \left[\int_{t_{2j-1}^{p}}^{t_{2j}^{p}} ds \int_{t_{2k}^{\overline{p}}}^{s} ds' L_1(s-s') \right]  \notag \\
&=\frac{i}{\pi} \left[\Xi^p_{2j-1}\Upsilon^{\overline{p}}_{2k} Q_1(t^p_{2j-1}-t^{\overline{p}}_{2k})+\Xi^p_{2j}\Upsilon^{\overline{p}}_{2k} Q_1(t^p_{2j}-t^{\overline{p}}_{2k})\right].
\end{align}

We finally recover the expression (\ref{Q_1_m_2}) of the main text. An analog computation permits to find back Eq. (\ref{Q_1_m_2}).
\section{Scaling regime}  

In the scaling regime $\Delta/\omega_c \ll 1$, it is possible to overcome the sign problem naturally arising in our method as shown in Ref. \onlinecite{stochastic}. Simplifications occur in Eqs. (\ref{Q_1_2}) and (\ref{Q_1_m_2}) as we can consider that  $ Q_1(t_j-t_k)=2 \pi  \alpha \tan^{-1} \left[\omega_c (t_j-t_k)\right]\simeq \pi^2  \alpha$. Then we have

\begin{align}
\sum_{k=0}^{2 n_p -1} \sum_{j : t_j^q>t_k^p} \Xi_j^q \Upsilon_k^p   Q_1(t_j^q-t_k^p) &=i\pi \alpha  \left[\sum_{j=1}^{2 n_q} \xi_j^q \eta^p_l \right] ,
\end{align}
for $q=p$ or $p=\overline{p}$. $\xi_j^q$ is the value of $\xi^q (t)$ in the interval $[t_j^q,t_{j+1}^q]$ and $\eta^p_l $ is the value of  $\eta^p (t)$ in the interval $[t_l^p,t_{l+1}^p]$. The integer $l$ is defined by $t_l^p<t_j^q \leq t_{l+1}^p$. In the case of $p=q$, we just have $l=j-1$. 

This expression does not depend on intermediate times, but only on the path taken. As a result, there is no need to introduce the time-dependent field $k$. After having introduced the field $h$ as in the main text, we finally recover Eqs. (\ref{eq:p1:two_spins}) and (\ref{eq:SSE:two_spins}) of the main text, with

\begin{small}
\[
V_{1}= \left( \begin{array}{ccccccc}

\begin{array}{cccc}
0&\frac{e^{-h}}{a}&-a e^{ h}&0 \\
a^2 e^{h}&0&0&-e^{h }\\
-\frac{e^{- h}}{a^2}&0&0&e^{- h }\\
0&-\frac{e^{-h}}{a}&a e^{h }&0
\end{array}  & &\begin{array}{cccc}
\frac{e^{-h}}{a}~~&0~~&0~~&0 \\
0~~&e^{-h}~~&0~~&0\\
0~~&0~~& e^{-h}~~&0\\
0~~&0~~&0~~&a e^{-h}
\end{array} & &\begin{array}{cccc}
-a e^{h}~&0~&0~&0 \\
0~&-e^{h}~&0~&0\\
0~&0~&-e^{h}~&0\\
0~&0~&0~&-\frac{e^{h}}{a}
\end{array} & &\textbf{(0)}\\
 & & & & & &\\
 
\begin{array}{cccc}
a^2 e^{h}~~~&0~~~&0~~~&0 \\
0~~~&a e^{h}~~~&0~~~&0\\
0~~~&0~~~&a e^{h}~~~&0\\
0~~~&0~~~&0~~& e^{h}
\end{array} & &\begin{array}{cccc}
0~&e^{- h }~&-e^{ h}~&0 \\
a e^{h }~&0~&0~&-\frac{e^{h}}{a}\\
-\frac{e^{- h}}{a}~&0~&0~&a e^{- h }\\
0~&-e^{- h }~&e^{h }~&0
\end{array} & &\textbf{(0)}& &\begin{array}{cccc}
-e^{h}~&0~&0~&0 \\
0~&-\frac{e^{h}}{a}~&0~&0\\
0~&0~&-\frac{e^{h}}{a}~&0\\
0~&0~&0~&-\frac{e^{h}}{a^2}
\end{array} \\
 & & & & & &\\
 
\begin{array}{cccc}
-\frac{e^{- h}}{a^2}&0&0&0 \\
0&-\frac{e^{-h}}{a}&0&0\\
0&0&-\frac{e^{- h}}{a}&0\\
0&0&0&-e^{-h}
\end{array}&  &\textbf{(0)}& & \begin{array}{cccc}
0&e^{- h }&- e^{ h }&0 \\
a e^{h }&0&0&-\frac{e^{h}}{a}\\
-\frac{e^{- h}}{a}&0&0&a e^{- h }\\
0&-e^{- h}&e^{h}&0
\end{array} & & \begin{array}{cccc}
e^{-h}~&0~&0&0 \\
0~&a e^{-h}~&0&0\\
0~&0~&a e^{-h}&0\\
0~&0~&0&a^2 e^{-h}
\end{array} \\
 & & & & & &\\
 
\textbf{(0)} &  &\begin{array}{cccc}
-\frac{e^{-h}}{a}&0&0&0 \\
0&-e^{-h}&0&0\\
0&0&-e^{-h}&0\\
0&0&0&-a e^{-h}
\end{array} & &\begin{array}{cccc}
a e^{h}~~~~&0~~~&0~~~~&0 \\
0~~~~& e^{h}~~~&0~~~~&0\\
0~~~~&0~~~&e^{h}~~~~&0\\
0~~~~&0~~~&0~~~~&\frac{e^{h}}{a}
\end{array} & & \begin{array}{cccc}
0&a e^{-h}&-\frac{e^{ h}}{a}&0 \\
e^{h }&0&0&-\frac{e^{h}}{a^2}\\
-e^{- h}&0&0&a^2 e^{- h }\\
0&-a e^{- h }&\frac{e^{h}}{a}&0
\end{array} \\

\end{array} \right),
\label{matrice_V}
\]

\end{small}
where $a=\exp(i\pi \alpha)$, $|\phi_i\rangle^T=|\phi_f\rangle^T=(1,0,0,0,0,0,0,0,0,0,0,0,0,0,0,0)$. It is also possible to compute for example the probability to arrive finally in the state $|+_z,-_z\rangle$. This can be done by taking $|\phi_f\rangle^T=(0,0,0,1,0,0,0,0,0,0,0,0,0,0,0,0)$. Similarly, one can compute the dynamics for another initial state. One can consider for example an initial density matrix given by Eq. (\ref{eq:density_matrix_two_spins}) of the main text. This corresponds to $|\phi_i\rangle^T=1/4(1,-1,-1,1,-1,1,1,-1,-1,1,1,-1,1,-1,-1,1)$.\\

We can use the scaling regime simplification exposed above, even when we do not have $\Delta/\omega_c \ll 1$. We write $Q_1(t)=\pi^2 \alpha+[Q_1(t)-\pi^2 \alpha]$ and we take into account the constant part as exposed above. The remaining part $[Q_1(t)-\pi^2 \alpha]$ is then decomposed into Fourier series.\\

\begin{figure*}[b!]
  \includegraphics[scale=0.4]{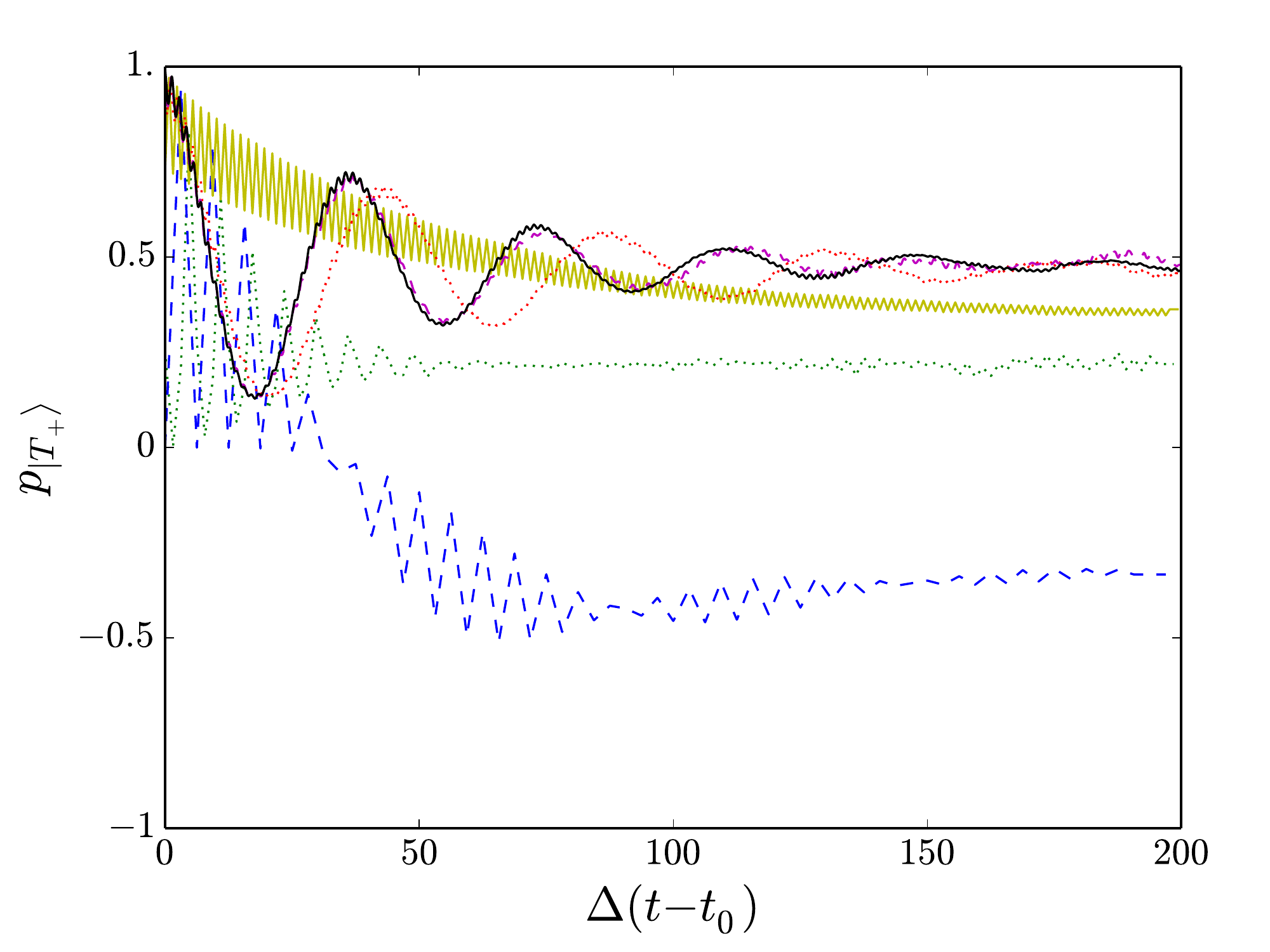}
  \caption{(Color online) Time evolution of $p_{|T_+\rangle} (t)$ for the dimer being initially in the state $|T_+\rangle$, for $N=6$ (dashed blue line), $N=7$ (dotted green line), $N=8$ (full yellow line), $N=9$ (dotted red line), $N=10$ (dashed purple line), and $N=11$ (full black line). Parameters are $\alpha=0.02$, $\omega_c=100$, and $K=0$.}
  \label{discretization_dynamics}
\end{figure*}

As we use a Fourier decomposition, we choose the same discretization step in time and in frequency, and take 2$^N$ points. In Fig.~\ref{discretization_dynamics}, we show the numerical convergence concerning the dynamics of $p_{|T_+\rangle} (t)$ for the dimer problem with initial condition $|T_+\rangle$ (see III B of the main text), with $\alpha=0.02$, $\omega_c=100$, $K=0$, for $N$ from $6$ to $11$. For $N>11$, all the curves give the same result (superposed to the full black curve).
In the regime $\alpha>\alpha_c/2$, one finds the existence of a ``sweet spot" which links the final time of the simulation and $\alpha$, for a given discretization.  




\section{Thermodynamic analysis of the action for the dissipative Ising model in transverse field}

The mean-field dynamics is not affected by the presence of the bath. This behavior can be understood thanks to a thermodynamic analysis of the action at low wave-vectors $q$ and low frequency $\omega$, which is dominated by the peaked contribution at $q=0$ of the long range Ising interaction. Using a mapping to a classical Ising model, it is possible to estimate the spin-spin coupling due to the environment, by focusing on the partition function (path integral approach) and tracing out the environmental modes\newline

  \begin{align}
\int D(b,b^*) e^{-S}=\exp\left\{\frac{1}{4 \pi} \int_0^{\beta} d\tau \int_0^{\beta} d\tau'  \sum_{j,r} \underbrace{ \int_0^{\infty}d\omega  J(\omega)\left[ e^{-\omega |\tau-\tau'|} +2 n_{\mathcal{B}}(\omega) \cosh \omega (\tau-\tau') \right] \cos\left(\omega\frac{x_j-x_r}{v_s}\right)}_{B(\tau-\tau',x_j-x_r)}\sigma_{j}(\tau) \sigma_{r}(\tau')\right\},
\label{final_2}
 \end{align}
 where the $\sigma_{j}$ are the classical spin variables corresponding to the eigenvalues of the quantum operators $\sigma^z_j$, and $\tau$ is the imaginary time. At zero temperature, we have
\begin{align}
B(\tau-\tau',x_j-x_r)=&Re\left[ \frac{2\pi \alpha \omega_c^2 }{\left(1+\omega_c|\tau-\tau'|+i \frac{x_j-x_r}{\xi} \right)^2} \right],
 \end{align}
  where $\xi=v_s/\omega_c$, then modifying the coupling between the spins. On the other hand, the direct Ising coupling is responsible for a coupling term of the form 
\begin{align}
C(\tau-\tau',x_j-x_r)=\frac{K}{M}\delta(\tau-\tau'),
 \end{align}
and the constant behavior in the space domain dominates in the low $q$, low $\omega$ expansion of the action. 

The mean-field coupling then dominates over the dissipative effects and we find back the characteristic features of the mean-field transition of the quantum Ising model in transverse field. This mean field behavior is valid as long as the direct Ising term $K$ is {\it not} zero.

\bibliographystyle{apsrev4-1}

\begin{thebibliography}{99}






\bibitem{haroche}
J.-M. Raimond, M. Brune and S. Haroche, Rev. Mod. Phys. \textbf{73}, 565 (2001).

\bibitem{leggett}
A. J. Leggett, S. Chakravarty, A. T. Dorsey, M. P. A. Fisher, A. Garg and W. Zwerger, Rev. Mod. Phys, \textbf{59}, 1 (1987).

\bibitem{weiss}
U. Weiss, Quantum dissipative systems, World Scientific, Singapore (2002).

\bibitem{KLH}
K. Le Hur, Annals of Phyics \textbf{323} 2208-2240 (2008).

\bibitem{Caldeira_Leggett}
A. O. Caldeira and A. J. Leggett, Physica 121A: 587 (1983).

\bibitem{Pierre}
S. Jezouin, M. Albert, F. D. Parmentier, A. Anthore,	 U. Gennser, A. Cavanna, I. Safi and F. Pierre, Nat. Commun. \textbf{4}, 1802 (2013).

\bibitem{Finkelstein}
H. T. Mebrahtu, I. V. Borzenets, D. E. Liu, H. Zheng, Y. V. Bomze, A. I. Smirnov, H. U. Baranger and G. Finkelstein, Nature \textbf{488}, p. 61 (2012).

\bibitem{InesSaleur}
I. Safi and H. Saleur, Phys. Rev. Lett. {\bf 93}, 126602 (2004). 

\bibitem{KLH2}
K. Le Hur,  Phys. Rev. Lett. {\bf 92}, 196804 (2004).

\bibitem{De_Gennes}
P. G. de Gennes Solid State Commun. \textbf{1}, 132 (1963).

\bibitem{Pfeuty}
P. Pfeuty, Annals of Physics \textbf{57}, 79-90 (1970).

\bibitem{sachdev}
S. Sachdev, Quantum phase transitions, Cambridge University Press (1999).

\bibitem{Pankov}
S. Pankov, S. Florens, A. Georges, G. Kotliar, and S. Sachdev, Phys. Rev. B \textbf{69}, 054426 (2004).

\bibitem{sachdev_werner_troyer}
S. Sachdev, P. Werner and M. Troyer, Phys. Rev. Lett. \textbf{92}, 237003 (2004).

\bibitem{werner_volker_troyer_chakravarty}
P. Werner, K. Volker, M. Troyer and S. Chakravarty, Phys. Rev. Lett. \textbf{94}, 047201 (2005).

\bibitem{DMFT}
A. Georges, G. Kotliar, W. Krauth, and M. J. Rozenberg, Rev. Mod. Phys. \textbf{68}, 13 (1996).

\bibitem{Anderson_Yuval_Hamann}
P. W. Anderson, G. Yuval, and D. R. Hamann, Phys. Rev. B \textbf{1}, 4464 (1970).

\bibitem{Blume_Emery_Luther}
M. Blume, V. J. Emery, and A. Luther, Phys. Rev. Lett. \textbf{25}, 450 (1970).


\bibitem{Schiro}
M. Schiro and M. Fabrizio, Phys. Rev. B {\bf 79}, 153302 (2009).

\bibitem{Millis}
P. Werner, T. Oka, M. Eckstein and J. Millis, Phys. Rev. B {\bf 81}, 035108 (2010).

\bibitem{Gull_Millis_Lichtenstein_Rubtsov}
E. Gull, A. J. Millis, A. I. Lichtenstein, A. N. Rubtsov, M. Troyer, and P. Werner, Rev. Mod. Phys., \textbf{83}, 349 (2011).

\bibitem{Schmidt_Werner_Muhlbacher_Komnik}
T. L. Schmidt, P. Werner, L. M\"{u}hlbacher, and A. Komnik, Phys. Rev. B \textbf{78}, 235110 (2008).

\bibitem{Peter_two_spins}
P. P. Orth, D. Roosen,  W. Hofstetter and K. Le Hur, Phys. Rev. B {\bf 82}, 144423 (2010).

\bibitem{TDNRG}
R. Bulla, H.-J. Lee, N.-H. Tong and M. Vojta, Phys. Rev. B {\bf 71}, 045122 (2005).

\bibitem{TDNRG_Anders_Schiller}
F. B. Anders and A. Schiller, Phys. Rev. B \textbf{74}, 245113 (2006).

\bibitem{TDNRG_1}
R. Bulla, T. A. Costi and T. Pruschke, Rev. Mod. Phys. 80, \textbf{395} (2008).

\bibitem{TDNRG_quench}
H. T. M. Nghiem and T. A. Costi, Phys. Rev. B \textbf{90}, 035129 (2014).


\bibitem{Dalibard_Castin_Molmer}
J. Dalibard, I. Castin and K. Molmer, Phys. Rev. Lett. \textbf{68}, 580 (1992).

\bibitem{2010stoch}
P. P. Orth, A. O. Imambekov and K. Le Hur, Phys. Rev. A \textbf{82}, 032118 (2010).

\bibitem{stochastic}
P. P. Orth, A. O. Imambekov and K. Le Hur, Phys. Rev. B \textbf{87}, 014305 (2013).

\bibitem{Rabi_article}
L. Henriet, Z. Ristivojevic, P. P. Orth and K. Le Hur, Phys. Rev. A \textbf{90}, 023820 (2014).


\bibitem{Stockburger_Mac}
J. T. Stockburger and C. H. Mac, J. Chem. Phys. \textbf{110}, 4983-4985 (1999).

\bibitem{Stockburger}
J. T. Stockburger and H. Grabert, Phys. Rev. Lett. \textbf{88}, 170407 (2002).

\bibitem{Stockburger_2}
J. T. Stockburger and H. Grabert, Chem. Phys. \textbf{296}, p. 159 (2004).


\bibitem{Schrodinger_langevin}
R. Katz and P. B. Gossiaux, arXiv:1504.08087 (2015).

\bibitem{Koch_morse}
W. Koch, F. Gro{\ss}mann, J. T. Stockburger, and J. Ankerhold, Phys. Rev. Lett. \textbf{100}, 230402 (2008).


\bibitem{recati_fedichev}
A. Recati, P. O. Fedichev, W. Zwerger, J. von Delft and P. Zoller, Phys. Rev. Lett. \textbf{94}, 040404 (2005).

\bibitem{orth_stanic_lehur}
P. P. Orth and I. Stanic and K. Le Hur, Phys. Rev. A {\bf 77}, 051601(R) (2008).

\bibitem{Carlos_scientific_reports}
C. Sabin, A. White, L. Hackermuller and I. Fuentes, Nature Scientific Reports, Vol. \textbf{4}, id. 6436 (2014).

\bibitem{Garst_Vojta}
M. Garst, S. Kehrein, T. Pruschke, A. Rosch and M. Vojta, Phys. Rev. B 69, 214413 (2004).

\bibitem{sougato}
D. P. S. McCutcheon, A. Nazir, S. Bose and A. J. Fisher, Phys. Rev. B {\bf 81}, 235321 (2010).

\bibitem{Winter_Rieger}
A. Winter and H. Rieger, Phys. Rev. B \textbf{90}, 224401 (2014).

\bibitem{Landau}
L. Landau, Physics of the Soviet Union \textbf{2}, 46 (1932).

\bibitem{Zener}
C. Zener, Proc. R. Soc. of London A \textbf{137}, 696 (1932).

\bibitem{Stueckelberg}
E. C. G. Stueckelberg, Helvetica Physica Acta \textbf{5}, 369 (1932).

\bibitem{Majorana}
E. Majorana, Nuovo Cimento \textbf{9}, 43 (1932).

\bibitem{Kibble}
T. W. B. Kibble, J. Phys. A \textbf{9}, 1387 (1976); Phys. Rep. \textbf{67}, 183 (1980).

\bibitem{Zurek}
W. H. Zurek, Nature (London) \textbf{317}, 505 (1985); Acta Phys. Pol. B \textbf{24}, 1301 (1993); Phys. Rep. \textbf{276}, 177 (1996).

\bibitem{dzarmaga}
J. Dziarmaga, Advances in Physics, vol. \textbf{59}, issue 6, pp. 1063-1189 (2010).

\bibitem{review_KBZ}
A. Del Campo and W. Zurek, Int. J. Mod. Phys. A \textbf{29}, 1430018 (2014).

\bibitem{damski}
B. Damski, Phys. Rev. Lett. \textbf{95}, 035701 (2005).

\bibitem{Garry_Camille}
G. Goldstein, C. Aron and C. Chamon,   Phys. Rev. B \textbf{92}, 174418 (2015).

\bibitem{Clerk}
P. Nalbach, S. Vishveshwara, and A. A. Clerk,  Phys. Rev. B \textbf{92}, 014306 (2015).

\bibitem{Hadzibabic}
N. Navon, A. L. Gaunt, R. P. Smith and Z. Hadzibabic, Science \textbf{347}, 167-170 (2015).

\bibitem{schaetz}
A. Friedenauer, H. Schmitz, J. T. Glueckert, D. Porras and T. Schaetz, Nat. Phys. \textbf{4}, 757 (2008).

\bibitem{Cirac_spinboson}
D. Porras, F. Marquardt, J. von Delft, and J.I. Cirac, Phys. Rev. A (R) \textbf{78}, 010101 (2008).

\bibitem{monroe}
R. Islam, E. Edwards, K. Kim, S. Korenblit, C. Noh, H. Carmichael, G.-D. Lin, L.-M. Duan, C.-C. J. Wang, J. Freericks, C. Monroe, Nat. Commun. \textbf{2}, 377 (2011).

\bibitem{Scelle}
R. Scelle, T. Rentrop, A. Trautmann, T. Schuster, and M. K. Oberthaler, Phys. Rev. Lett. \textbf{111}, 070401 (2013).

\bibitem{Sortais}
Y. R. P. Sortais, H. Marion, C. Tuchendler, A. M. Lance, M. Lamare, P. Fournet, C. Armellin, R. Mercier, G. Messin, A. Browaeys, and P. Grangier, Phys. Rev. A \textbf{75}, 013406 (2007).

\bibitem{rydberg_1}
L. B\'{e}guin, A. Vernier, R. Chicireanu, T. Lahaye, and A. Browaeys, Phys. Rev. Lett. \textbf{110}, 263201 (2013).

\bibitem{rydberg_2}
A. Grankin, E. Brion, E. Bimbard, R. Boddeda, I. Usmani, A. Ourjoumtsev and P. Grangier, New J. Phys. \textbf{16} 043020 (2014).

\bibitem{rydberg_3}
M. Marcuzzi, E. Levi, S. Diehl, J. P. Garrahan and I. Lesanovsky, Phys. Rev. Lett. \textbf{113}, 210401 (2014).

\bibitem{Giamarchi}
T. Giamarchi, Quantum Physics in One Dimension, Oxford, Oxford University Press 2004.

\bibitem{Si}
Q. Si, Chapter of the book ``Understanding Quantum Phase Transitions'', ed. Lincoln D. Carr (CRC Press/Taylor \& Francis, Boca Raton, 2010).

\bibitem{RKKY}
M. A. Ruderman and C. Kittel, Phys. Rev. {\bf 96}, 99 (1954); T. Kasuya, Prog. Theor. Phys. {\bf 16}, 45 (1956); K. Yosida, Phys. Rev. {\bf 106}, 893 (1957).

\bibitem{Senellart}
A. Dousse, L. Lanco, J. Suffczynski, E. Semenova, A. Miard, A. Lemaitre, I. Sagnes, C. Roblin, J. Bloch and P. Senellart, Phys. Rev. Lett. {\bf 101}, 267404 (2008).

\bibitem{Majer}
 J. M. Chow, J. M. Gambetta, Jens Koch, B. R. Johnson, J. A. Schreier, L. Frunzio, D. I. Schuster, A. A. Houck, A. Wallraff, A. Blais, M. H. Devoret, S. M. Girvin, and R. J. Schoelkopf, Nature {\bf 449}, 443-447 (2007).

\bibitem{Kontos}
M. R. Delbecq,  L.E. Bruhat, J.J. Viennot, S. Datta, A. Cottet and T. Kontos, Nat. Commun. {\bf 4}, 1400 (2013).

\bibitem{FV}
R. P. Feynman and F. L. Vernon, Ann. Phys. (N.Y.) \textbf{24}, 118 (1963).

\bibitem{Grabert_Schramm_Ingold}
H. Grabert, P. Schramm, G. L. Ingold, Phys. Rep. \textbf{168}, 115 (1988).

\bibitem{Lesovik}
G. B. Lesovik, A. O. Lebedev and A. O. Imambekov, JETP Lett. \textbf{75}, 474 (2002).

\bibitem{Tu_Zhang}
M. W. Y. Tu and W.-M. Zhang, Phys. Rev. B \textbf{78}, 235311 (2008).

\bibitem{Zhang_Nori}
W.-M. Zhang, P.-Y. Lo, H.-N. Xiong, M. W.-Y. Tu, and F. Nori, Phys. Rev. Lett. \textbf{109}, 170402 (2012).

\bibitem{de_Vega_review}
I. de Vega, D. Alonso, arXiv:1511.06994 (2015).

\bibitem{Lesage_Saleur}
F. Lesage, and H. Saleur, Phys. Rev. Lett. \textbf{80}, 4370 (1998).

\bibitem{KLHQPT}
K. Le Hur, chapter in the book  ``Understanding Quantum Phase Transitions'', edited by Lincoln D. Carr (Taylor and Francis, Boca Raton, 2010).

\bibitem{Wang_Thoss}
H. Wang, and M. Thoss, New J. Phys. \textbf{10}, 115005 (2008).

\bibitem{Kashuba_Schoeller}
O. Kashuba, M. Kennes, M. Pletyukhov, V. Meden, and H. Schoeller, Phys. Rev. B \textbf{88}, 165133 (2013).

\bibitem{essler}
P. Calabrese, F. H. L. Essler, and M. Fagotti, Phys. Rev. Lett. \textbf{106}, 227203 (2011).

\bibitem{gambasi}
L. Foini,  L. F. Cugliandolo and A. Gambassi, J. Stat. Mech. P09011 (2012).

\bibitem{delcampo}
A. Del Campo and W. H. Zurek, Int. J. Mod. Phys. A \textbf{29}, 1430018 (2014).

\bibitem{roux_kollath}
J-S Bernier, D. Poletti, P. Barmettler, G. Roux and C. Kollath, Phys. Rev. A \textbf{85}, 033641 (2012).

\bibitem{sciolla_biroli}
B. Sciolla and G. Biroli, Phys. Rev. B 88, 201110(R)  (2013).

\bibitem{rancon}
A. Rancon, Chen-Lung Hung, Cheng Chin, and K. Levin Phys. Rev. A {\bf 88} 031601(R) (2013).

\bibitem{Silbey_Harris}
R. Silbey, and R. A. Harris, J. Chem. Phys. \textbf{80}, 2615 (1984).


\bibitem{Florens}
S. Bera, A. Nazir, A. W. Chin, H. U. Baranger, S. Florens, Phys. Rev. B \textbf{90}, 075110 (2014).

\bibitem{synchronization}
A. S. Pikovsky, M. Rosenblum, and J. Kurths, Synchronization: A Universal Concept in Nonlinear Science (Cambridge University Press, New York, 2001).

\bibitem{synchronization_fuchs}
Y. Liu, F. Pi\'{e}chon, and J. N. Fuchs, Europhys. Lett. \textbf{103}, 17007 (2013).

\bibitem{synchronization_salomon}
M. Delehaye, S. Laurent, I. Ferrier-Barbut, S. Jin, F. Chevy, C. Salomon, arXiv:1510.06709 (2015).

\bibitem{kayanuma_1}
M. Wubs, K. Saito, S. Kohler, P. Hanggi and Y. Kayanuma, Phys. Rev. Lett. \textbf{97}, 200404 (2006).

\bibitem{kayanuma_2}
K. Saito, M. Wubs, S. Kohler, Y. Kayanuma and P. Hanggi, Phys. Rev. B \textbf{75}, 214308 (2007).

\bibitem{su3}
M. N. Kiselev, K. Kikoin and M. B. Kenmoe, Europhys. Lett. \textbf{104}, 57004 (2013).

\bibitem{LZSM_shevshenko}
S. N. Shevchenko, S. Ashhab and F. Nori, Phys. Rept. \textbf{492}, 1 (2010).

\bibitem{Garrahan}
I. Lesanovsky,  M. van Horssen, M. Guta and J. P. Garrahan, Phys. Rev. Lett. {\bf 110}, 150401 (2013).

\bibitem{Marco}
M. Schir\' o,  C. Joshi, M. Bordyuh, R. Fazio, J. Keeling, and H. E. T\" ureci, arXiv:1503. 04456 (2015).

\bibitem{Sanchez}
E. Sanchez-Burillo, D. Zueco, J. J. Garcia-Ripoll and L. Martin-Moreno, Phys. Rev. Lett. {\bf 113}, 263604 (2014).

\bibitem{Keeling}
G. Kulaitis, F. Kr\"{u}ger, F. Nissen and J. Keeling, Phys. Rev. A {\bf 87}, 013840 (2013).


\bibitem{sengupta_powell_sachdev}
K. Sengupta, S. Powell and S. Sachdev, Physical Review A \textbf{69}, 5 (2004).


\bibitem{Bulla}
F. Anders, R. Bulla and M. Vojta, Phys. Rev. Lett. {\bf 98}, 210402 (2007).

\bibitem{Doucet}
K. Le Hur, P. Doucet-Beaupr\' e and W. Hofstetter  Phys. Rev. Lett. {\bf 99}, 126801 (2007).

\bibitem{review}
K. Le Hur, L. Henriet, A. Petrescu, K. Plekhanov, G. Roux and M. Schiro, arXiv:1505.00167.

\bibitem{Montambaux}
L.-K. Lim, J.-N. Fuchs and G. Montambaux, Phys. Rev. Lett. {\bf 108}, 175303 (2012)  and Phys. Rev. Lett. {\bf 112}, 155302 (2014).

\bibitem{DemlerKondo}
J. Bauer, C. Salomon and E. Demler,  Phys. Rev. Lett. {\bf 111}, 215304 (2013).


\bibitem{Brandes_course}
T. Brandes, Chapter 7 of UMIST-Bradford Lectures on Background to Quantum Information Theory (2004).








\end{thebibliography}

\end{document}